\documentclass[10pt,journal,compsoc]{IEEEtran}

\ifCLASSOPTIONcompsoc
  \usepackage[nocompress]{cite}
\else
  \usepackage{cite}
\fi

\ifCLASSINFOpdf
\else
\fi

\usepackage{graphicx,enumerate}
\usepackage{caption}
\usepackage{subcaption}
\usepackage{graphicx}
\graphicspath{{figures/}}
\usepackage{mathtools,commath,amsmath,amssymb,amsfonts,bm}\interdisplaylinepenalty=2500
\usepackage{amsthm}
\usepackage{algpseudocode,algorithm}
\usepackage{tabularx,adjustbox}
\usepackage{todonotes}
\usepackage{breqn}
\usepackage{color,cite}
\clearpage
\extrafloats{100}
\RequirePackage{tcolorbox}
\tcbuselibrary{breakable}
\tcbset{parbox=false, left=1ex, right=1ex}

\usepackage{multirow}
\usepackage{xspace}
\usepackage{url}
\newcolumntype{C}[1]{>{\hsize=#1\hsize\centering\arraybackslash}X}

\def\vec#1{\mathchoice{\mbox{\boldmath$\displaystyle#1$}}
  {\mbox{\boldmath$\textstyle#1$}}
  {\mbox{\boldmath$\scriptstyle#1$}}
  {\mbox{\boldmath$\scriptscriptstyle#1$}}}

\newcommand{\etal}{\textit{et~al.}\xspace}
\newcommand{\ie}{\textit{i.e.},\xspace}
\newcommand{\eg}{\textit{e.g.},\xspace}
\newcommand{\mech}{\ensuremath{\mathcal{M}}\xspace}

\newtheorem{lemma}{Lemma}

\newtheorem{definition}{Definition}

\algnewcommand\algorithmicinput{\textbf{Input:}}
\algnewcommand\Input{\item[\algorithmicinput]}
\algnewcommand\algorithmicoutput{\textbf{Output:}}
\algnewcommand\Output{\item[\algorithmicoutput]}
\algnewcommand\algorithmictier{\textbf{Step:}}
\algnewcommand\Tier{\item[\algorithmictier]}

\begin{document}
\title{Towards Fair and Privacy-Preserving Federated Deep Models}
\author{Lingjuan~Lyu$^{*}$,~\IEEEmembership{Member,~IEEE},
        Jiangshan~Yu, Karthik~Nandakumar,~\IEEEmembership{Senior Member,~IEEE},
        Yitong~Li, Xingjun~Ma, 
        Jiong~Jin,~\IEEEmembership{Member,~IEEE},
        Han Yu$^{*}$,
        and
        Kee Siong Ng
\IEEEcompsocitemizethanks{\IEEEcompsocthanksitem 
L. Lyu is with The Department of Computer Science, National University of Singapore. E-mail: lyulj@comp.nus.edu.sg.
%lingjuanlvsmile@gmail.com. 
\IEEEcompsocthanksitem J. Yu is with the Faculty of Information Technology, Monash University, Clayton, Australia. E-mail:jiangshan.yu@monash.edu.
\IEEEcompsocthanksitem K. Nandakumar is with IBM Singapore Lab, 018983. E-mail:nkarthik@sg.ibm.com.
\IEEEcompsocthanksitem Y. Li and X. Ma are with the School of Computing and Information Systems, The University of Melbourne, Melbourne, Australia, 3010. E-mail: yitongl4@student.unimelb.edu.au; xingjun.ma@unimelb.edu.au.
\IEEEcompsocthanksitem J. Jin is with the School of Software and Electrical Engineering, Swinburne University of Technology, Melbourne, Australia. E-mail: jiongjin@swin.edu.au.
\IEEEcompsocthanksitem Han Yu is with the School of Computer Science and Engineering, Nanyang Technological University, Singapore. E-mail: han.yu@ntu.edu.sg.
\IEEEcompsocthanksitem Kee Siong Ng is with the Software Innovation Institute, Australian National University, Australia. E-mail: keesiong.ng@anu.edu.au.}
\thanks{$^{*}$Corresponding authors.}
}

\markboth{IEEE Transactions on Parallel and Distributed Systems}%
{Lyu \MakeLowercase{\textit{\etal}}: Towards Fair and Privacy-Preserving Federated Deep Models}

\IEEEtitleabstractindextext{\begin{abstract}
The current standalone deep learning framework tends to result in overfitting and low utility. This problem can be addressed by either a centralized framework that deploys a central server to train a global model on the joint data from all parties, or a distributed framework that leverages a parameter server to aggregate local model updates. Server-based solutions are prone to the problem of a single-point-of-failure. In this respect, collaborative learning frameworks, such as federated learning (FL), are more robust. Existing federated learning frameworks overlook an important aspect of participation: fairness. All parties are given the same final model without regard to their contributions. 
To address these issues, we propose a decentralized Fair and Privacy-Preserving Deep Learning (FPPDL) framework to incorporate fairness into federated deep learning models. In particular, we design a local credibility mutual evaluation mechanism to guarantee fairness, and a three-layer onion-style encryption scheme to guarantee both accuracy and privacy. Different from existing FL paradigm, under FPPDL, each participant receives a different version of the FL model with performance commensurate with his contributions. Experiments on benchmark datasets demonstrate that FPPDL balances fairness, privacy and accuracy. It enables federated learning ecosystems to detect and isolate low-contribution parties, thereby promoting responsible participation.
\end{abstract}

}

\maketitle

\IEEEdisplaynontitleabstractindextext

\IEEEpeerreviewmaketitle

%%%%%%%%%%%%%%%%%%%%%%%%%%%%%%%%%%%%%%%%%%%%%%555
\IEEEraisesectionheading{\section{Introduction}\label{sec:introduction}}
\IEEEPARstart{D}{eep} learning has become an important technology to deal with challenging real-world problems such as image classification and speech recognition. Empirical evidence has shown that deep learning models can benefit significantly from large-scale datasets~\cite{krizhevsky2012imagenet}. However, large-scale datasets are not always available for a new domain, due to the significant time and effort required for data collection and annotation~\cite{wang2018iterative,ma2018dimensionality}. 
Moreover, training complex deep networks on large-scale datasets is computationally expensive and may not be feasible for a single party in practice. Therefore, there is a high demand to perform deep learning in a collaborative manner among a group of parties. 

This trend is motivated by the fact that the data owned by a single party may be very homogeneous, resulting in overfitting which negatively impacts accuracy when the model is applied to previously unseen data, \ie poor generalizability. Utilizing data from diverse parties to train deep models can help mitigate this problem. However, collaborative model training may not be viable due to privacy concerns. Federated learning (FL), which incorporates privacy preservation techniques into collaborative model training, offers a potential solution to this challenge~\cite{FL2019}.

In the current federated learning paradigm~\cite{mcmahan2016federated}, all participants receive the same federated model at the end of collaborative model training regardless of their contributions. This makes the paradigm vulnerable to free-riding participants. For example, several banks may want to work together to build model to predict the creditworthiness of small and medium enterprises. However, but larger banks with more data maybe reluctant to train their local model based on high quality local data for fear of smaller banks benefiting from the shared FL model and eroding its market share~\cite{FL2019}. Without the guarantee of privacy and the promise of collaborative fairness, participants with high quality and large datasets may be discouraged from joining federated learning, thereby negatively affect the formation of a healthy FL ecosystem. Existing research on fairness mostly focuses on protecting sensitive attributes or reducing the variance of the prediction distribution across participants~\cite{cummings2019compatibility,jagielski2018differentially}. The problem of treating federated learning participants fairly remains open~\cite{FL2019}.

In this paper, we address the problem of treating FL participants fairly based on their contributions to build a healthy FL ecosystem. We refer to the proposed framework as the decentralized Fair and Privacy-Preserving Deep Learning (FPPDL) framework. Unlike existing work such as~\cite{Yu-et-al:2020AIES} which uses monetary rewards to incentivize good behaviour, our proposed solution fundamentally changes the current FL paradigm so that participants may not receive the same FL model in the end. Instead, each of them will receive a final FL model with performance reflecting their individual contributions to the federation. FPPDL does not require participants to trust each other or any third party. It records all operations, including uploading and downloading \textit{differentially private artificial samples} and \textit{encrypted model updates}, as transactions through blockchain technology. Through mutual evaluations of local credibility that considers the relative contribution of each party during both initial benchmarking and privacy-preserving collaborative deep learning, FPPDL achieves collaborative fairness. For privacy preservation, instead of leveraging differential privacy at the cost of utility, we propose a three-layer onion-style encryption scheme to guarantee accuracy and privacy.

To the best of our knowledge, this paper is the first to achieve collaborative fairness in federated learning through adjusting the level of performance of the version of the FL model allocated to each participant based on his contribution. Extensive experiments based on two benchmark datasets under three realistic settings demonstrate that FPPDL achieves high fairness, delivers comparable accuracy to existing centralized and distributed deep learning frameworks, and outperforms standalone deep learning.

In terms of the threat model, FPPDL adopts an honest-but-curious setting: each party is assumed to be curious in inferring sensitive information of other parties; and yet, it is assumed to be honest in operations. This setting is reasonable as the main incentive for parties to participant in the collaborative system is to get better local models compared to their standalone models without any collaboration. Moreover, in our scenario parties are considered as organisations such as financial or biomedical institutions acting with responsibilities by laws. However, we also discuss how our local credibility mutual evaluation mechanism can help prevent certain malicious behaviours of the insider attacker, and resist the outsider attacker in Section~\ref{sec:Discussion}.

The rest of this paper is organized as follows. Section~\ref{sec:Related_work} reviews the existing deep learning frameworks, and the related literatures on privacy-preserving collaborative deep learning, and fairness in federated learning which are major problems we aim to tackle. Section~\ref{sec:FPPDL} presents technical details of the proposed FPPDL framework. Section~\ref{sec:Performance} evaluates the performance of FPPDL in terms of accuracy and fairness for different SGD frameworks under different settings, followed by discussions in Section~\ref{sec:Discussion}. Section~\ref{sec:Conclusion} concludes the paper and points out potential future research directions. 

%%%%%%%%%%%%%%%%%%%%%%%%%%%%%%%%%%%%%%%%
\section{Related Work}
\label{sec:Related_work}
In this section, we review relevant literature on deep learning frameworks, privacy preservation and fairness in federated learning to position our research in relation to existing research.

% %%%%%%%%%%%%%%%%%%%%%%%%%%%%%%%%%%%%%
\begin{figure*}[!htp]
\centering
        \begin{subfigure}[ht]{0.23\textwidth} \includegraphics[width=3cm,height=2.3cm]{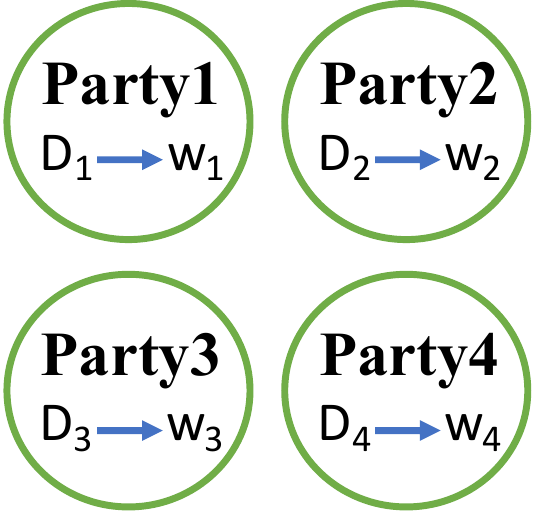}
				\label{fig:standalone}
        \end{subfigure}
        \begin{subfigure}[ht]{0.23\textwidth} \includegraphics[width=3.5cm,height=2.3cm]{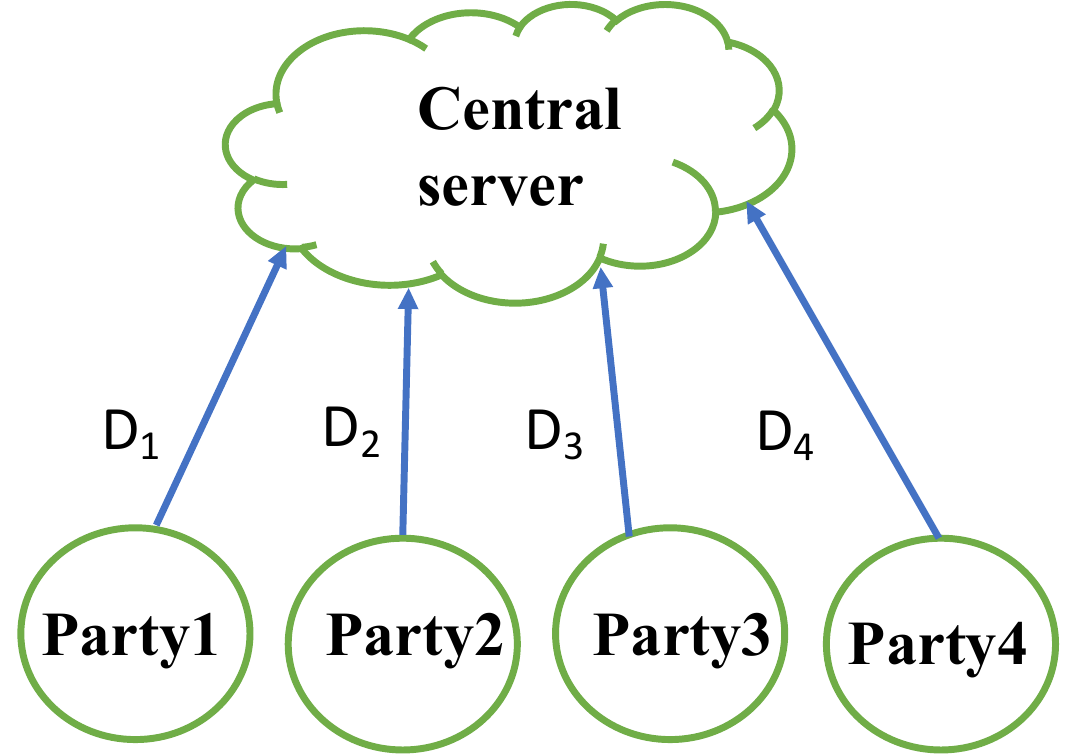}
				\label{fig:centralized}
        \end{subfigure}
        \begin{subfigure}[ht]{0.23\textwidth} \includegraphics[width=3.5cm,height=2.3cm]{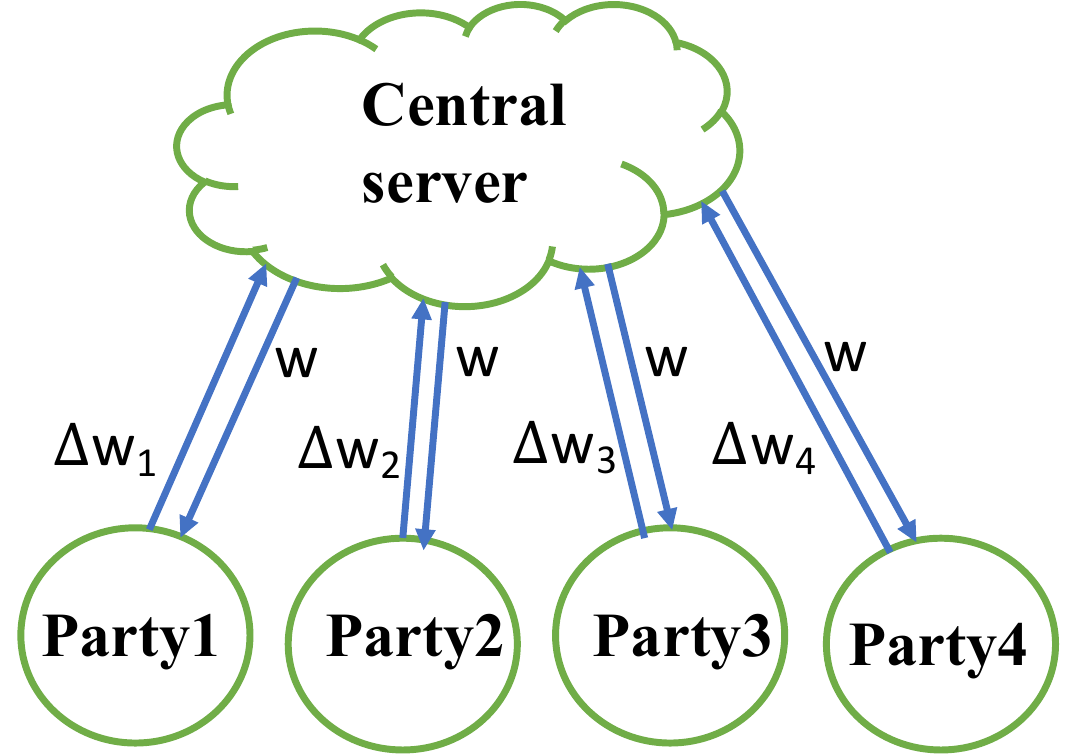}
				\label{fig:distributed}
        \end{subfigure}
        \begin{subfigure}[ht]{0.23\textwidth} \includegraphics[width=3.5cm,height=2.3cm]{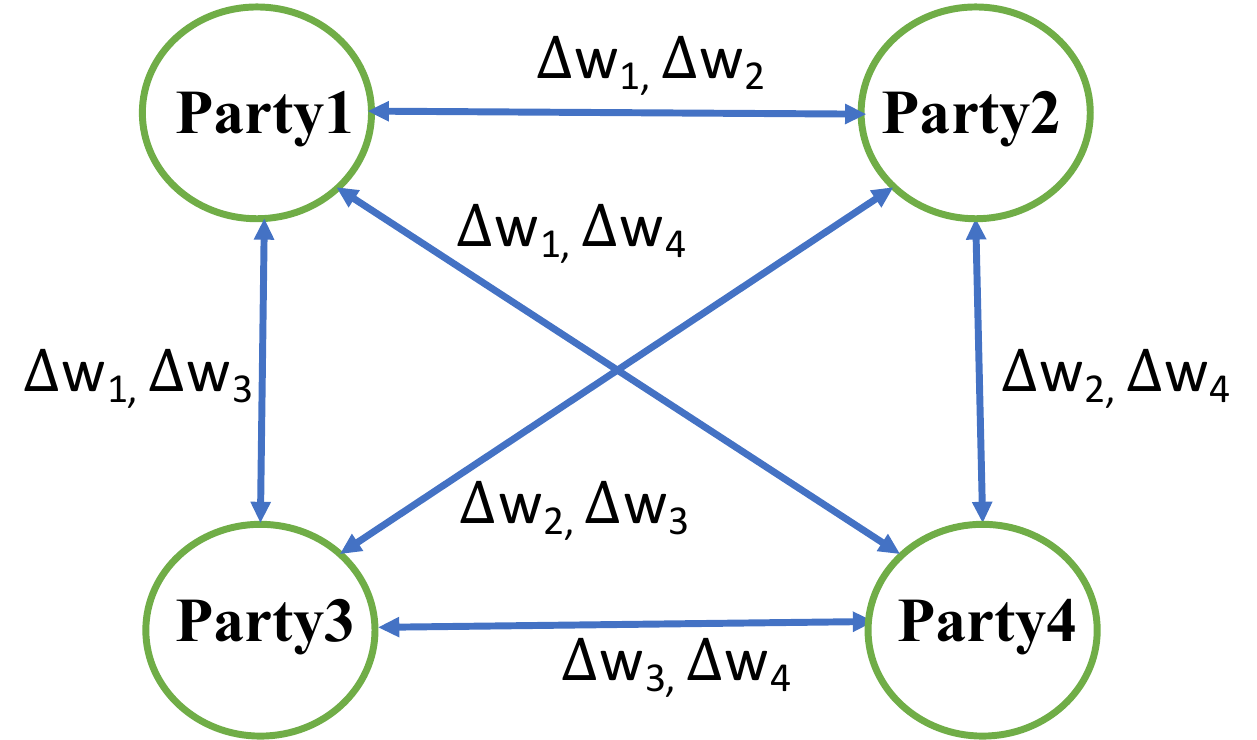}
				\label{fig:decentralized}
        \end{subfigure}
        \caption{(a): Standalone framework. (b): Centralized framework. (c): Distributed framework. (d): Decentralized framework.}
\label{fig:model}
\end{figure*}
% %%%%%%%%%%%%%%%%%%%%%%%%%%%%%%%%%%%%%
\subsection{Overview of Deep Learning Frameworks}
\label{sec:frameworks}
In general, deep learning frameworks fall into the following categories: \textit{Standalone framework}; server-based frameworks including \textit{Centralized framework} and \textit{Distributed framework}; and \textit{Decentralized framework}. In particular, in distributed framework and decentralized framework, parties are all involved in the global or consensus model improvement process. Hence, we refer to them as collaborative deep learning frameworks. A comparison among different deep learning frameworks is provided in Table~\ref{tbl:cc}.

\textit{\textbf{Standalone framework}}: Parties individually train standalone models on their local training data without any collaboration (Fig.~\ref{fig:model}(a)). However, standalone models might fail to generalise to the unseen data. 

\textit{\textbf{Centralized framework}}: Participants pool their data into a centralized server to train a global model (Fig.~\ref{fig:model}(b)). This centralized framework is very effective, but it violates data privacy as  all participants' data are exposed to the server. 

\textit{\textbf{Distributed framework}}: Dean \etal~\cite{dean2012large} first introduced the concept of distributed deep learning, where parties collaboratively train a model by sharing local model updates with a parameter server. Distributed learning had been extensively studied in~\cite{zinkevich2010parallelized,shokri2015privacy,mcmahan2016federated}. 

It should be noted that both the centralized framework and the distributed framework require a central server to mediate the training process, which makes them susceptible to the following issues: (1) Party policies: due to privacy concern, parties may not want to cede control to an untrusted server; (2) Single-point-of-failure: if the central server fails or is shut down for maintenance, the whole network stops working.

\textit{\textbf{Decentralized framework}}: the above issues in the central server-based frameworks can be addressed by a decentralized framework~\cite{Kuo2016ModelChain,chen2018machine,kim2019efficient,zhu2018blockchain,weng2019deepchain,kuo2019fair}, which parallelizes the computation among all parties (Fig.~\ref{fig:model}(d)). In particular, Kuo \etal~\cite{Kuo2016ModelChain} first proposed a decentralized machine learning framework: ModelChain, which integrates Blockchain with privacy-preserving machine learning.~\cite{zhu2018blockchain,weng2019deepchain} investigated privacy-preserving deep learning on blockchain. ~\cite{kuo2019fair} studied the problem of fairness in load sharing in blockchain-based privacy-preserving learning, which is different from the collaborative fairness in our work. Specifically, their decentralized architecture is developed under a less secure setting, where one site can access the models of all the other sites.~\cite{chen2018machine,kim2019efficient} utilized differential privacy for privacy-preserving machine learning on blockchain. However, ~\cite{kim2019efficient} had pointed out that the proposed algorithms in~\cite{chen2018machine} cannot guarantee privacy-preserving properties correctly as they did not consider composition for a repeated additive-noise mechanism.

In summary, existing collaborative frameworks (distributed or decentralized) focus on how to learn a global consensus model with higher accuracy than individual standalone models. In reality, some parties may contribute more compared with other parties, while others may contribute nearly nothing or even negatively. The reason is that data owned by different parties may be of different quality, and there may exist unpredictable random errors during data collection and storage. On the other hand, parties may choose only to use a limited part of its data for collaborative model training.
 
\begin{table*}[ht]
 \centering\caption{\label{tbl:cc}Comparing different deep learning frameworks.}
 \begin{tabularx}{\textwidth}{|p{3cm}|X|c|X|X|X|}
 \hline
 Frameworks & Standalone & Centralized~\cite{gilad2016cryptonets,ohrimenko2016oblivious} & Distributed~\cite{zinkevich2010parallelized,shokri2015privacy,mcmahan2016federated,mohassel2017secureml} & Decentralized~\cite{Kuo2016ModelChain,weng2019deepchain,zhu2018blockchain,kuo2019fair,chen2018machine,kim2019efficient} & Decentralized (our FPPDL)\\
 \hline\hline
    Architecture & Fig.~\ref{fig:model}(a) & Fig.~\ref{fig:model}(b) & Fig.~\ref{fig:model}(c) & Fig.~\ref{fig:model}(d) & Fig.~\ref{fig:model}(d) \\
 \hline
 Global model
    & No
    & Yes
    & Yes
    & Depends 
    & Depends \\
\hline
 Local models
    & Yes
    & No
    & Yes
    & Yes 
    & Yes \\
\hline
 Collaborative Fairness
    & NA
    & NA
    & No
    & No
    & Yes \\
 \hline
 Quality control
    & NA
    & No
    & No
    & No
    & Yes \\
 \hline
 \end{tabularx}
\end{table*}

\subsection{Privacy Preserving Collaborative Learning}
As pointed out by Shokri \etal~\cite{shokri2015privacy}, centralized deep learning commonly comes with many privacy concerns. Specifically, all the sensitive training data are revealed to a third party; data owners have no control over the learning objective; the learned model is not directly available to data owners. To mitigate these privacy risks, Gilad-Bachrach \etal~\cite{gilad2016cryptonets} developed CryptoNets to run deep learning on homomorphically encrypted data. However, CryptoNets assumes that neural network model has been trained beforehand, hence their system is mainly used to provide encrypted outputs to users. In contrast, SecureML~\cite{mohassel2017secureml} conducts privacy-preserving learning via secure multiparty computation (SMC), where data owners need to process, encrypt and/or secret-share their data among two non-colluding servers in the initial setup phase. SecureML allows data owners to train various models on their joint data without revealing any information beyond the outcome. However, such an approach incurs high computational and communication costs~\cite{lyu2017privacy,lyu2019fog}. 

The most relevant work is Distributed Selective Stochastic Gradient Descent (DSSGD)~\cite{shokri2015privacy}. To preserve privacy, instead of explicitly sharing training data, each party computes and shares (with the PS) its local model gradients based on local training data, while updating its local model by downloading the most-updated parameters from the PS. To further mitigate privacy leakage from the shared model updates,
each party adds noise to local model updates to ensure per-parameter differential privacy. However, their system requires a central parameter server to mediate training process. Therefore, it suffers from the common issues in the central server-based frameworks. 

The disadvantages of having a centralized parameter server can be summarized as follows:
\begin{enumerate}
    \item Privacy leakage: As evidenced in \cite{aono2018privacy}, local data information may be leaked to an honest-but-curious PS, even if only a small portion of local model updates is released to the PS. In particular, a PS can infer the true data or label of the participates with non-negligible probability for the local neural network with only one neuron. The above observations similarly hold for general neural networks, with both cross-entropy and squared-error cost functions. Even for general neural networks with regularization, the released local gradients can still reveal the truth value.
    \item Vulnerability to active adversaries: Most distributed learning frameworks assume that all the parties are honest. In reality, if a party turns out to be malicious, it can sabotage the learning process by spoofing random samples to infer information about the victim party's private data. 
\end{enumerate}

A special case of distributed deep learning is federated learning~\cite{FL2019}. In FL, to preserve privacy of individual model updates, Bonawitz \etal~\cite{bonawitz2017practical} proposed a practical secure aggregation protocol, which is proven to be secure under the honest-but-curious and active adversary settings, even if an arbitrarily chosen subset of users drop out at any time. In particular, secure multi-party computation (SMC) is leveraged to compute sums of model parameter updates from individual users' devices in a secure manner, which comes at the cost of extra computation and communication overheads. Another more efficient method is to use differential privacy by enabling the server to add the tailored noise to the weighted-average user updates to guarantee user-level privacy~\cite{mcmahan2018learning}. However, the default trusted Google server is entitled to see all users' update clearly, aggregate individual updates and add noise to the aggregation. Thus, their method is not preferred when the server is not a trusted party. We also remark that when the server is untrusted, the weighted aggregation becomes unrealistic, as the server may not know the data size of each party for weight calculation. Instead, the proposed FPPDL allows each party to integrate other parties' updates based on their local credibility and sharing level. 

\subsection{Fairness in Federated Learning}
Existing approach for promoting collaborative fairness among federated learning participants is based on incentive schemes. In general, participants shall receive payoffs that is commensurate with their contributions. Equal division is an example of egalitarian profit-sharing \cite{Yang-et-al:2017IEEE}. Under this scheme, the available total payoff at a given round is equally divided among all participants. Under the Individual profit-sharing scheme \cite{Yang-et-al:2017IEEE}, each participant $i$'s own contribution to the collective (assuming the collective only contains $i$) is used to determine his share of the total payoff.

The Labour Union game \cite{Gollapudi-et-al:2017} profit-sharing scheme determines a participant's share of the total payoff based on his marginal contribution to the utility of the collective formed by his predecessors (i.e. each participant's marginal contribution is computed based on the same sequence as they joined the federation). The Fair-value game scheme \cite{Gollapudi-et-al:2017} is a marginal loss-based scheme. Under this scheme, a participant's share of the total payoff is determined by the sequence following which the participants leave a federation. The Shapley game profit-sharing scheme \cite{Gollapudi-et-al:2017} is also a marginal contribution-based scheme. Unlike the Labour Union game, Shapley game aims to eliminate the effect of the participants joining the collective in different sequences in order to more fairly estimate their marginal contributions to the collective. Thus, it averages the marginal contribution for each participant under all different permutations of him joining the collective relative to other participants. This approach is computationally expensive.

For gradient-based federated learning approaches, the gradient information can be regarded as a type of data. However, in these cases, output agreement-based rewards are hard to apply as mutual information requires a multi-task setting which is usually not present in such cases. Thus, among these three categories of schemes, model improvement is the most relevant way of designing rewards for federated learning. There are two emerging federated learning incentive schemes focused on model improvement.

A scheme which pays for marginal improvements brought about by model updates was proposed in \cite{Richardson-et-al:2019}. The sum of improvements might result in overestimation of contribution. Thus, the proposed approach also includes a model for correcting the overestimation issue. This scheme ensures that payment is proportional to model quality improvement, which means the budget for achieving a target model quality level is predictable. It also ensures that data owners who submit model updates early receive a higher reward. This motivates them to participate even in early stages of the federated model training process. 

In addition to the contributions made by participants, \cite{Yu-et-al:2020AIES} proposed a joint objective optimization-based approach to take costs and waiting time into account in order to achieve additional notions of fairness when distributing payoffs to FL participants. Different from the aforementioned approaches, the proposed FPPDL framework does not utilize monetary payoffs to achieve fair treatment of FL participants. Instead, it allocates each of them a different version of the FL model with performance commensurate with his contributions. This represents a alternative paradigm to existing federated learning in which all participants receive the same final FL model.

%- - - - - - - -
\section{Preliminaries}
In this section, we introduce key technologies which form the building blocks of the proposed FPPDL framework, including differential privacy, homomorphic encryption, and blockchain.

\subsection{Differential Privacy}
\label{sec:dp}
Differential privacy~\ref{def:dp}~\cite{dwork2014algorithmic} trades off privacy and accuracy by perturbing the data in a way that is (i) computationally efficient, (ii) does not allow an attacker to recover the original data, and (iii) does not severely affect utility. 
\begin{definition}\label{def:dp}
A randomized mechanism \mech: $\mathcal{D} \to \mathcal{R}$ with domain $\mathcal{D}$ and range $\mathcal{R}$ satisfies $(\epsilon,\delta)$-differential privacy if for all two neighbouring inputs $D,D'\in\mathcal{D}$ that differ in one record and for any measurable subset of outputs $S \subseteq \mathcal{R}$ it holds that
\begin{eqnarray*}
\Pr\{\mech(D)\in S\} &\leq& \exp(\epsilon) \cdot \Pr\{\mech(D')\in S\} +
\delta\enspace.
\end{eqnarray*}
Furthermore \mech is said to preserve (pure) $\epsilon$-differential privacy if $\delta=0$.
\end{definition}
The formal definition of differential privacy has two parameters: privacy budget $\epsilon$ measures the privacy leakage; and $\delta$ bounds the probability that the privacy loss exceeds $\epsilon$. The values of $(\epsilon,\delta)$ are accumulated as the algorithm repeatedly accesses the private data~\cite{abadi2016deep}. 

\subsection{Homomorphic Encryption} 
Homomorphic encryption is a form of encryption that is widely used to derive the aggregate in a secure manner. Existing homomorphic encryption techniques include fully homomorphic encryption, somewhat homomorphic encryption and partially homomorphic encryption. Fully homomorphic encryption can support arbitrary computation on ciphertexts, but is less efficient~\cite{gentry2009fully}. On the other hand, somewhat homomorphic encryption and partially homomorphic encryption only support a limited number of operations~\cite{damgaard2012multiparty}. 

However, all these techniques generally result in longer ciphertext than the plaintext, incurring extra communication costs. To address this issue in this paper, we take inspirations from stream ciphers~\cite{canteaut2018stream} to develop efficient homomorphic-ciphertext compression, which also allows additive homomorphic operation over ciphertexts encrypted under different parties' key streams. More details are provided in Section~\ref{sec:Encryption}. 

\subsection{Blockchain}
Blockchain is a decentralized (\ie a peer-to-peer, non-intermediated) system that is maintained by all the participants in the system. There are two types of blockchains, namely permissionless blockchain and permissioned blockchain. With permissionless blockchains, such as Bitcoin~\cite{nakamoto2008bitcoin}, participants can join and leave at anytime and the number of participants is not pre-defined nor fixed. With permissioned blockchains (a.k.a. consortium blockchains), such as IBM's Hyperledger Fabric, participants require permissions from the system to join or leave. The set of participants are normally predefined~\cite{DBLP:journals/corr/abs-1908-08316}.

For the application with a relatively stable set of participants, a permissioned blockchain is preferred. It can serve as a distributed key-value store, where a fault tolerance (a.k.a. Byzantine agreement) scheme is required for reaching consensus on the global state. Blockchain is well known for its transparency, accountability and robustness -- data and all operations are recorded on the blockchain in an append-only manner and are accessible by all the participants. Intuitively, the incremental characteristic of federated deep learning makes it suitable for leveraging Blockchain. However, a reasonable approach to integrate Blockchain with privacy-preserving deep learning needs to be developed.

\section{The FPPDL Framework}
\label{sec:FPPDL}
%%%%%%%%%%%%%%%%%%%%%%%%%%%%%%%%%%%%%%%
This section describes the design of our proposed decentralized Fair and Privacy-Preserving Deep Learning (FPPDL) framework, and an investigation of Blockchain as the decentralized architecture for FPPDL. Tables~\ref{tbl:symbs} presents the list of symbols used in this paper and their meanings for easy readability.

%%%%%%%%%%%%%%%%%%%%%%%%%%%%%%%%%%%%%%%

\begin{table}[ht]
    \centering\caption{\label{tbl:symbs}List of Symbols.}
    \begin{tabularx}{\linewidth}{c|X}
    \hline    \hline
    Symbol & Meaning \\
    \hline
    $D_i$, $M_i$ & local training data and local model of party $i$\\
    $SD_i$ & $\mu$ DPGAN samples randomly chosen by party $i$\\ 
    $p_i$, $d_i$ & points and gradients download budget of party $i$ \\
    $c_i^j$, ${c_i^j}'$ & local credibility and updated local credibility of party $j$ given by party $i$\\
    $u_i$ & number of DPGAN samples released by party $i$\\
    $d_i^j$ & number of meaningful gradients of party $j$ released to party $i$\\
    $\lambda_j$ & sharing level of party $j$ \\
    $sacc_j, acc_j$ & standalone and final model accuracy of party $j$ \\ 
    $\Delta \vec{w_j}$ & gradient vector of party $j$ \\
    $\Delta \vec{\tilde{w}_j^i}$ & masked gradient vector of party $j$ shared with party $i$ by filling the remaining $|\Delta \vec{w_j^i}|-d_i^j$ gradients with 0\\
    $\vec{w_i}$ & parameter of party $i$ at previous round\\
    $\vec{w_i}'$ & updated parameter of party $i$ at current round \\
    $n$ & number of participating parties \\
    $c_{th}$ & lower bound of the credibility threshold  \\
    $C$ & credible party set with local credibility above $c_{th}$ agreed by $2/3$ parties \\
    $m_j$ & number of matches between majority labels and party $j$'s predicted labels  \\
    $(sk'_i, pk'_i)$ &  party $i$'s key pair for signing and verification, respectively \\
    $k_i$ & party $i$'s keystream used in the first layer of three-layer onion-style encryption \\
    $fsk$ & fresh symmetric encryption key used in the second layer of three-layer onion-style encryption \\
    $(sk_i,pk_i)$ & party $i$'s key pair for decryption and encryption in the third layer of three-layer onion-style encryption\\
    $Enc$ & homomorphic encryption \\
    $Senc$ & symmetric key encryption \\
    $Aenc$ & public key encryption \\
    $E$ & number of local training epochs in each round\\
    $B, lr$ & local batch size, local learning rate \\
    \hline    \hline
    \end{tabularx}
\end{table}

\subsection{Design Objectives}
\label{sec:goals}
\subsubsection{Privacy Preservation} 
In FPPDL, we assume parties do not trust each other or any third party. Hence, parties may not be willing to share their information when training a joint model without the promise of privacy protection. Under FPPDL, instead of sharing the original data or model parameters, each party leverages Differentially Private GAN (DPGAN) to publish differentially private local samples for mutual evaluation during the initial benchmarking phase. Then, they encrypt the shared gradients using the proposed three-layer onion-style encryption scheme to preserve privacy during collaborative deep model training. 

\subsubsection{Fairness}
Since our focus here is to distribute different variants of the final FL model to participants based on their contributions, the notion of fairness most relevant for our purpose is \textit{Fairness through Awareness}. Under this notion, individuals who are similar with respect to a similarity metric defined for a particular task should receive a similar outcome~\cite{Mehrabi-et-al:2019}. A party with high-contribution should be rewarded more than a party with low-contribution party. Moreover, we clarify that the low-contribution parties are not malicious, \ie they follow the protocol honestly and aim to benefit from other parties' data, but contribute lowly, or even nearly nothing or negatively. Under the collaborative model training scenario, we define collaborative fairness as:
\begin{definition}\label{def:fairness}
Collaborative fairness. In collaborative learning systems, a high-contribution party is deserved to be rewarded with a better performing local model than a low-contribution party. Specially, in IID setting, fairness can be quantified by the correlation coefficient between the contributions by different parties and their respective final model accuracies. 
\end{definition}

Considering these two goals, we design a local credibility mutual evaluation mechanism to enforce fairness in FPPDL, where participants trade their information in an "earn-and-pay" way using their "points".
The local credibility and points of each participant are initialized through an initial benchmarking phase, and updated through privacy-preserving collaborative deep model training. 

The basic idea is that participants can earn points by contributing their information to other participants. Then, they can use the earned points to trade information with other participants. Thus, participants are encouraged to upload more samples or gradients to earn more points (as long as it is within the limit of their sharing levels), and use these points to download more gradients from others. All trades are recorded as immutable transactions in a blockchain, providing transparency and auditability. In particular, FPPDL ensures fairness during download and upload processes as follows:
\begin{itemize}
\item \textbf{Download}: Since one party might contribute differently to different parties, the credibility of this party might be different from the view of different parties. Therefore, each party $i$ records a private local credibility list for all parties sorted in descending order of their credibility values. The higher the credibility of party $j$ in party $i$'s credibility list, the more likely party $i$ will download gradients from party $j$, and consequently, more points will be rewarded to party $j$ by party $i$.
\item \textbf{Upload}: Once a party receives download request for its local gradients, it can determine how many meaningful gradients to send back based on both the download request from the requester and its own sharing level.
\end{itemize}

\subsection{Blockchain-based Architecture}
To develop a decentralized architecture for FPPDL, we incorporate the privacy-preserving deep learning algorithm into a private Blockchain using Blockchain 2.0, which is only available to the participating parties. Compared with the current server-based architecture, FPPDL inherits the peer-to-peer architecture of Blockchain, allowing each party to remain modular while interoperating with others. In addition, instead of ceding control to the central server, each party keeps full control of its own data. Moreover, Blockchain provides the native ability to automatically coordinate the joining and departure of each party, further facilitating independence and modularity of the federation. Blockchains, with no single point of failure, also enhances robustness. Here, we design two types of blocks in the Blockchain for FPPDL, namely the {\it init block} and the {\it operation block}. 

\textbf{An init block} initializes benchmarking of the usefulness of each party's training data, as a set of init transactions. An init transaction contains the initial points that the transaction creator earned, its contributed DPGAN samples, and its public key that will be used for authenticating future transactions. The genesis block (\ie the first block) of the Blockchain is an init block, which contains the initial points and local credibility values to all the participants according to their relative contributions, as stated in Algorithm \ref{Algorithm:benchmarking}. If any party joins or adds new data during later updates, a new init block will be created and added to the existing Blockchain. 

\textbf{An operation block} contains a set of transactions defining the UPLOAD operation and/or DOWNLOAD operation. All UPLOAD and DOWNLOAD transactions are signed by their creator using the private key associated with the public key recorded in the init transaction. An UPLOAD operation commits that a data owner has uploaded local model gradients to the party who sent a download request. A DOWNLOAD operation states that a participant is committed an order to request some local model updates from other participants. Upon receiving a DOWNLOAD transaction, Blockchain miners verify its signature, check if the requester has enough balance points to download the number of requested gradients, and record the verified transactions in an operation block. Once the DOWNLOAD transaction is recorded in the Blockchain, the requested local model gradients will be encrypted and uploaded by the owner to a publicly accessible storage, and re-encrypted using the recipient's public key defined in the DOWNLOAD transaction. 

\begin{figure}[!htp]
\centering 
\includegraphics[scale=0.3]{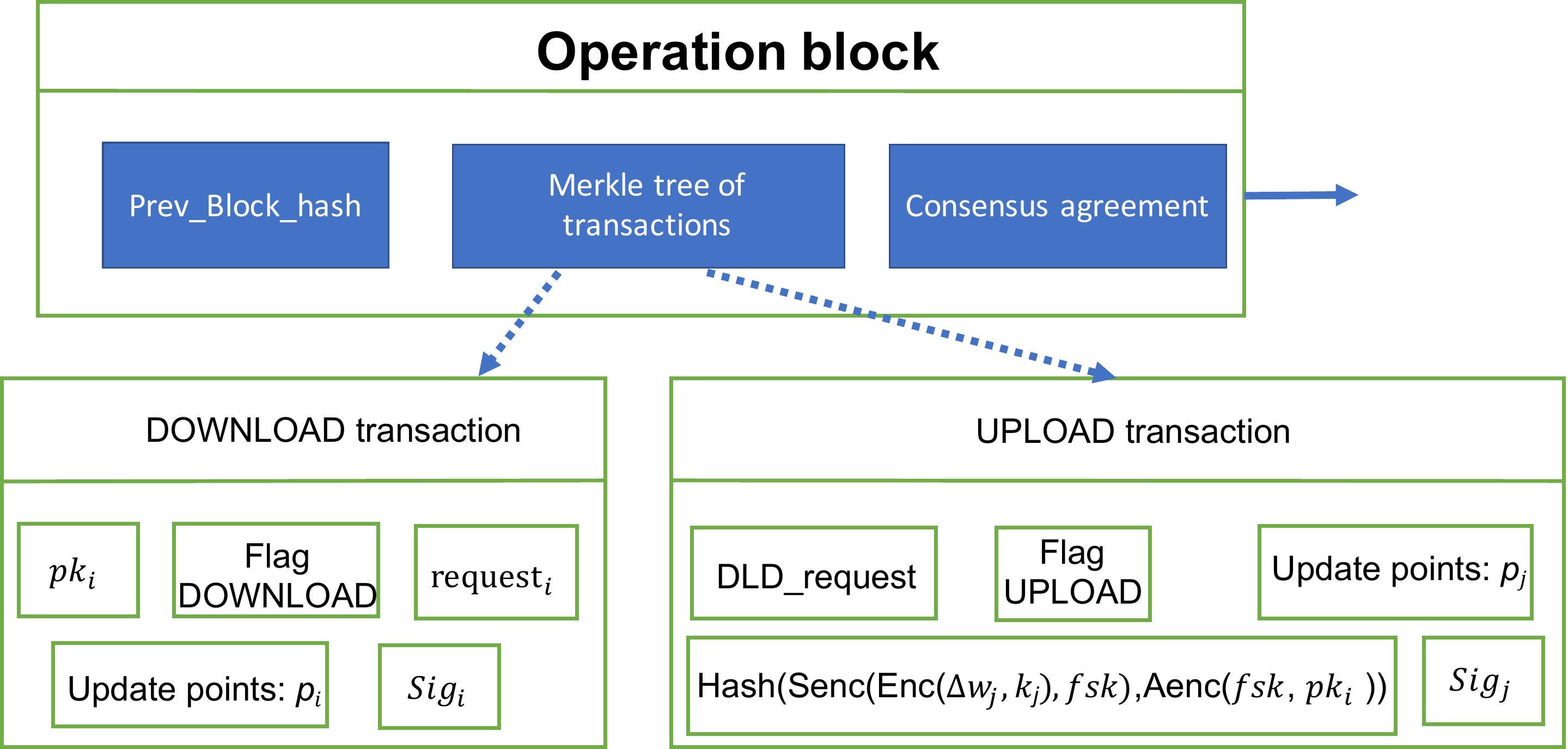}
\caption{\textbf{An example structure of the operation block}. It mainly contains three key components, namely, the hash value Prev\_Block\_hash of the previous block, a set of UPLOAD/DOWNLOAD transactions organized as a Merkle tree, and the consensus agreement of this block. In particular, a Prev\_Block\_hash links the current block to the previous one, and the request in the UPLOAD transaction acts as a reference to the associated DOWNLOAD transaction. $pk_i$ in the DOWNLOAD transaction is the public key that will be used in the last layer of our three-layer onion-style encryption scheme, $request_i$ is a unique request ID of this transaction and will be referenced in the corresponding UPLOAD transaction via DLD\_request, and $Sig_i$ is the signature on this transaction. $Enc$, $Senc$, and $Aenc$ refer to homomorphic encryption, symmetric key encryption, and public key encryption, respectively.}
\label{fig:block_update}
%\vspace{-0.2cm}
\end{figure}

The privacy of local model gradients is protected through a three-layer onion-style encryption scheme (see Section~\ref{sec:CL}). The first layer encrypts the local model gradients through our proposed symmetric key based homomorphic encryption (Algorithm \ref{algo:Encryption}), which allows each party to learn the aggregate of the received gradients without revealing individual gradients, \ie party obliviousness. The second and third layers present a standard hybrid encryption process: the second layer uses a freshly generated symmetric key $fsk$ to re-encrypt the first layer ciphertext, and the third layer encrypts $fsk$ with the requester party $i$'s public key $pk_i$. In this way, we minimize the required computational cost incurred by the asymmetric key based encryption. The commitment of the uploaded encrypted local model gradients (\eg hash value of the ciphertext, as presented in Fig~\ref{fig:block_update}) will be included in the UPLOAD transaction. 

\begin{figure}[!htp]
\centering 
\includegraphics[scale=0.35]{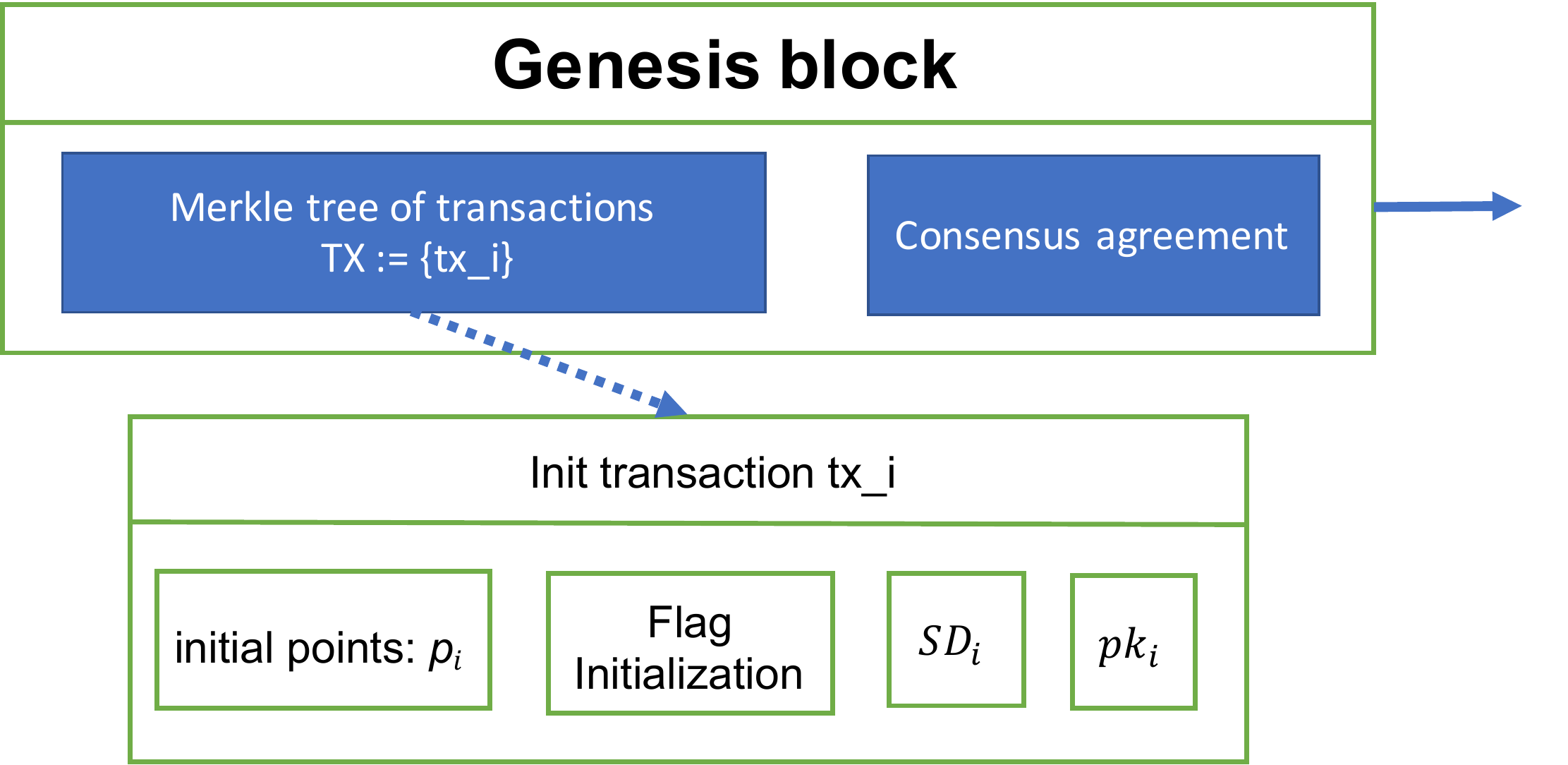}
\caption{\textbf{An example structure of the genesis block}. It mainly contains two key components, one is a set of init transactions organized as leaves of a Merkle tree; and the other one is the consensus agreement reached by the participants through the underlying consensus protocol (\eg PBFT or PoS), which is specific to the deployed Blockchain. The $pk_i'$ in the init transaction is a signature verification key of party $i$.}
\label{fig:block_initialization}
%\vspace{-0.2cm}
\end{figure}

In our private Blockchain, only the requester who pays can read the plaintext. Others can verify that this transaction has happened, but cannot read the plaintext. When a requester blames a data uploader, the data uploader reveals the plaintext as evidence. In this case, the requester will be forced to pay a fine that it deposits when filing a claim if it is shown to be an dishonest claim. Once an UPLOAD transaction is recorded in the Blockchain, the points will be automatically transferred from the requester to the uploader. An example of INITIALIZE and DOWNLOAD transaction stored by the Blockchain are shown in Fig.~\ref{fig:block_initialization} and Fig.~\ref{fig:block_update}, respectively. For our application scenario, we expect a relatively stable and small set of participants, such as financial institutions acting with legal liabilities, which falls under the umbrella of horizontal federated learning (HFL) involving business participants~\cite{lyu2020threats}. This allows us to adopt a permissioned blockchain.

%%%%%%%%%%%%%%%%%%%%%%%%%%%%%%%%%%%%%
%%%%%%%%%%%%%%%%%%%%%%%%%%%%%%%%%%%%%%%

%%%%%%%%%%%%%%%%%%%%%%%%%%%%%%%%%%%%%
%%%%%%%%%%%%%%%%%%%%%%%%%%%%%%%%%%%%%%%

%%%%%%%%%%%%%%%%%%%%%%%%%%%%%%%%%%%%%%%
\begin{algorithm}[ht]
\caption{Initial Benchmarking}\label{Algorithm:benchmarking}
\small
\begin{algorithmic}
  \State \textbf{Input: number of participating parties $n$, C=\{1,\ldots,n\}}
  \State \textbf{Output: local credibility and points of all parties}
  \State 1: \textbf{Pre-train aprior models}: Each party $i$ trains standalone model $M_i$ and local DPGAN based on its local training data.
   \State 2: \textbf{Sharing level initialization}: During initialization, party $i$ randomly selects and releases $u_i$ artificial samples generated by local DPGAN to any party $j$, sharing level is autonomously determined as $\lambda_i=u_i/|D_i|$, where $|D_i|$ is local training data size of party $i$.
   \State 3: \textbf{Local credibility initialization}: Party $j$ labels the received artificial samples by its local model $M_j$, then returns the predicted labels back to party $i$. Meanwhile, party $i$ also labels its own DPGAN samples using $M_i$. Afterwards, party $i$ applies majority voting to all the predicted labels, then initializes the local credibility of party $j$ as $c_i^j=\frac{m_j}{u_i}$, where $m_j$ is the number of matches between majority labels and party $j$'s predicted labels, and $u_i$ is the number of DPGAN samples released by party $i$. The detailed explanation is elaborated in Section~\ref{sec:credit_initialization}.
   \State 4: \textbf{Local credibility normalization}: 
      $c_i^j=\frac{c_i^j}{{\textstyle\sum}_{j \in C  \setminus i} c_i^j}$
      \If{$c_i^j<c_{th}$}
    \State party $i$ reports party $j$ as a low-contribution party
     \EndIf
    \State 5: \textbf{Credible party set}: If the majority of parties report party $j$ as low-contribution, Blockchain removes party $j$ from the credible party set $C$ and all parties run step 4 again.
 \State 6: \textbf{Points initialization to download gradients}: $p_i=\lambda_i*|\vec{w_i}|*(n-1)$.
\end{algorithmic}
\end{algorithm}
%%%%%%%%%%%%%%%%%%%%%%%%%%%%%%%%%%%%%%%

\section{Implementation of FPPDL}
\label{sec:FPPDL}
\begin{figure*}[!htp]
\centering 
\includegraphics[scale=0.46]{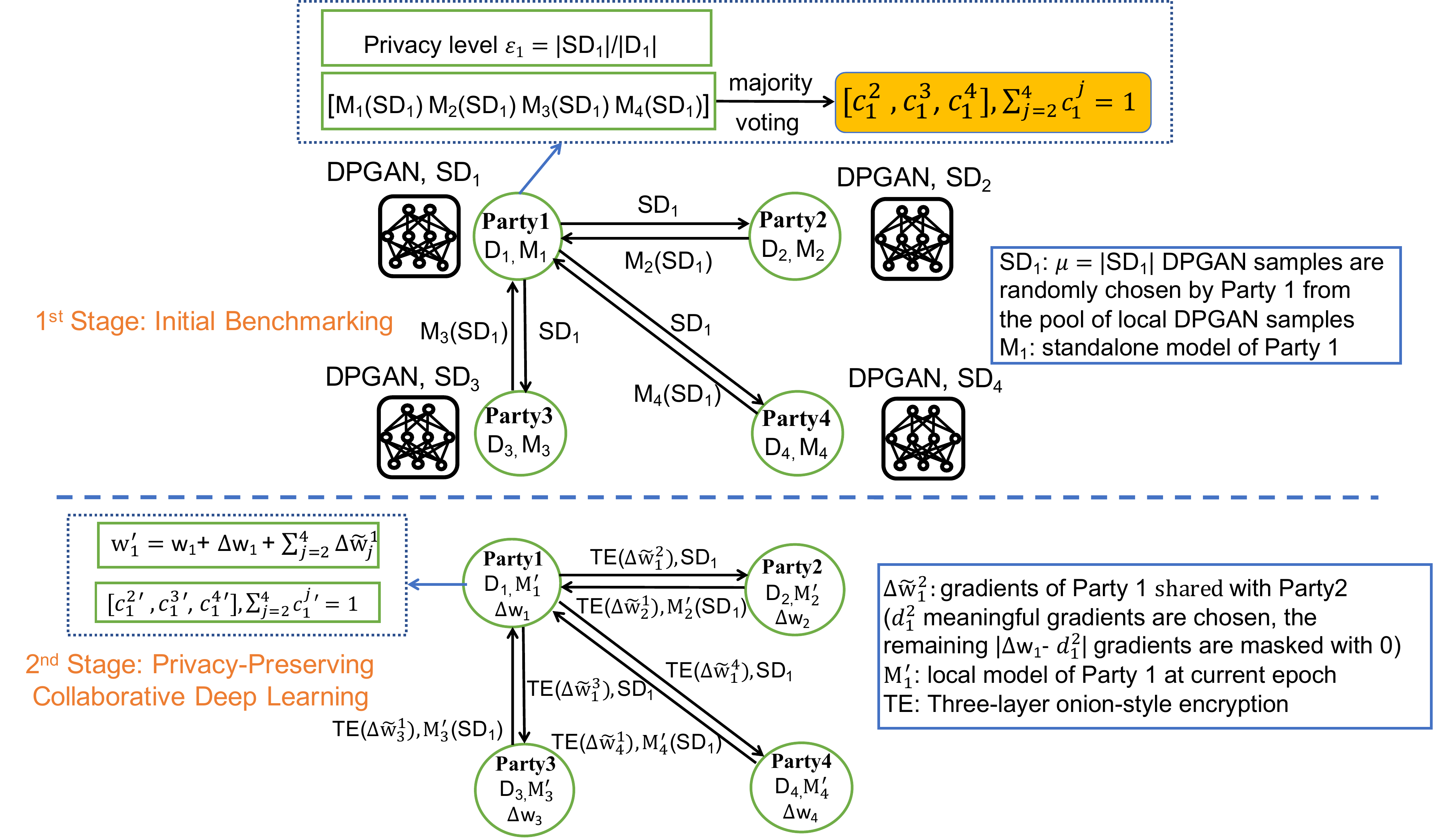}
\caption{Two-stage implementation of FPPDL.}
\label{fig:FPPDL_flow}
%\vspace{-0.2cm}
\end{figure*}

%%%%%%%%%%%%%%%%%%%%%%%%%%%%%%%%%%%%%%%
This section details the two-stage implementation of FPPDL to enforce both fairness and privacy. These include how to initialize local credibility values, sharing levels and points through initial benchmarking, and how to update local credibility values and points in the privacy-preserving collaborative deep learning phase, followed by the quantification of fairness. The two-stage implementation is shown in Fig.~\ref{fig:FPPDL_flow}.

\subsection{Initial Benchmarking}\label{sec:benchmarking}
The proposed initial benchmarking algorithm aims to assess the quality of local training data of each participant via mutual evaluation without looking at the raw data before collaborative model training starts. The algorithm works as follows: each participant trains a DPGAN based on its local training data to generate artificial samples. However, these generated samples will not disclose the true sensitive examples, as well as the true distribution of data, but only a few implicit density estimation within a modest privacy budget used in DPGAN. Each participant publishes individually generated artificial samples based on its individual sharing level without releasing labels. All the other participants produce predictions for the received artificial samples using their pre-trained standalone models and send the predicted labels back to the party who generated these samples. 

The aim of sharing artificial DPGAN samples is two-fold:
\begin{enumerate}
\item To obtain prior information about individual models before collaborative learning starts. If a participant does not have a reasonable amount of training data to produce a decent model, it will perform poorly during the initial evaluation phase. Therefore, other participants will be cautious when sharing gradients with it.
\item To obtain a rough estimate of data distribution of other participants. Two participants can mutually benefit only if their data distributions are different, but with some degree of overlap. Suppose that two participants A and B have published almost identical artificial samples, it means that their training data distributions are almost identical. In this case, the updates from B are unlikely to increase the accuracy of model A and vice versa. Therefore, during the subsequent communication rounds, A and B should avoid downloading updates from each other. Other participants can choose to download updates from either A or B but not both.
On the other hand, suppose that participants A and B have completely different data distributions, the updates from B are unlikely to increase the accuracy of model A and vice versa. Thus, during the subsequent rounds, A and B should also avoid downloading updates from each other. Furthermore, suppose that A's data distribution is different from that of all the other participants, all these participants should try to avoid A. This automatically takes care of the scenario where a honest participant publishes some gradients, while all the other honest participants assign very low credibility to the publisher. In this case, the data distribution of the publisher is completely different from that of the other participants. Hence it is reasonable to reduce the credibility of the publisher. 
\end{enumerate}

We next describe the detailed procedures of initial benchmarking in Algorithm~\ref{Algorithm:benchmarking}, including: local credibility initialization, and sharing level and points initialization.

\subsubsection{Local Credibility Initialization}\label{sec:credit_initialization}
For local credibility initialization, each party compares the majority voting of all the combined labels with a particular party's predicted labels to evaluate the effect of this party. It relies on the fact that the majority voting of all the combined labels reflects the outcome of the majority of parties, while the predicted labels of party $j$ only reflects the outcome of party $j$. 

For example, in the case of party $i$ initializing local credibility list for other parties, party $i$ broadcasts its DPGAN samples to other parties, who label these samples using their pre-trained standalone models, and send the corresponding predicted labels back to party $i$. Meanwhile, party $i$ also labels its own artificial samples using its pre-trained standalone model, then combines all parties' predicted labels as a label matrix with total $n$ columns, where each column corresponds to one party's predicted labels. Party $i$ then initializes the local credibility of party $j$ as $c_i^j=\frac{m_j}{u_i}$, where $m_j$ is the number of matches between the majority labels and party $j$'s predicted labels, and $u_i$ is the number of DPGAN samples released by party $i$. Afterwards, party $i$ normalizes $c_i^j$ within [0,1]. 

If the majority of parties report that the local credibility of one party is lower than the threshold $c_{th}$, implying a potentially low-contribution party, it will be banned from the local credibility lists of all parties. Here, $c_{th}$ is mainly used to detect and isolate the low-contribution party, and it should be agreed by the majority of parties. However, it should not be too small or too large as fairness and accuracy may be affected. If it is too small, it might allow low-contribution party to into the collaborative learning system without being detected and isolated. If it is too large, it might ban most participants from the system. In the following update process, party $i$ is more likely to download gradients from more credible participants, while download less, even ignoring those published by less credible parties. 

\subsubsection{Sharing Level and Points Initialization}
Sharing level is denoted by the the upper bound of the number of samples or gradients one party can share with others. 
Based on the number of artificial samples $u_i$ that party $i$ publishes at the beginning, a suitable sharing level of party $i$ can be automatically estimated as $\lambda_i=u_i/|D_i|$, where $D_i$ is the local training data of party $i$. Points are initialized as follows:
\begin{equation}\label{eq:points_initialization}
	p_i=\lambda_i*|\vec{w_i}|*(n-1)
\end{equation}
where $\lambda_i$ is the sharing level of party $i$ (\ie the higher, the more data one party would like to share), $|\vec{w_i}|$ is the number of model parameters, and $n$ is the number of parties. The points gained from initial benchmarking will be used to download gradients in the following collaborative learning process, and the number of gradients $i$ can downloaded depends on both the local credibility and sharing level of the party from which it is requesting.

\begin{algorithm*}[ht]
\caption{Privacy-Preserving Collaborative Deep Learning}\label{Algorithm:global_credibility_update}
\small
\begin{algorithmic} 
\State \textbf{Input: $C$, $c_i^j$, $p_i$, $p_j$, $d_i$, $\lambda_j$, $\vec{w_i}$, $\Delta \vec{w_j}$}
\State \textbf{Output: updated points $p_j'$, $p_i'$, parameters $\vec{w_i}'$, and local credibility ${c_i^j}'$}
      \State 1: \textbf{Trade gradients as per download requests, local credibility, and sharing level; Party points update}: In each communication round, party $i$ aims to download total $d_i=p_i$ gradients from all parties in $C$, while party $j \in C \setminus i$ can at most provide $\lambda_j \times |\Delta \vec{w_j}|$ gradients, one point is consumed/rewarded for each download and upload. Each party $i$ updates local model parameters based on the gradients of party $j \in C \setminus i$ as follows:
      
\For{$j \in C \setminus i$}
\State $d_i^j=min(c_i^j*d_i, \lambda_j*|\Delta \vec{w_j}|)$, $p_j'=p_j+d_i^j$, $p_i'=p_i-d_i^j$
\State $\Delta \vec{w_j^i}=\Delta \vec{w_j}$, party $j$ first chooses $d_i^j$ meaningful gradients from $\Delta \vec{w_j^i}$ according to ``largest values" criterion: sort gradients in $\Delta \vec{w_j^i}$ and choose $d_i^j$ of them, starting from the largest, then masks the remaining $|\Delta \vec{w_j^i}|-d_i^j$ gradients with 0 as $\Delta \vec{\tilde{w}_j^i}$.
\EndFor
 \State 2: \textbf{Three-layer onion-style encryption}: 
 Party $j$ follows Algorithm~\ref{algo:Encryption} to encrypt the masked gradients $\Delta \vec{\tilde{w}_j^i}$ with its keystream $k_j$ as $c = Enc(\Delta \vec{\tilde{w}_j^i}, k_j)$, and re-encrypts the encrypted gradients $c$ with a fresh symmetric encryption key $fsk$ as $Senc(c,fsk)$, the symmetric encryption key of the second layer is encrypted in the third layer by the receiver party $i$'s public key $pk_i$ as $Aenc(fsk,pk_i)$. Finally, the two-layer encrypted gradients $Senc(c,fsk)$ and the encrypted fresh symmetric encryption key $Aenc(fsk,pk_i)$ are sent to party $i$; 
 \State 3: \textbf{Parameter update}: party $i$ uses the paired secret key $sk_i$ to decrypt the received encrypted fresh symmetric encryption key as $fsk$, then uses $fsk$ to decrypt the two-layer encrypted gradients as $c = Enc(\Delta \vec{\tilde{w}_j^i}, k_j)$, finally decrypts the sum of all the received gradients using homomorphic property and updates local parameters by integrating all its plain gradients $\Delta {\vec{w_i}}$ as: $\vec{w_i}'=\vec{w_i}+\Delta {\vec{w_i}}+Dec({\textstyle\sum}_{j \in C \setminus i} Enc(\Delta \vec{\tilde{w}_j^i},k_j),-k_i)=\vec{w_i}+\Delta \vec{w_i}+{\textstyle\sum}_{j \in C \setminus i} \Delta \vec{\tilde{w}_j^i}$, where $\vec{w_i}$ is party $i$'s local parameters at previous communication round.
  \State 4: \textbf{Local credibility update}: party $i$ randomly selects and releases $u_i$ artificial samples to any party $j$ for labelling, mutual evaluation is repeated by following Step 3 of Algorithm \ref{Algorithm:benchmarking} to calculate local credibility of party $j$ at current communication round as ${c_i^j}'$. Party $i$ updates local credibility of party $j$ by integrating its historical credibility as: ${c_i^j}'=0.2*c_i^j+0.8*{c_i^j}'$, 
  where $c_i^j$ is the local credibility of party $j$ at previous communication round.
  
  \State 5: \textbf{Local credibility normalization}: 
      ${c_i^j}'=\frac{{c_i^j}'}{{\textstyle\sum}_{j \in C} {c_i^j}'}$
       	\If{${c_i^j}'<c_{th}$}
	\State party $i$ reports party $j$ as a low-contribution party
     	\EndIf
    \State 6: \textbf{Credible party set}: If the majority of parties report party $j$ as low-contribution, Blockchain removes party $j$ from credible party set $C$ and all parties run Step 5 again.
\vspace{2mm}
\end{algorithmic}
 \end{algorithm*}

\subsubsection{Differentially Private GAN (DPGAN)}
During initial benchmarking, although each party only releases a small amount of unlabeled samples, it may still disclose privacy of local training data. The approach of generating samples under differential privacy with generative adversarial network (GAN) offers a solution to this problem. Under FPPDL, we train a \emph{Differentially Private GAN} (DPGAN) by adding tailored noise to the gradients during DPGAN learning~\cite{zhang2018differentially} at each party. 

In the context of a GAN, the discriminator is the only component that accesses the private real data. Therefore, we only need to train the discriminator under differential privacy. The differential privacy guarantee of the entire GAN directly follows because the computations of the generator are simply post-processing from the discriminator. The main idea follows the post-processing property of differential privacy~\cite{dwork2014algorithmic}, as stated in Lemma~\ref{lemma:post-processing}. 

To counter the stability and scalability issues of training DPGAN models, we apply multi-fold optimization strategies, including weight clustering, adaptive clipping and warm starting, which significantly improve both training stability and utility~\cite{zhang2018differentially}. Unlike PATE~\cite{papernot2016semi}, where privacy loss is proportional to the amount of data needed to be labeled in public test data, differentially private generator can generate infinite number of samples for the intended analysis, while rigorously guaranteeing $(\epsilon,\delta)$-differential privacy of training data. Without loss of generality, we exemplify DPGAN in the context of the improved WGAN framework~\cite{arjovsky2017wasserstein} and let each party generates a total of 1,000 artificial samples. As demonstrated in~\cite{zhang2018differentially}, DPGAN is able to synthesize both grey and RGB image with inception scores fairly close to the real data and samples generated by regular GANs without any privacy protection. 

\begin{lemma}
\label{lemma:post-processing}
Let algorithm $\mathcal{A}: \mathbb{R}^n \to \mathbb{R}$ be a randomized algorithm that is ($\epsilon,\delta$)-differentially private. Let $f: \mathbb{R} \to \mathbb{R}^'$ be an arbitrary randomized mapping. Then 
% \begin{equation*}
$f \circ \mathcal{A}: \mathbb{R}^n \to \mathbb{R}^'$
% \end{equation*}
is ($\epsilon,\delta$)-differentially private.
\end{lemma}

Meanwhile, it is well-known that larger amount of training data causes less privacy loss, and allows for more iterations within a moderate privacy budget~\cite{abadi2016deep}. Due to the scarcity of training data of each party, data augmentation is exploited to expand local data size of each party to 100 times, which allows DPGAN to generate realistic samples within a moderate privacy budget. In particular, we augment original data with rotation range of 1 and width shift range and height shift range of 0.01. 

\begin{figure}[!htp]
\centering 
\includegraphics[scale=0.4]{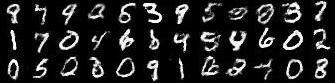}
\caption{Generated DPGAN samples with $\epsilon=4,\delta=10^{-5}$ using the augmented 60000 MNIST examples of one party who owns 600 original MNIST examples.}
\label{fig:dpgan_samples}
%\vspace{-0.2cm}
\end{figure}

In our study, we use moments accountant described in~\cite{abadi2016deep} to track the spent privacy over the course of training. Our DPGAN is able to generate realistic  MNIST samples with $\epsilon=4$ and $\delta=10^{-5}$, as shown in Fig.~\ref{fig:dpgan_samples}. Note that each party can individually train DPGAN and generate massive DPGAN samples offline without affecting collaboration.
 
\subsection{Privacy-Preserving Collaborative Deep Learning}\label{sec:CL}
Algorithm~\ref{Algorithm:global_credibility_update} summarizes the steps for the privacy-preserving collaborative deep learning in each communication round, including how to update points as per upload/download, how to preserve privacy of individual model updates using three-layer onion-style encryption followed by parameter and local credibility update, and credible party set maintenance by the Blockchain system. In particular, the gradients download budget of party $i$, \ie $d_i$, is closely related with how many points $p_i$ party $i$ has in each communication round. More concretely, $d_i$ should not exceed $p_i$, otherwise, party $i$ will not have enough points to pay for the gradients provided by other parties. Moreover, $d_i$ can be dynamically determined based on the existing points $p_i$ in each communication round. For simplicity, we initialize $d_i=p_i$ in each communication round, but how many gradients can be downloaded will be dependent on both the local credibility list of the requester and sharing levels of the requested parties, which can be referred to Section~\ref{sec:goals}. In the following sections, we will focus on the most important details for parameter update, three-layer onion-style encryption, and local credibility update.

\subsubsection{Parameter Update with Homomorphic Encryption}
\label{sec:Encryption}
Sharing gradients can prevent direct exposure of the local data, but may indirectly disclose local data information. 
To further prevent potential privacy leakage from sharing gradients and facilitate gradients aggregation during the collaborative learning process, we use additive homomorphic encryption such that each party can only decrypt the sum of all the received encrypted gradients. Specifically, Vernam cipher or one-time pad (OTP) has been mathematically proved to be completely secure, which cannot be broken given enough ciphertext and time. Therefore, we use simple and provably secure OTP for additively homomorphic encryption that allows efficient aggregation of encrypted data~\cite{castelluccia2005efficient,lyu2018ppfa}. The main idea of forming the ciphertext is to combine the keystream with the plaintext digits. Meanwhile, rather than XOR operation typically found in stream ciphers, which is unsecured under the frequency analysis attacks, our encryption scheme uses modular addition (+), and is hence very efficient~\cite{castelluccia2005efficient}. The security relies on two important features: (1) the keystream changes from one message to another; and (2) all the operations are performed modulo a large integer $M$~\cite{castelluccia2005efficient}. 

The detailed procedure for homomorphic encryption is presented in Algorithm~\ref{algo:Encryption}. In practice, if $p=\max(x_i)$, $M$ is derived as $M = 2^{\lceil log_2(p \times n)\rceil}$. All computations in the remainder of this paper are modulo $M$ unless otherwise stated. However, all the original floating-point values need to be mapped to the integer domain by using Scaling, Rounding, Unscaling (SRU) algorithm~\cite{lyu2018ppfa}. A pseudorandom keystream $k$ can be generated by a secure pseudo random function (PRF) by implementing a secure stream cipher, such as Trivium~\cite{de2008trivium}, keyed with each party's keystream $k_{i}$ and a unique message ID. For encryption purpose, the secret keys are pre-computed through a trusted setup, which can be performed by a trusted dealer or through a standard SMC protocol. 

For example, a trusted key managing authority can generate these keystreams in each communication round, but the generated keystreams cannot be used more than once. The trusted setup generates non-zero random shares of 0: ${\textstyle\sum}_{i \in C} k_i = 0$, such that each participant $i \in C$ obtains a keystream $k_i$. Note that if the Blockchain removes party $j$ from the credible party set $C$, a new credible party set $C$ should be constructed. 

\begin{algorithm}[ht]
\caption{Homomorphic Encryption Scheme}\label{algo:Encryption}
%\small
 \begin{algorithmic}
\State \textbf{Setup}
	\State 1: A trusted dealer randomly generates $|C|$ keystreams: $k_1,\ldots,k_{|C|} \in [0, M-1]$, such that ${\textstyle\sum}_{i \in C} k_i$ (mod $M$)= 0, where $M$ is a large integer. 
	\State 2: Party $i$ obtains keystream $k_i$.

\vspace{2mm}

\State \textbf{Enc($m$, $k$)}
  \State 1: Represent message $m$ as integer $m \in [0, M-1]$.
  \State 2: Let $k$ be a randomly generated keystream, where $k \in [0, M-1]$.
  \State 3: Compute $c = Enc(m, k) =m+k$.

\vspace{2mm}
 
\State \textbf{Dec($c$, $k$)}
  \State 1: $Dec(c, k) =c-k$.

\vspace{2mm}

\State \textbf{AggrDec($k_i$)}
	\State 1: Let $c_j=Enc(m_j, k_j)$, where $j \in C \setminus i$.%, j \neq i$.
	\State 2: Party $i$ uses $-k_i={\textstyle\sum}_{j \in C \setminus i} k_j$ to decrypt the aggregation of other parties as follows:
	$Dec({\textstyle\sum}_{j \in C \setminus i} c_j, -k_i) ={\textstyle\sum}_{j \in C \setminus i} c_j-{\textstyle\sum}_{j \in C \setminus i} k_j={\textstyle\sum}_{j \in C \setminus i} m_j$.	
\vspace{2mm}
\end{algorithmic}
\end{algorithm}

Model parameter of party $i$ is updated as per gradients-encrypted SGD as follows:
\begin{align*}
\vec{w_i}'&=\vec{w_i}+\Delta \vec{w_i}+Dec({\textstyle\sum}_{j \in C \setminus i} Enc(\Delta \vec{\tilde{w}_j^i},k_j),-k_i) \\
&=\vec{w_i}+\Delta \vec{w_i}+{\textstyle\sum}_{j \in C \setminus i} \Delta \vec{\tilde{w}_j^i}
\end{align*}
where $Enc$ and $Dec$ correspond to encryption and decryption operations in Algorithm~\ref{algo:Encryption}, $\vec{w_i}$ is the local parameters of party $i$ at previous round, $\Delta \vec{\tilde{w}_j^i}$ is the masked gradient vector of party $j$ shared with party $i$, where only $d_i^j$ gradients are meaningful, \ie $d_i^j$ elements of total $|\Delta \vec{\tilde{w}_j^i}|$ elements are kept intact, while the remaining $|\Delta \vec{\tilde{w}_j^i}|-d_i^j$ elements are nullified as 0. The second equality follows the homomorphic addition property, thus participant $i$ can get the updated $\vec{w_i}'$ correctly after decryption, without having access to either $\Delta \vec{\tilde{w}_j^i}$ or $\Delta \vec{w_j}$. FPPDL ensures party obliviousness by ensuring that each participant knows nothing but the sum of its received gradients in each communication round, and cannot infer any information about other participants' data.

\subsubsection{Three-layer Onion-style Encryption}
However, as all parties need to store different encrypted gradients that are meant to be sent to different parties on Blockchain for commitment, all the encrypted gradients are also accessible to all parties. Applying public-key encryption on top of homomorphic encryption for authentication~\cite{lyu2018ppfa} can address this problem. However, as the released gradient vector is high-dimensional, encrypting gradient vector is both computation and communication expensive.

Therefore, we propose a three-layer onion-style encryption scheme. The first layer protects local model gradients by using symmetric key keystream $k_j$ for homomorphic encryption, as presented in Algorithm~\ref{algo:Encryption}. The second layer and the third layer are classic hybrid encryption, as used in OpenPGP~\cite{callas2007openpgp} for instance. In particular, in the second layer, a fresh symmetric encryption key $fsk$ will be generated and used to re-encrypt the ciphertext of the first layer, and then the fresh symmetric key is encrypted by using the receiver's public key $pk_i$ in the third layer. In this way, the encryption of high-dimensional data becomes very effective, and the receiver could be authenticated as well: only the receiver who has the corresponding secret key $sk_i$ paired with the public key $pk_i$ can decrypt the two-layer encrypted gradients committed on the Blockchain. 

\subsubsection{Local Credibility Update}
Instead of using the standalone models as in the local credibility initialization, during each round of collaborative learning, each party randomly selects and shares a subset of DPGAN samples as per individual sharing level, then calculates the local credibility of other parties based on the returned labels, which are evaluated by using its updated local model at current round. The mutual evaluation follows the same procedure as in Step 3 of Algorithm \ref{Algorithm:benchmarking}. Finally, local credibility of each party is updated by integrating its historical local credibility as per Step 4 of Algorithm~\ref{Algorithm:global_credibility_update}. In this way, local credibility of each party can be adaptively updated, reflecting more accurately how one party contributes to different parties during collaborative learning. 

\subsection{Quantification of Fairness}
In collaborative learning system, collaborative fairness should be quantified from the point of view of the whole system. In this work, we quantify collaborative fairness through the correlation coefficient between party contributions (\ie standalone model accuracy which characterizes the learning capability of each party on its own local data, and sharing level, which characterizes the sharing willingness of each party) and party rewards (\ie final model accuracies of different parties). 

Specifically, we take party contributions as the X-axis, which represents the contributions of different parties from the system view. In particular, in Setting 2, we characterize different parties' contributions by their sharing levels and standalone model accuracies, as the party who is less private and has local data with better generalization empirically contributes more. In Setting 1 and Setting 3, we characterize different parties' contributions by their standalone model accuracies, as the party who has local data with better generalization empirically contributes more. Moreover, in Setting 3, the party with more local data typically yields higher standalone model accuracy in IID scenarios. In summary, the X-axis can be expressed by Equation~\ref{eq:x_axis}, where $\lambda_j$ and $sacc_j$ denote the sharing level and standalone model accuracy of party $j$ respectively: 
\begin{equation}\label{eq:x_axis}
\vec{x}=
\resizebox{0.93\hsize}{!}{
$\Big\{\begin{array}{cl}
\{\frac{\lambda_1}{\textstyle\sum \lambda_j},\cdots,\frac{\lambda_n}{\textstyle\sum \lambda_j}\}+\{\frac{sacc_1}{\textstyle\sum sacc_j},\cdots,\frac{sacc_n}{\textstyle\sum sacc_j}\},  & \mbox{Setting 2} \\
\{sacc_1,\cdots,sacc_n\}, & \mbox{Setting 1\&3} 
\end{array}
$}
\end{equation}

Similarly, we take party rewards (\ie final model accuracies of different parties) as the Y-axis, as expressed by Equation~\ref{eq:y_axis}, where $acc_j$ denotes the final model accuracy of party $j$:
\begin{equation}\label{eq:y_axis}
\vec{y}=\{acc_1,\cdots,acc_n\}
\end{equation}

As the Y-axis measures local model performance of different parties after collaboration, it is expected to be positively correlated with the X-axis to deliver good fairness. Hence, we formally quantify collaborative fairness in Equation~\ref{eq:fairness}:
\begin{equation}\label{eq:fairness}
r_{xy}=\frac{{\textstyle\sum}_{i=1}^n (x_i-\bar{x})(y_i-\bar{y})}{(n-1)s_xs_y}
\end{equation}
where $\bar{x}$ and $\bar{y}$ are the sample means of $\vec{x}$ and $\vec{y}$, $s_x$ and $s_y$ are the corrected standard deviations. The range of fairness is within [-1,1], with higher values implying good fairness. Conversely, negative coefficient implies poor fairness.

\section{Experimental Evaluation}
\label{sec:Performance}
In this section, we evaluate the performance of the proposed FPPDL framework by comparing it against the state of the art on real-world datasets.

\subsection{Datasets}
We implement experiments on two benchmark image datasets. The first is the MNIST dataset\footnote{\url{http://yann.lecun.com/exdb/mnist/}} for handwritten digit recognition consisting of 60,000 training examples and 10,000 test examples. Each example is a 32x32 gray-level image~\cite{shokri2015privacy}, with digits locating at the center of the image. The second is the SVHN dataset\footnote{\url{http://ufldl.stanford.edu/housenumbers/}} of house numbers obtained from Google's street view images, which contains 600,000 training examples, from which we use 100,000 for training and 10,000 for testing. Each example is a 32x32 centered image with three channels (RGB). SVHN is more challenging as most of the images are noisy, and contain distractors at the sides. The size of the input layer of neural networks for MNIST and SVHN are 1024 and 3072, respectively. The objective is to classify the input as one of 10 possible digits within [``0''-``9''], thus the size of the output layer is 10. We normalize the training examples by subtracting the average and dividing by the standard deviation of training examples. For reproducibility purposes, our code will be made available here: https://github.com/lingjuanlv/FPPDL.

\subsection{Baselines}
We demonstrate the effectiveness of our proposed FPPDL framework by comparison with the following three frameworks. In all frameworks, stochastic gradient descent (SGD) is applied to each party.

\begin{enumerate}
    \item \textit{Standalone} framework: which assumes parties train standalone models on local training data without any collaboration. This framework delivers maximum privacy, but minimum utility, because each party is susceptible to falling into local optima when training alone.
    \item \textit{Centralized} framework: which allows a trusted server to have access to all participants' data in the clear, and train a global model on the combined data using standard SGD. Hence, it is a privacy-violating framework.
    \item \textit{Distributed} framework: which enables parties to train independently and concurrently, and chooses a fraction of parameters to be uploaded at each iteration. In particular, as shown in~\cite{shokri2015privacy}, Distributed Selective SGD (DSSGD) achieves even higher accuracy than the centralized SGD because updating only a small fraction of parameters at each round acts as a regularization technique to avoid overfitting. Hence, we take DSSGD for the analysis of the distributed framework. As DSSGD with round robin parameter exchange protocol results in the highest accuracy~\cite{shokri2015privacy} and facilitates fairness calculation, we follow the round robin protocol for DSSGD, where participants run SSGD sequentially, each downloads a fraction of the most updated parameters from the server, runs local training, and uploads selected gradients; the next party follows in the fixed order. Gradients are uploaded according to the ``largest values'' criterion. 
\end{enumerate}

\subsection{Experiment Setup}
\label{sec:Setup}
For local model architecture, we consider two popular neural network architectures: \emph{multi-layer perceptron} (MLP) and \emph{convolutional neural network} (CNN), which are the same as in~\cite{shokri2015privacy}. 
For local model training, we set the learning rate as 0.001, learning rate decay as 1e-7, and mini-batch size as 1. In addition, to reduce the impact of different initializations and avoid non-convergence, each party is initialized with the same parameter $w_0$, then local training is run on individual training data to update local model parameter $\vec{w_i}$. To boost fairness, we let each party individually train 10 epochs before collaborative learning starts. For all experiments, we empirically set the local credibility threshold as 
$c_{th}=\frac{1}{|C|-1}*\frac{2}{3}$ via grid search, where $|C|$ is the number of alive parties, \ie credible parties in the system. Next, we investigate three realistic IID settings as follows: 

\textbf{Setting 1: Same sharing level, same data size:} in the first case, sharing level of each party is set as $0.1$, \ie each party only releases $10\%$ meaningful gradients during collaboration. For each party, we randomly sample $1\%$ of the entire database as the local training data of each party, \ie $600$ examples for MNIST and $1000$ examples for SVHN, this setting is the same as Shokri \etal~\cite{shokri2015privacy} when the upload rate of each party equals 0.1;

\textbf{Setting 2: Different sharing level, same data size:} in the second case, sharing level of each party is randomly sampled from $[0.1,0.5]$, and parties release meaningful gradients as per individual sharing level during collaboration. For each participant, we randomly sample $1\%$ of the entire database as local training data as above. 

\textbf{Setting 3: Different data size, same sharing level:} in the third case, we simulate the case where different parties have different data size. In particular, for MNIST dataset, we randomly partition total \{2400, 9000, 18000, 30000\} examples among \{4,15,30,50\} parties respectively. Similarly, for SVHN dataset, total \{4000, 15000, 30000, 50000\} examples are randomly partitioned among \{4,15,30,50\} parties respectively. The sharing level of each party is fixed to $0.1$. 

\textbf{Remark}. 
In Setting 1 and Setting 2, the purpose of allocating $600$ MNIST examples or $1000$ SVHN examples for each party is to fairly compare with Shokri \etal~\cite{shokri2015privacy}, in which each party is allocated with 600 MNIST examples or $1000$ SVHN examples (small number of local examples to simulate data scarity which necessitates collaboration). Therefore, for MNIST, we simulate the total examples of 2400 (4 parties) up to 30,000 (50 parties). For larger datasets like 300,000 examples, it would require 500 parties, imposing heavy requirement on real deployment, while delivering similar analysis as in Sec.~\ref{sec:results}. We also remark that our Setting 2 and Setting 3 are relatively conservative, by increasing the contribution diversity among parties, for example, sampling sharing level from [0,1] instead of [0,0.5], partitioning data size among parties in a more imbalanced way, our FPPDL can definitely results in higher fairness.

\subsection{Experimental Results}
\label{sec:results}
\begin{table}[ht]
\caption{Fairness of distributed framework and our FPPDL over MNIST dataset, with different model architectures, different party numbers (P-$k$) and different settings as described in Section~\ref{sec:Setup}.}
\label{tbl:MNIST_fairness}
\centering
\begin{tabularx}{\linewidth}{|p{0.25cm}|c|c|c|c|c|c|X|X|}
\hline
\multirow{2}{*}{} & \multicolumn{4}{c|}{Setting 2} & \multicolumn{4}{c|}{Setting 3}
\tabularnewline
\cline{2-9}
 & \multicolumn{2}{c|}{Distributed} & \multicolumn{2}{c|}{FPPDL} & \multicolumn{2}{c|}{Distributed} & \multicolumn{2}{c|}{FPPDL}
\tabularnewline
\hline
 & CNN & MLP & CNN & MLP & CNN & MLP & CNN & MLP
\tabularnewline
\hline
\textit{P4}  &-0.68 &0.30 &\textbf{0.89} & \textbf{0.92} &-0.97 &0.05 & \textbf{0.98} &\textbf{0.96}
\tabularnewline
\hline
\textit{P15} &0.20 &-0.15 & \textbf{0.76} &\textbf{0.82}  &0.03 &-0.07  & \textbf{0.90} &\textbf{0.83}
\tabularnewline
\hline
\textit{P30} &-0.02 &0.02 &\textbf{0.79}  &\textbf{0.85} &0.13 &0.01 & \textbf{0.75} &\textbf{0.63}
\tabularnewline
\hline
\textit{P50} &-0.16 &-0.05 & \textbf{0.75} &\textbf{0.67}  &0.14 &-0.07 & \textbf{0.72} &\textbf{0.60}
\tabularnewline
\hline
\end{tabularx}
\end{table}

\begin{table}[ht]
\caption{Fairness of distributed framework and our FPPDL over SVHN dataset, with different model architectures, different party numbers (P-$k$) and different settings.}
\label{tbl:SVHN_fairness}
\centering
\begin{tabularx}{\linewidth}{|p{0.25cm}|c|c|c|c|c|c|X|X|}
\hline
\multirow{2}{*}{} & \multicolumn{4}{c|}{Setting 2} & \multicolumn{4}{c|}{Setting 3}
\tabularnewline
\cline{2-9}
 & \multicolumn{2}{c|}{Distributed} & \multicolumn{2}{c|}{FPPDL} & \multicolumn{2}{c|}{Distributed} & \multicolumn{2}{c|}{FPPDL}
\tabularnewline
\hline
 & CNN & MLP & CNN & MLP & CNN & MLP & CNN & MLP
\tabularnewline
\hline
\textit{P4}  &0.27 &0.26 &\textbf{0.78} &\textbf{0.76} &0.28 &0.20 &\textbf{0.97} &\textbf{0.93}
\tabularnewline
\hline
\textit{P15} &0.16 &0.19 &\textbf{0.77} &\textbf{0.71}  &-0.13 &0.16  &\textbf{0.87} &\textbf{0.88}
\tabularnewline
\hline
\textit{P30} &-0.14 &0.12 &\textbf{0.68} &\textbf{0.65}  &-0.15 &-0.27 &\textbf{0.67} &\textbf{0.78}
\tabularnewline
\hline
\textit{P50} &-0.25 &-0.37 &\textbf{0.67} &\textbf{0.66}  &-0.23 &0.15 &\textbf{0.65} &\textbf{0.69}
\tabularnewline
\hline
\end{tabularx}
\end{table}

For collaborative fairness comparison, we only analyze our FPPDL and the distributed framework using DSSGD, neglecting centralized framework and standalone framework, because parties cannot get access to the trained global model in the centralized framework, while parties do not collaborate in the standalone framework. Table~\ref{tbl:MNIST_fairness} and Table~\ref{tbl:SVHN_fairness} list the calculated fairness of the distributed framework and our FPPDL over MNIST and SVHN datasets, with different architectures, different party numbers and different settings, as detailed in Section~\ref{sec:Setup}. In particular, we omit the results for setting 1 with the same sharing level and same data size, as fairness is a less concerned problem in this setting. All the fairness results for setting 2 and setting 3 are averaged over five trails to reduce the impact of different initialization in each trail. 

As is evidenced by the high positive values of fairness, with most of them above 0.5, FPPDL achieves reasonably good fairness, confirming the intuition behind fairness: the party who is less private and has more training data delivers higher accuracy. In contrast, the distributed framework exhibits bad fairness with significantly lower values than that of FPPDL in all cases, and even negative values in some cases, manifesting the lack of fairness in the distributed framework. This is because in the distributed framework, all the participating parties can derive similarly well models, no matter how much one party contributes.

\begin{figure*}[!htp]
\centering
        \begin{subfigure}[ht]{0.23\textwidth}
                \includegraphics[width=4cm,height=3.8cm]{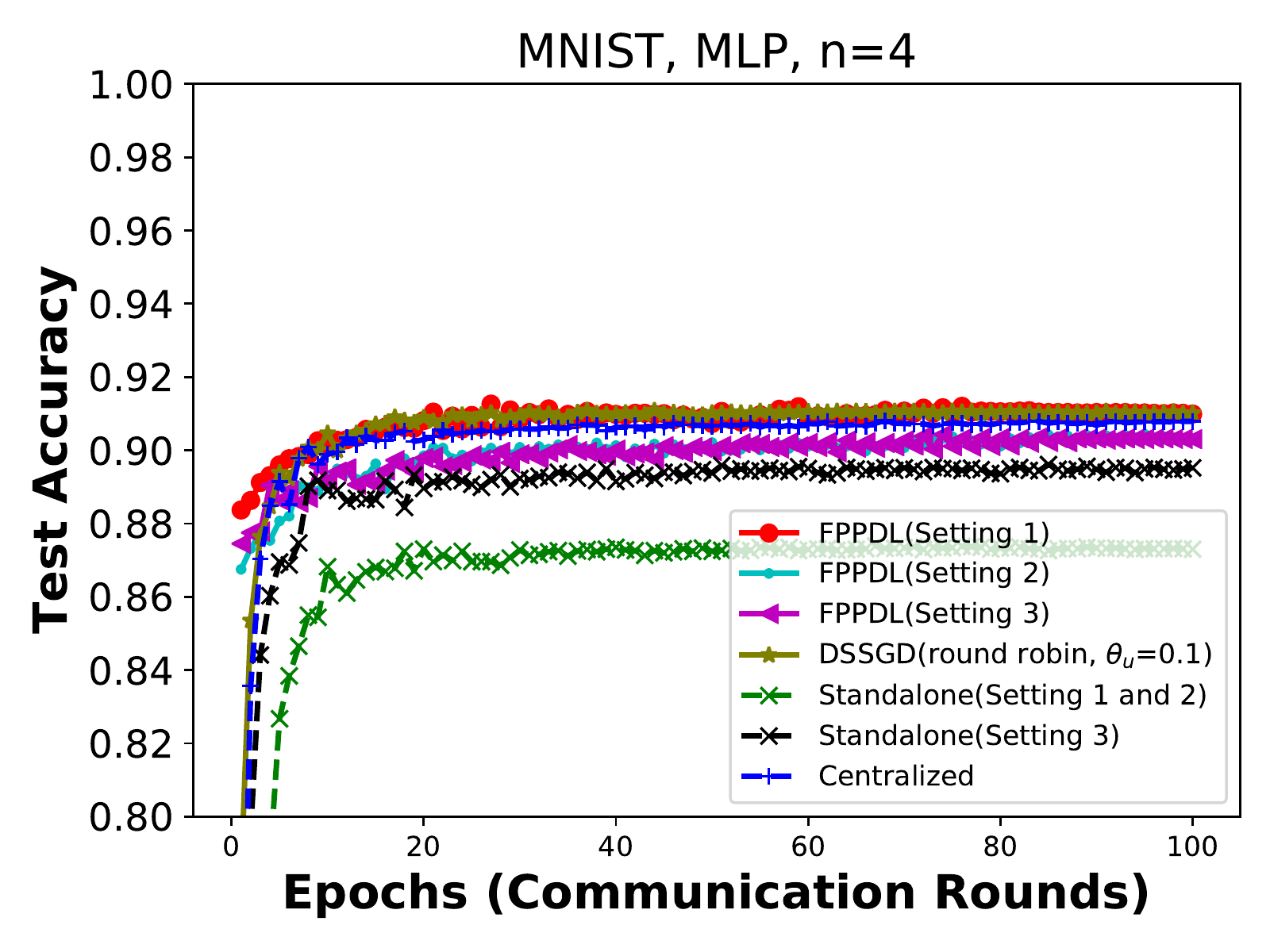}\label{fig:mnist_party4_epoch100_deep}
        \end{subfigure}
        \begin{subfigure}[ht]{0.23\textwidth}
                \includegraphics[width=4cm,height=3.8cm]{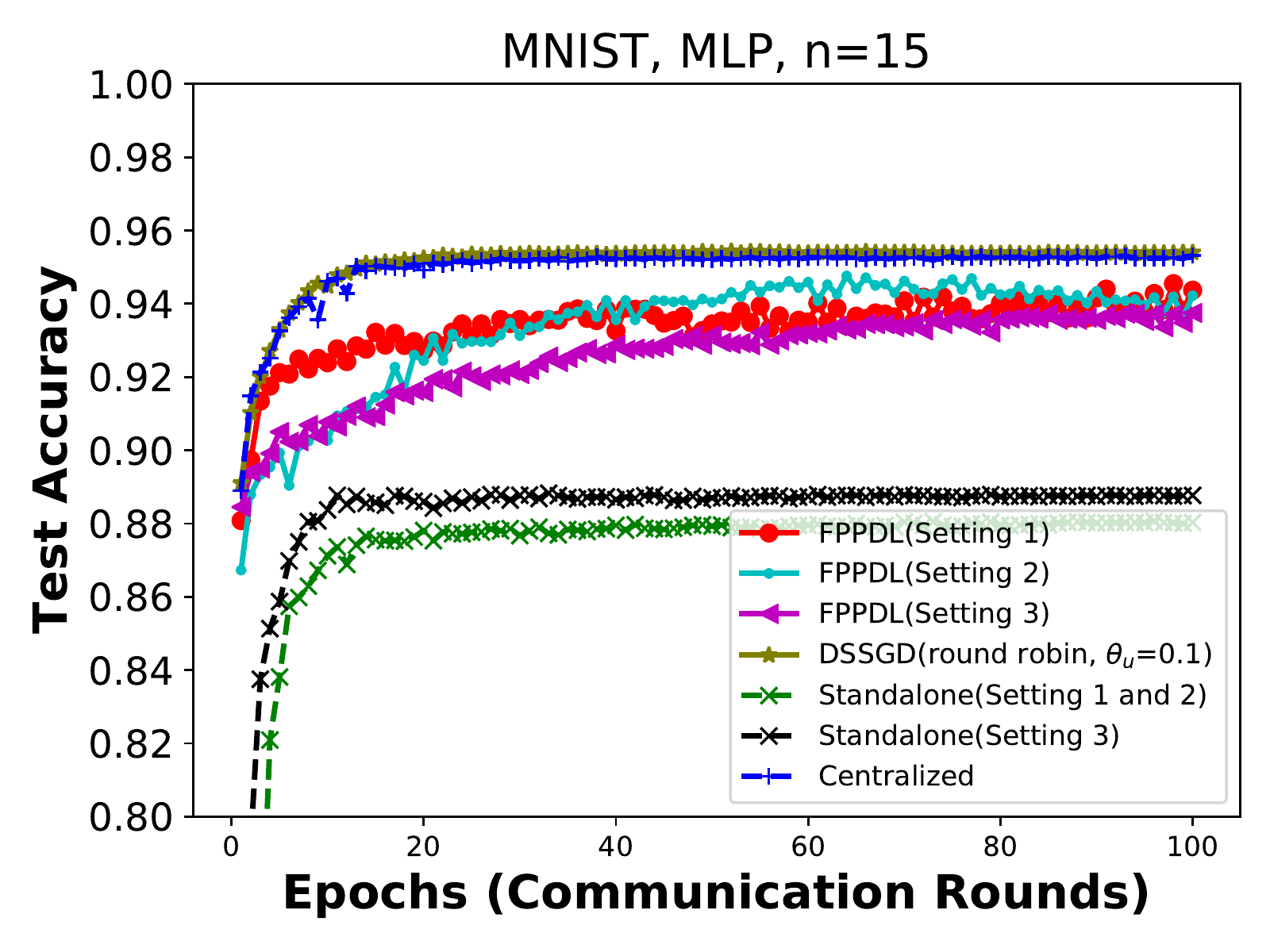}\label{fig:mnist_party15_epoch100_deep}
        \end{subfigure}
        \begin{subfigure}[ht]{0.23\textwidth}
                \includegraphics[width=4cm,height=3.8cm]{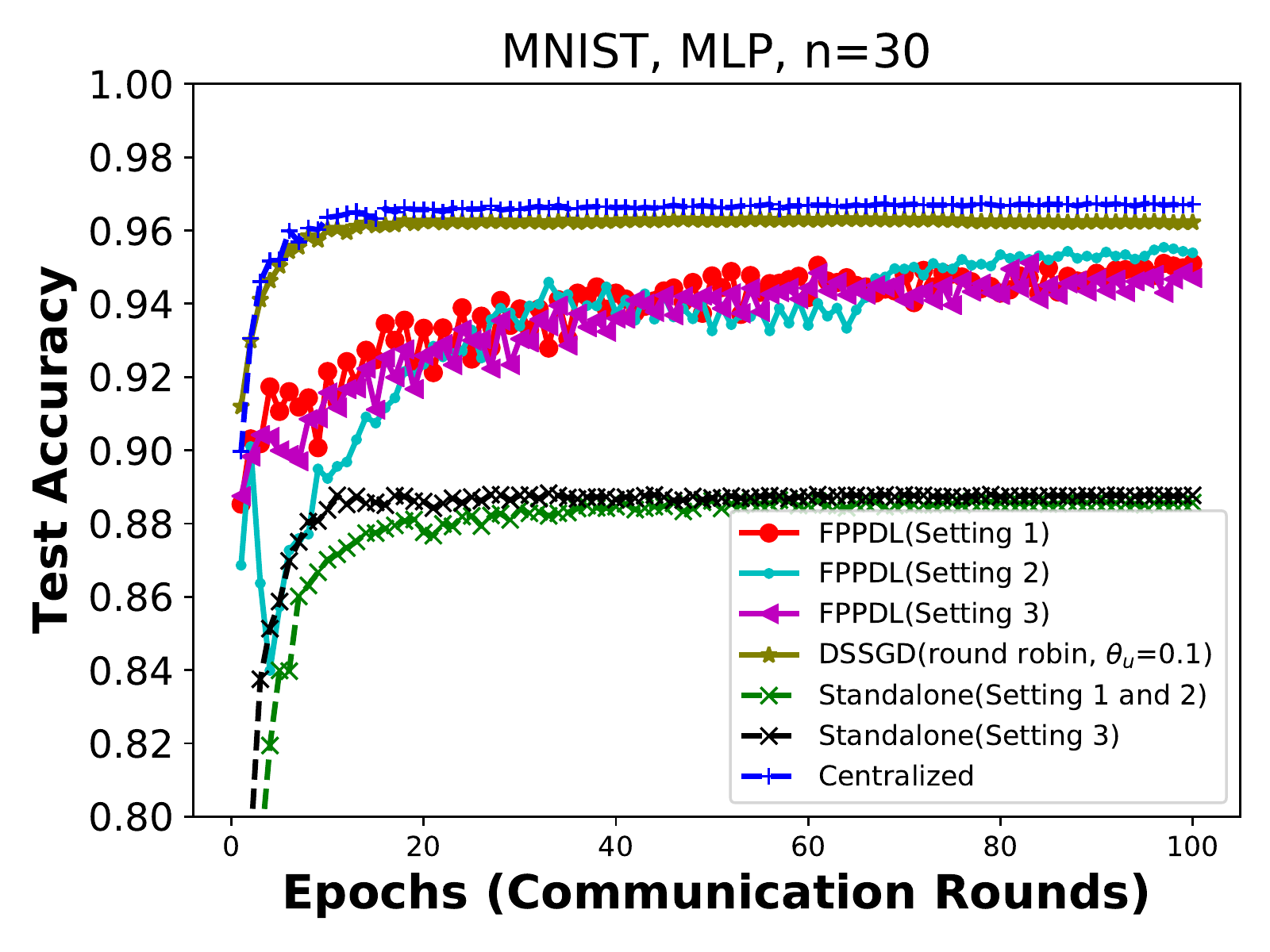}\label{fig:mnist_party30_epoch100_deep}
        \end{subfigure}
        \begin{subfigure}[ht]{0.23\textwidth}
                \includegraphics[width=4cm,height=3.8cm]{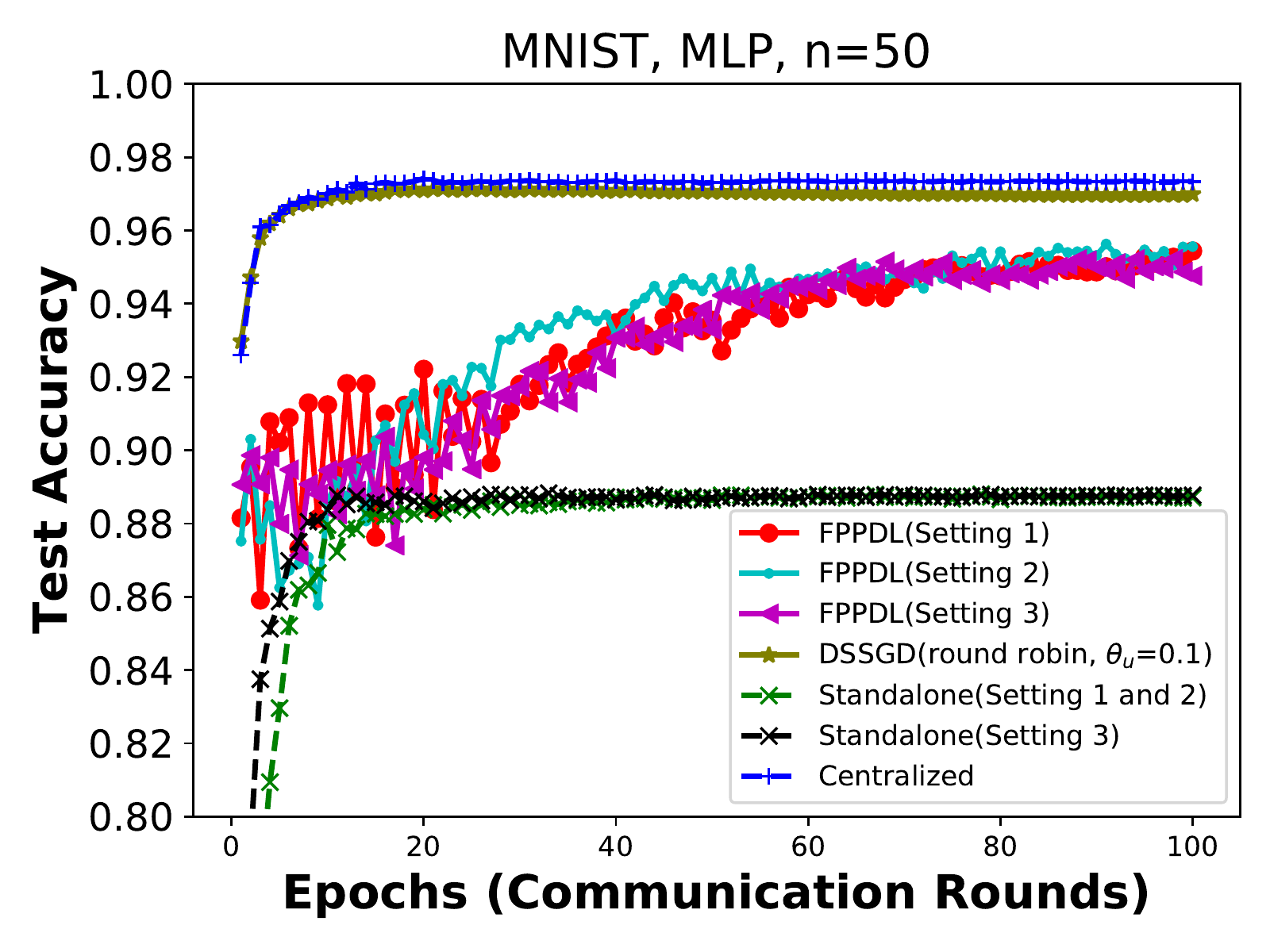}\label{fig:mnist_party50_epoch100_deep}
        \end{subfigure}
        \caption{System convergence for MNIST MLP. Collaboration involves different number of parties in $\{4,15,30,50\}$.}
\label{fig:mnist_epoch100_deep}
\end{figure*}

% \vskip -2\baselineskip plus -1fil

\begin{figure*}[!htp]
\centering
        \begin{subfigure}[ht]{0.23\textwidth}
                \includegraphics[width=4cm,height=3.8cm]{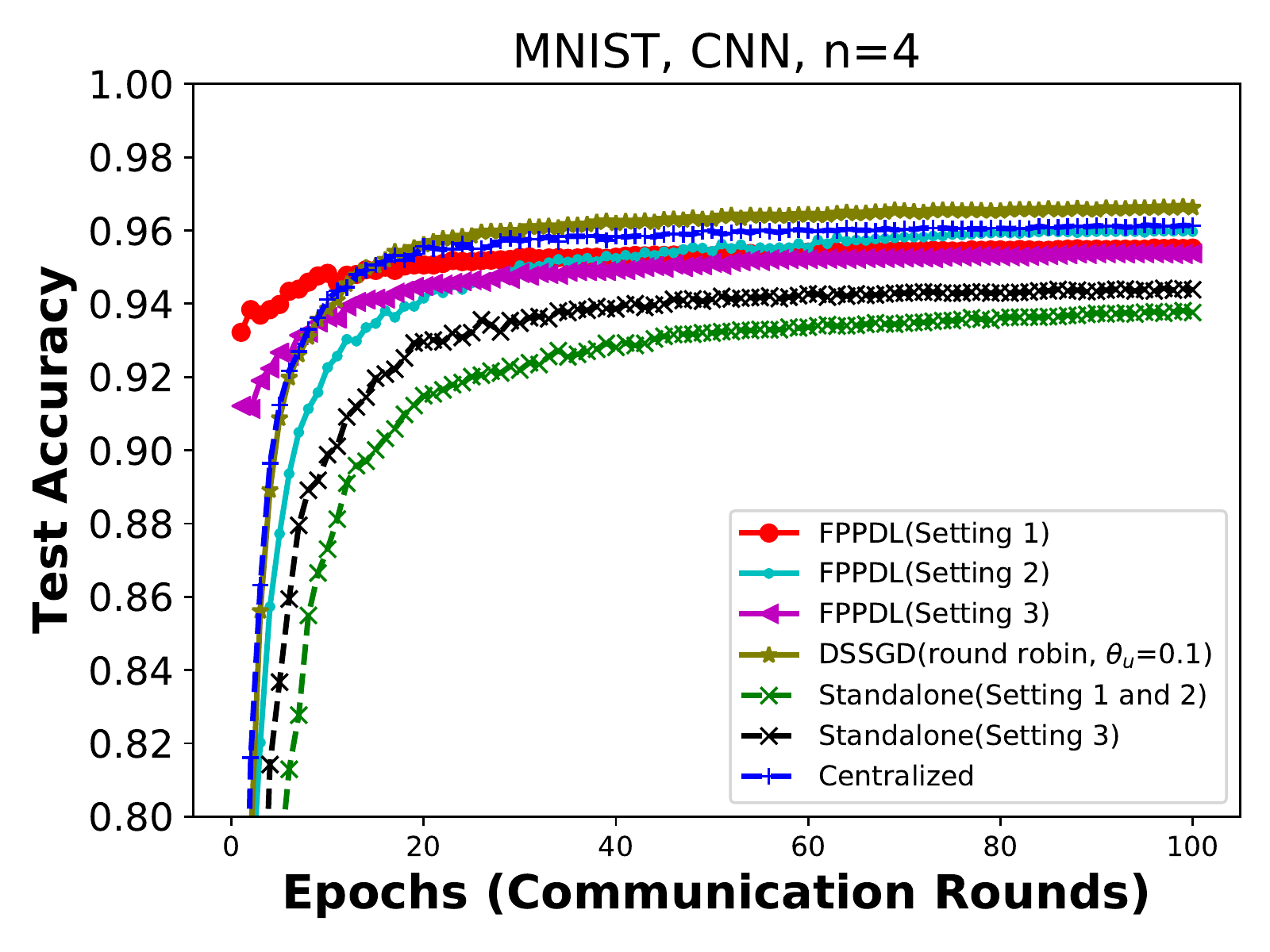}\label{fig:mnist_party4_epoch100_cvn}
        \end{subfigure}
        \begin{subfigure}[ht]{0.23\textwidth}
                \includegraphics[width=4cm,height=3.8cm]{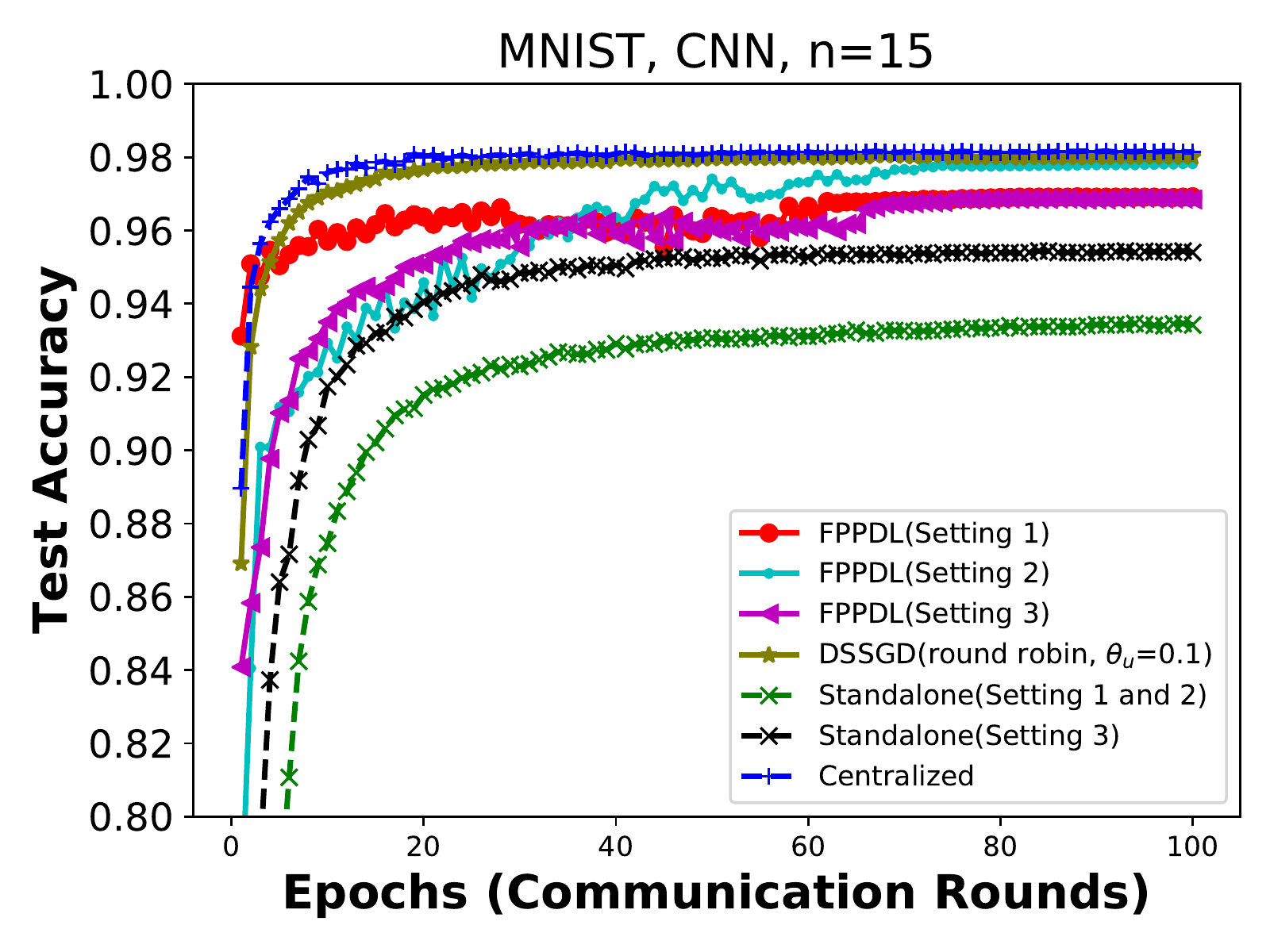}\label{fig:mnist_party15_epoch100_cvn}
        \end{subfigure}
        \begin{subfigure}[ht]{0.23\textwidth}
                \includegraphics[width=4cm,height=3.8cm]{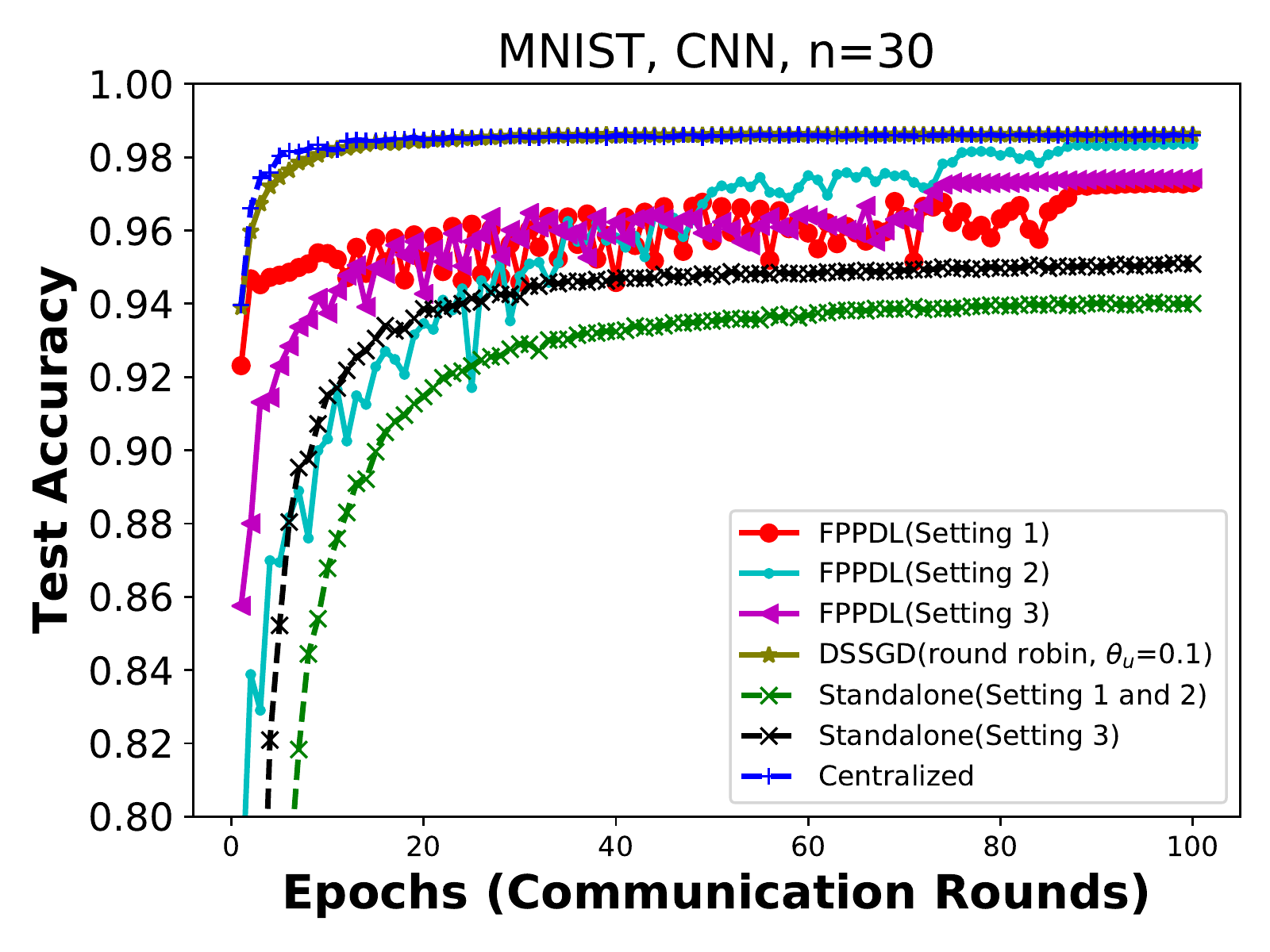}\label{fig:mnist_party30_epoch100_cvn}
        \end{subfigure}
        \begin{subfigure}[ht]{0.23\textwidth}
                \includegraphics[width=4cm,height=3.8cm]{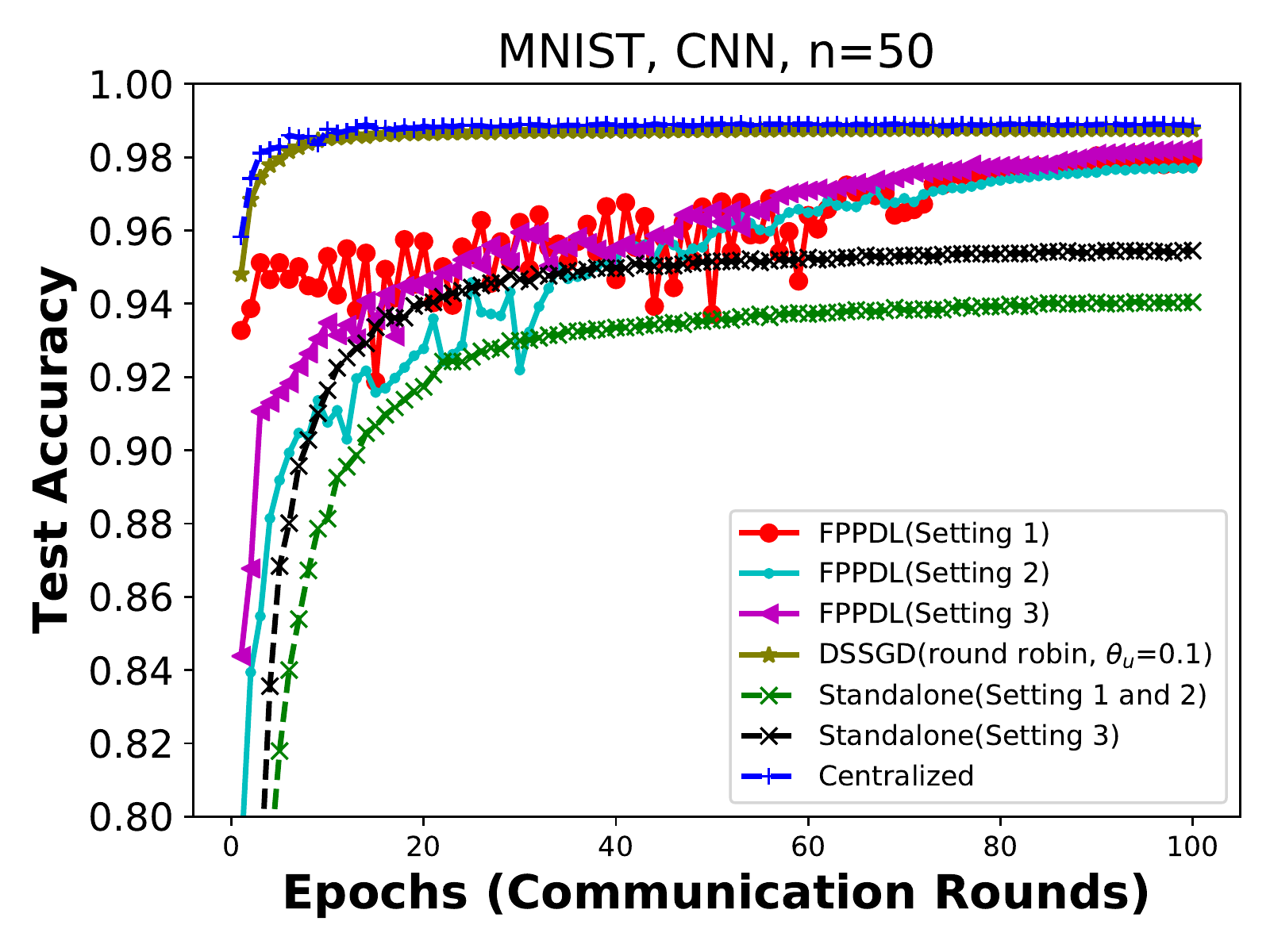}\label{fig:mnist_party50_epoch100_cvn}
        \end{subfigure}
        \caption{System convergence for MNIST CNN. Collaboration involves different number of parties in $\{4,15,30,50\}$.}
\label{fig:mnist_epoch100_cvn}
\end{figure*}

\begin{figure*}[t]
\centering
        \begin{subfigure}[ht]{0.23\textwidth}
                \includegraphics[width=4cm,height=3.8cm]{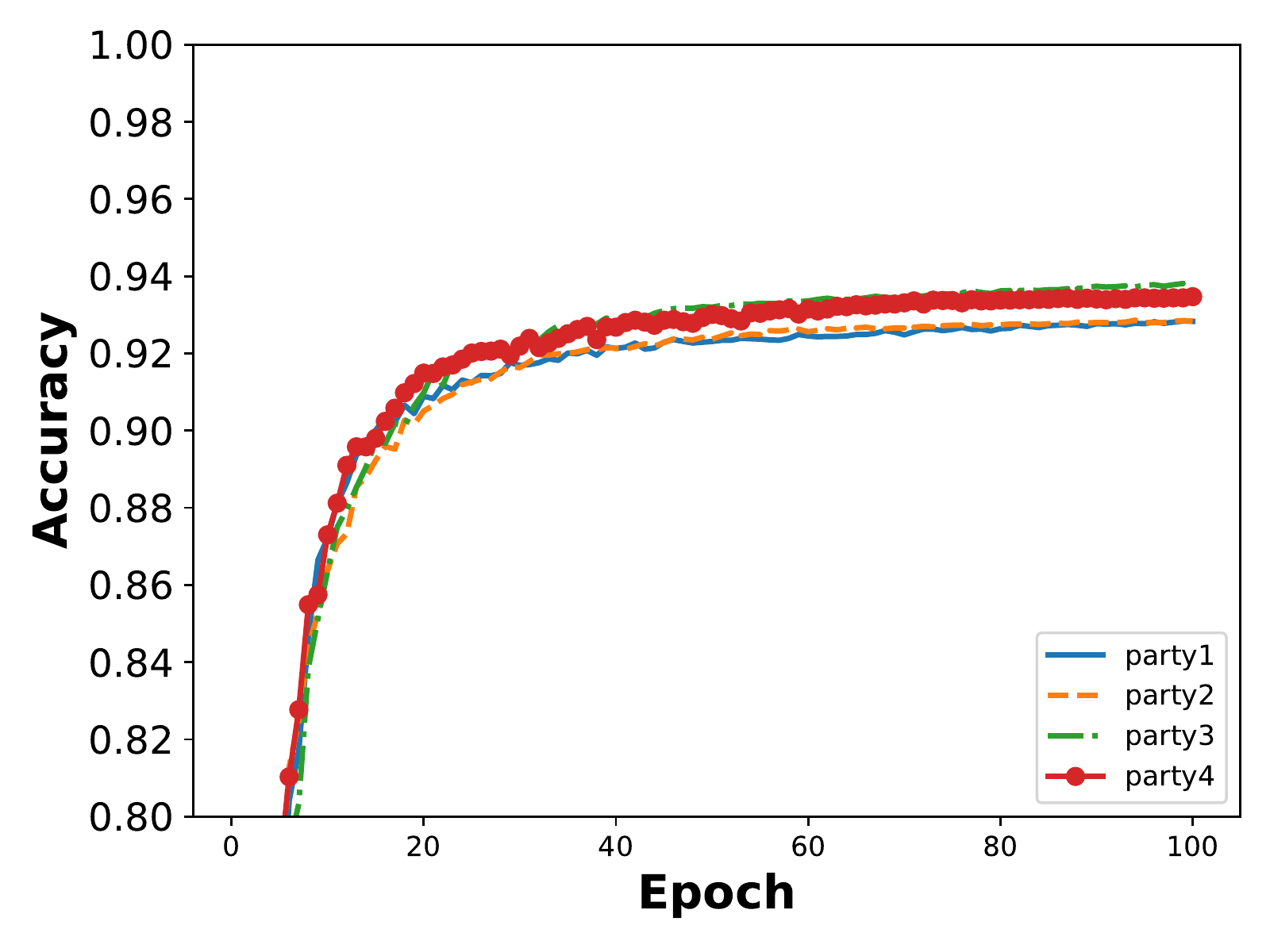}\label{fig:FPPDL_mnist_cvn_p4e100_setting1_standalone}
                \subcaption{Standalone Setting 2}
        \end{subfigure}
        \begin{subfigure}[ht]{0.23\textwidth}
                \includegraphics[width=4cm,height=3.8cm]{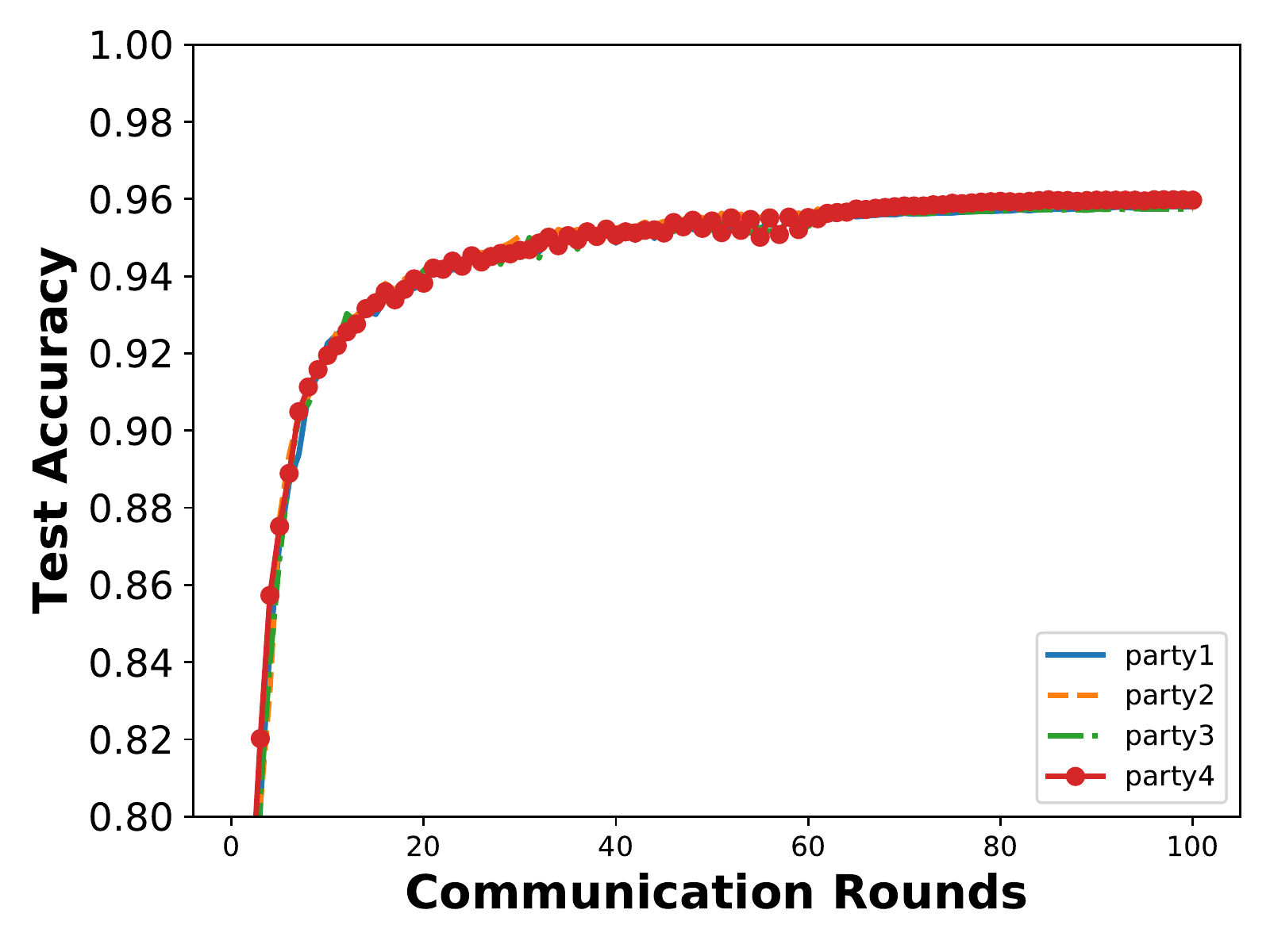}\label{fig:FPPDL_mnist_cvn_p4e100_setting2_pretrain1}
                \subcaption{FPPDL Setting 2}
        \end{subfigure}
        \begin{subfigure}[ht]{0.23\textwidth}
                \includegraphics[width=4cm,height=3.8cm]{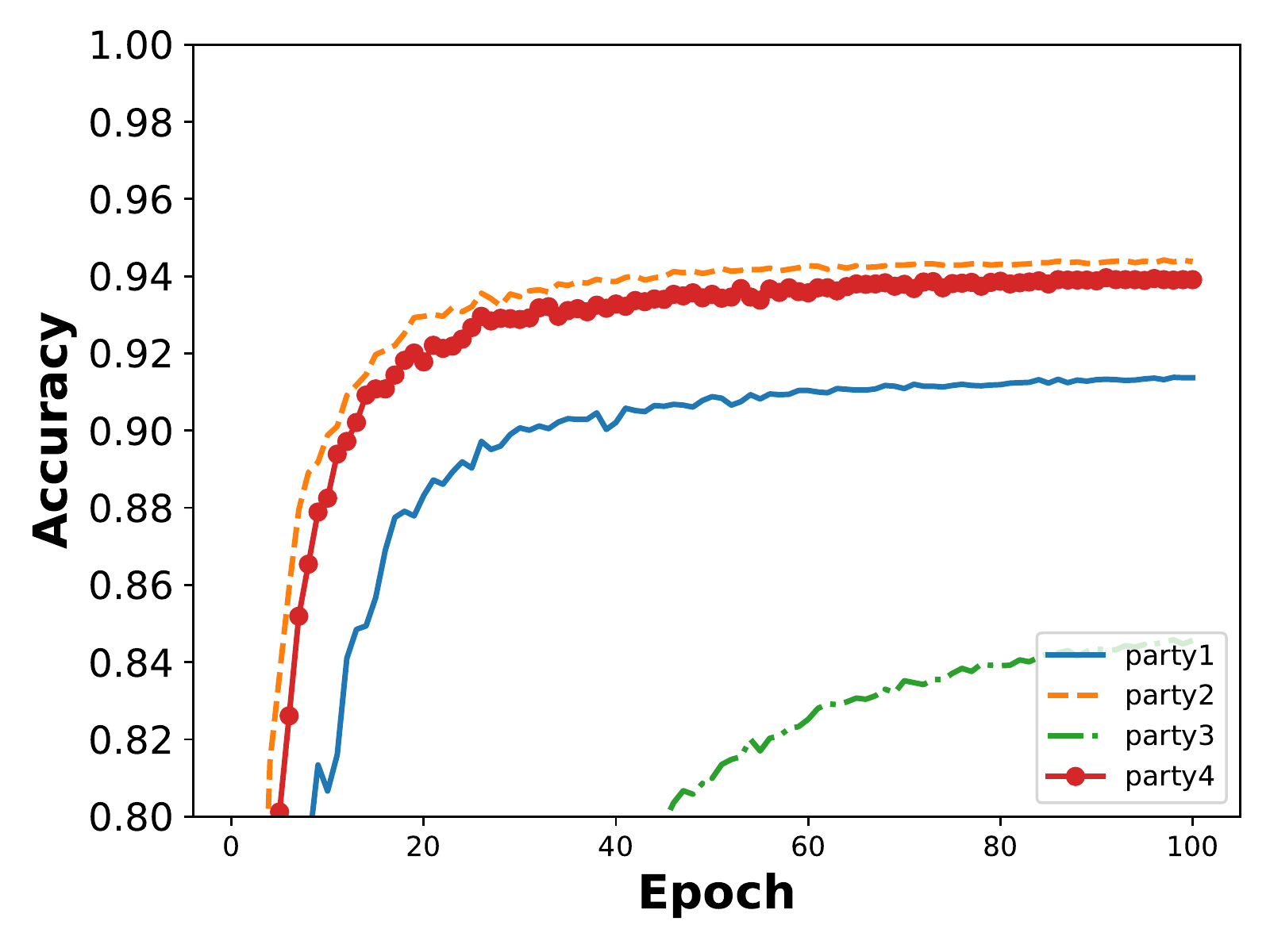}\label{fig:FPPDL_mnist_cvn_p4e100_setting3_standalone}
                \subcaption{Standalone Setting 3}
        \end{subfigure}
        \begin{subfigure}[ht]{0.23\textwidth}
                \includegraphics[width=4cm,height=3.8cm]{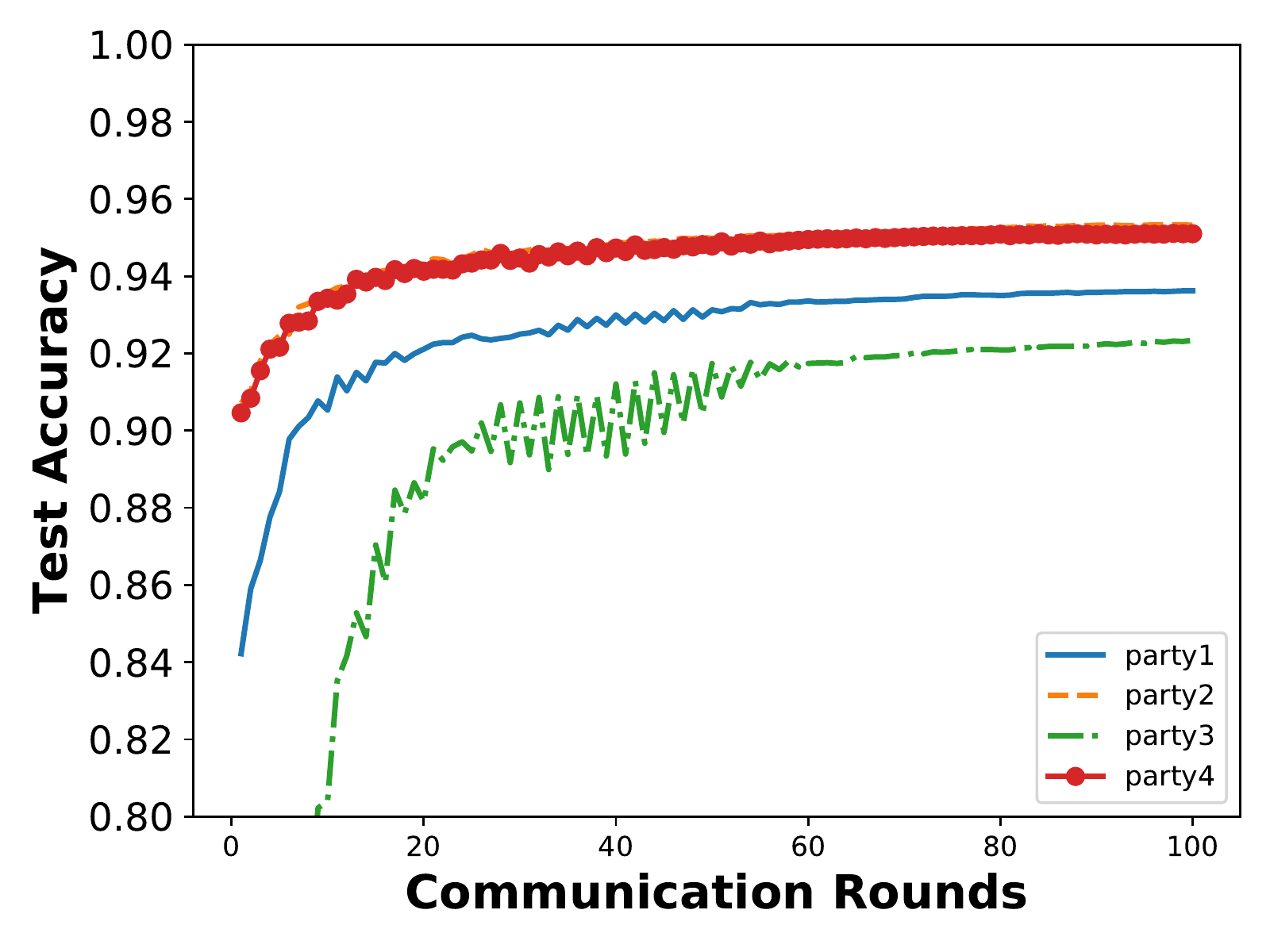}\label{fig:FPPDL_mnist_cvn_p4e100_setting3_pretrain1}
                \subcaption{FPPDL Setting 3}
        \end{subfigure}
        \caption{Individual convergence for MNIST CNN using Standalone framework and our FPPDL (P4, B=1, E=1, lr=0.001).}
\label{fig:mnist_p4_cvn_convergence}
\end{figure*}

\begin{figure*}[!htp]
\centering
        \begin{subfigure}[ht]{0.23\textwidth}
                \includegraphics[width=4cm,height=3.8cm]{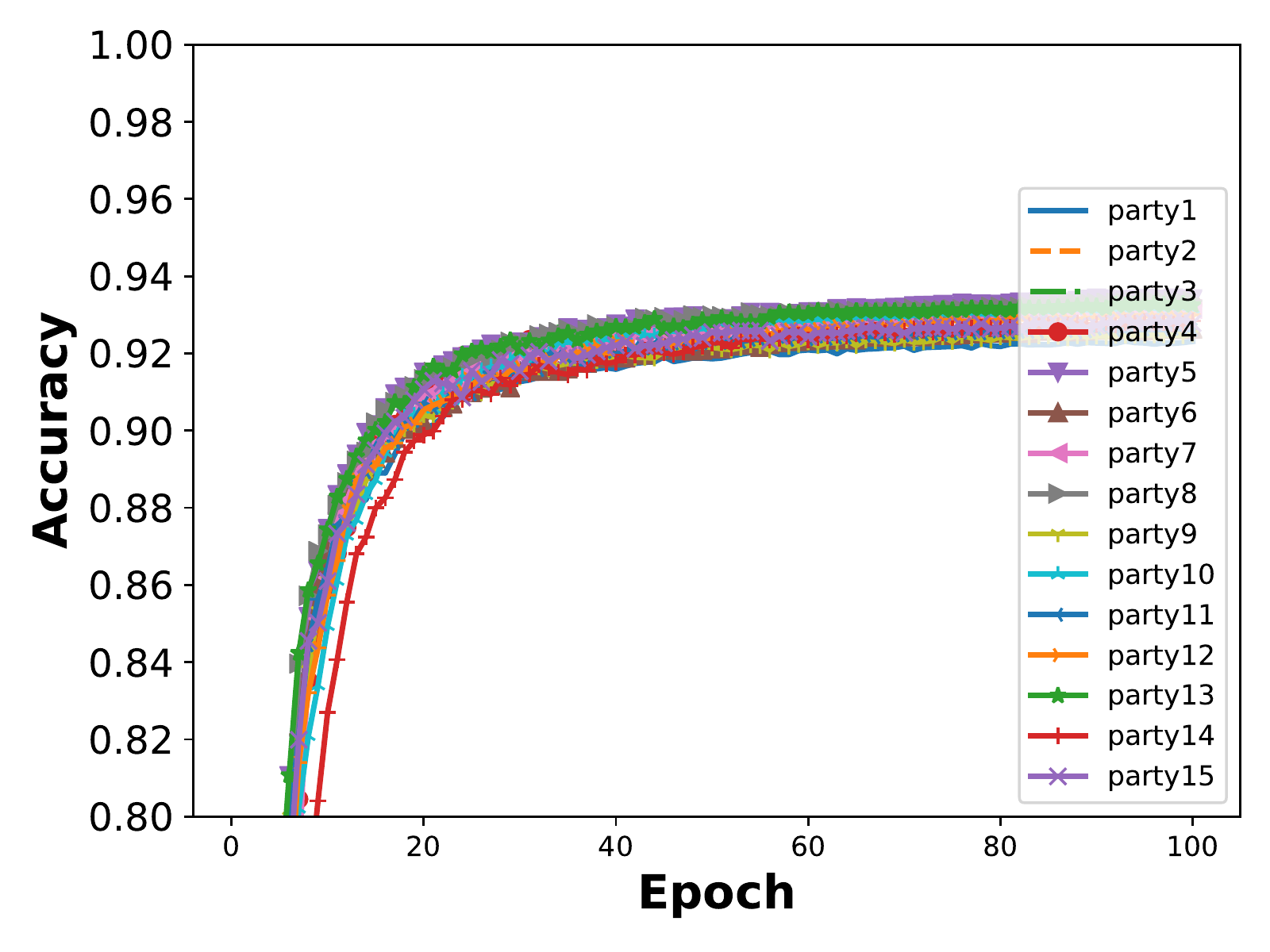}\label{fig:FPPDL_mnist_cvn_p15e100_setting1_standalone}
                \subcaption{Standalone Setting 2}
        \end{subfigure}
        \begin{subfigure}[ht]{0.23\textwidth}
                \includegraphics[width=4cm,height=3.8cm]{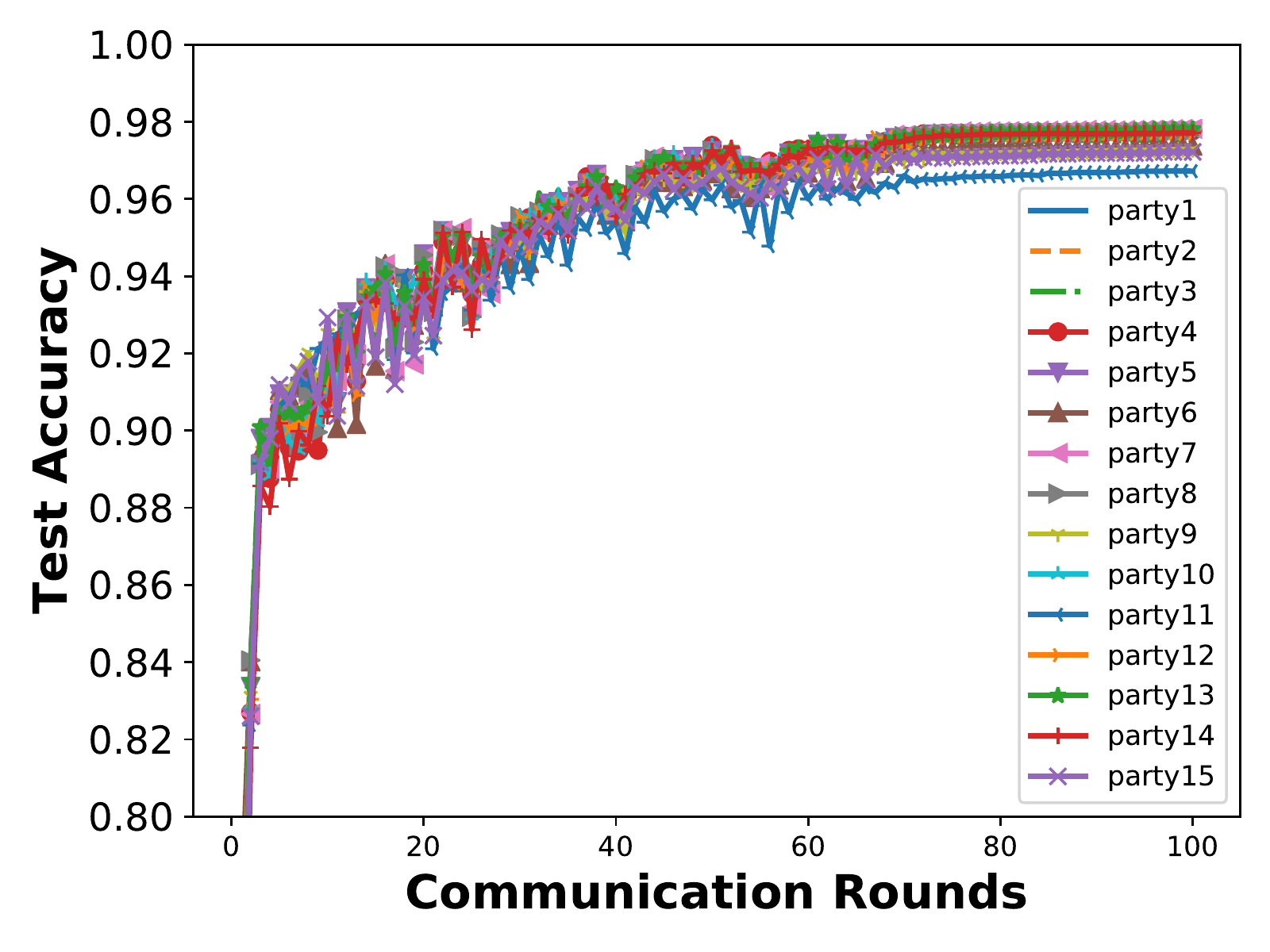}\label{fig:FPPDL_mnist_cvn_p15e100_setting2_pretrain1}
                \subcaption{FPPDL Setting 2}
        \end{subfigure}
        \begin{subfigure}[ht]{0.23\textwidth}
                \includegraphics[width=4cm,height=3.8cm]{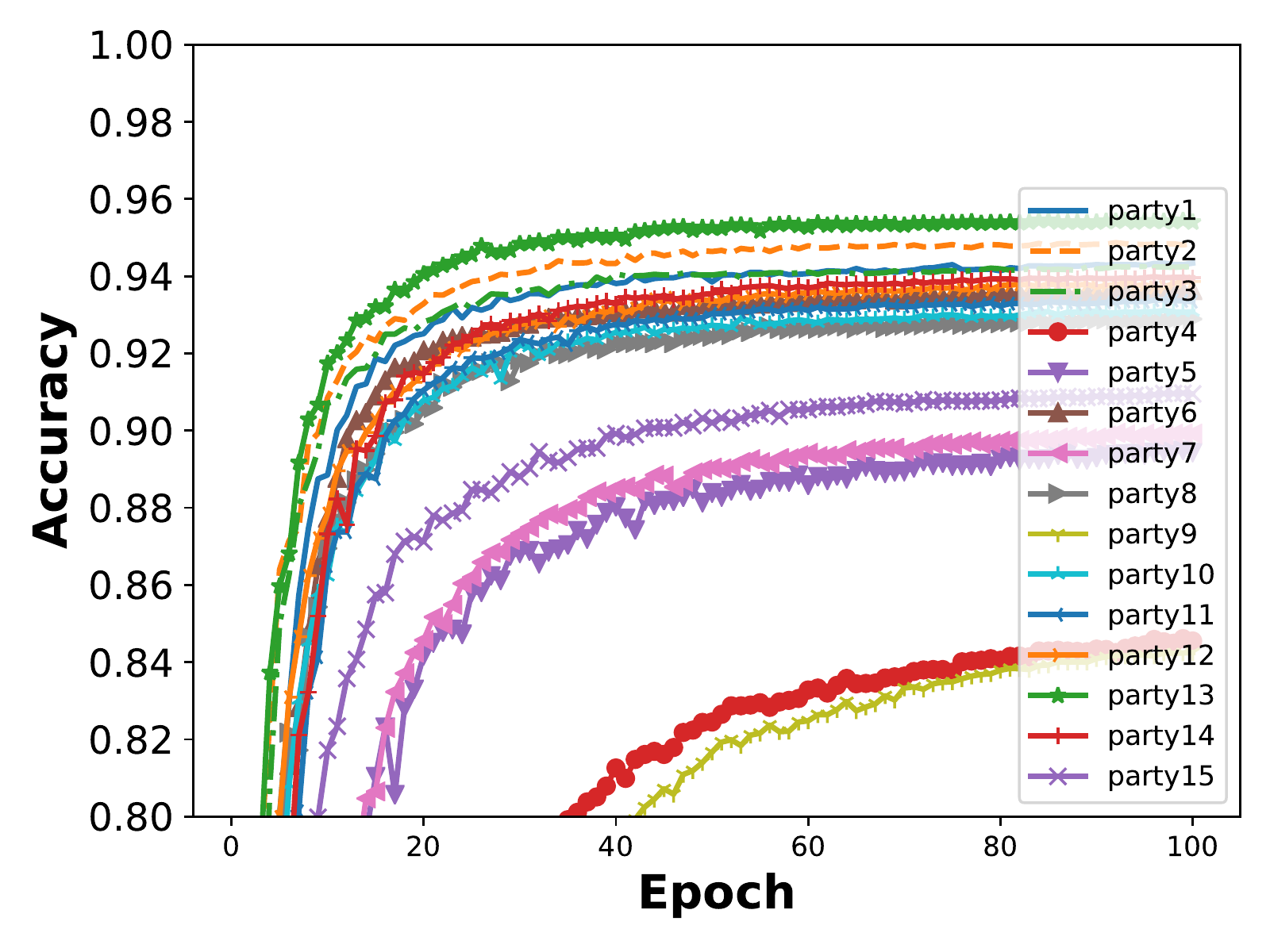}\label{fig:FPPDL_mnist_cvn_p15e100_setting3_standalone}
                \subcaption{Standalone Setting 3}
        \end{subfigure}
        \begin{subfigure}[ht]{0.23\textwidth}
                \includegraphics[width=4cm,height=3.8cm]{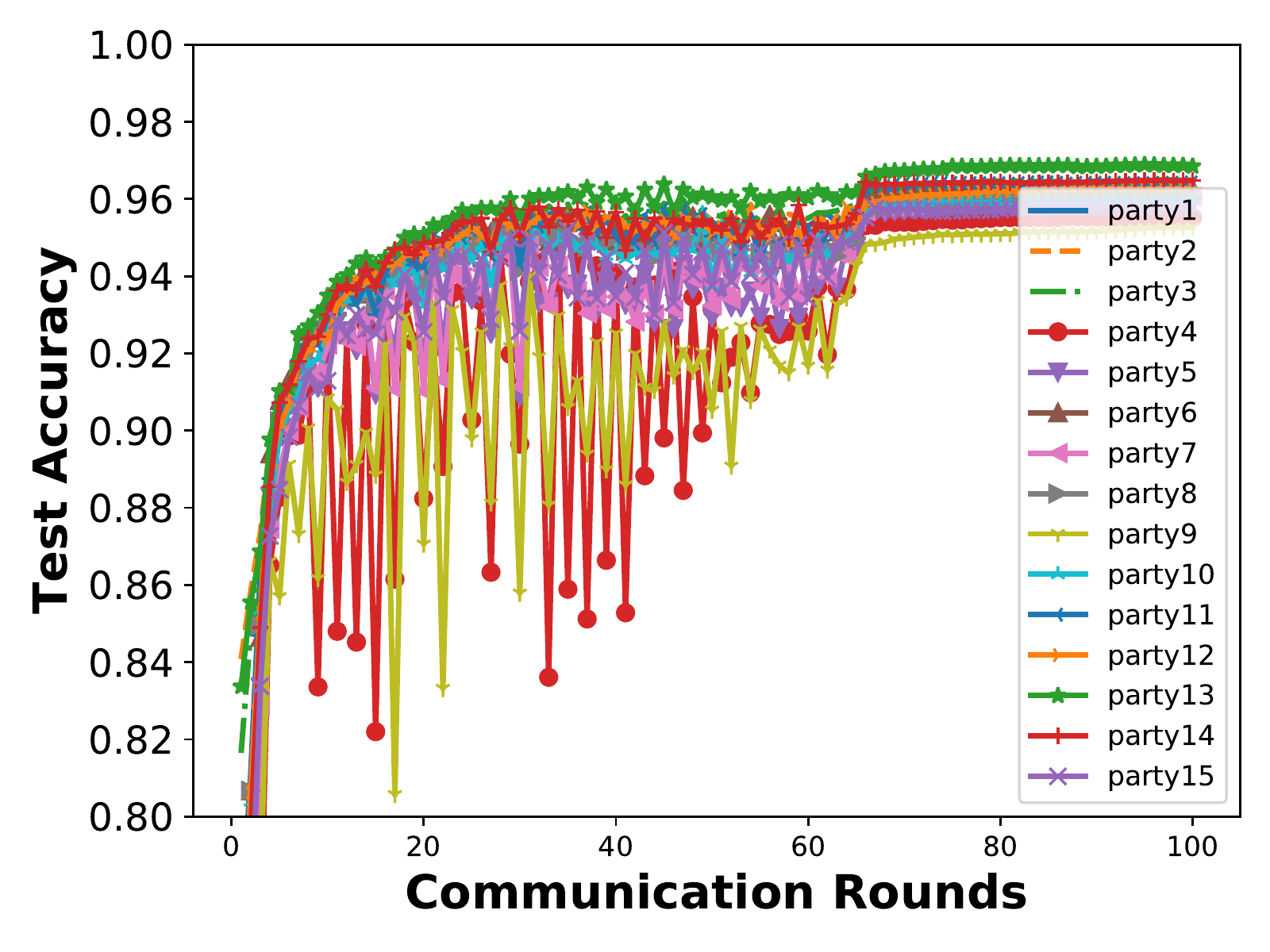}\label{fig:FPPDL_mnist_cvn_p15e100_setting3_pretrain1}
                \subcaption{FPPDL Setting 3}
        \end{subfigure}
        \caption{Individual convergence for MNIST CNN using Standalone framework and our FPPDL (P15, B=1, E=1, lr=0.001).}
\label{fig:mnist_p15_cvn_convergence}
\end{figure*}

\begin{figure*}[!htp]
\centering
        \begin{subfigure}[ht]{0.23\textwidth}
                \includegraphics[width=4cm,height=3.8cm]{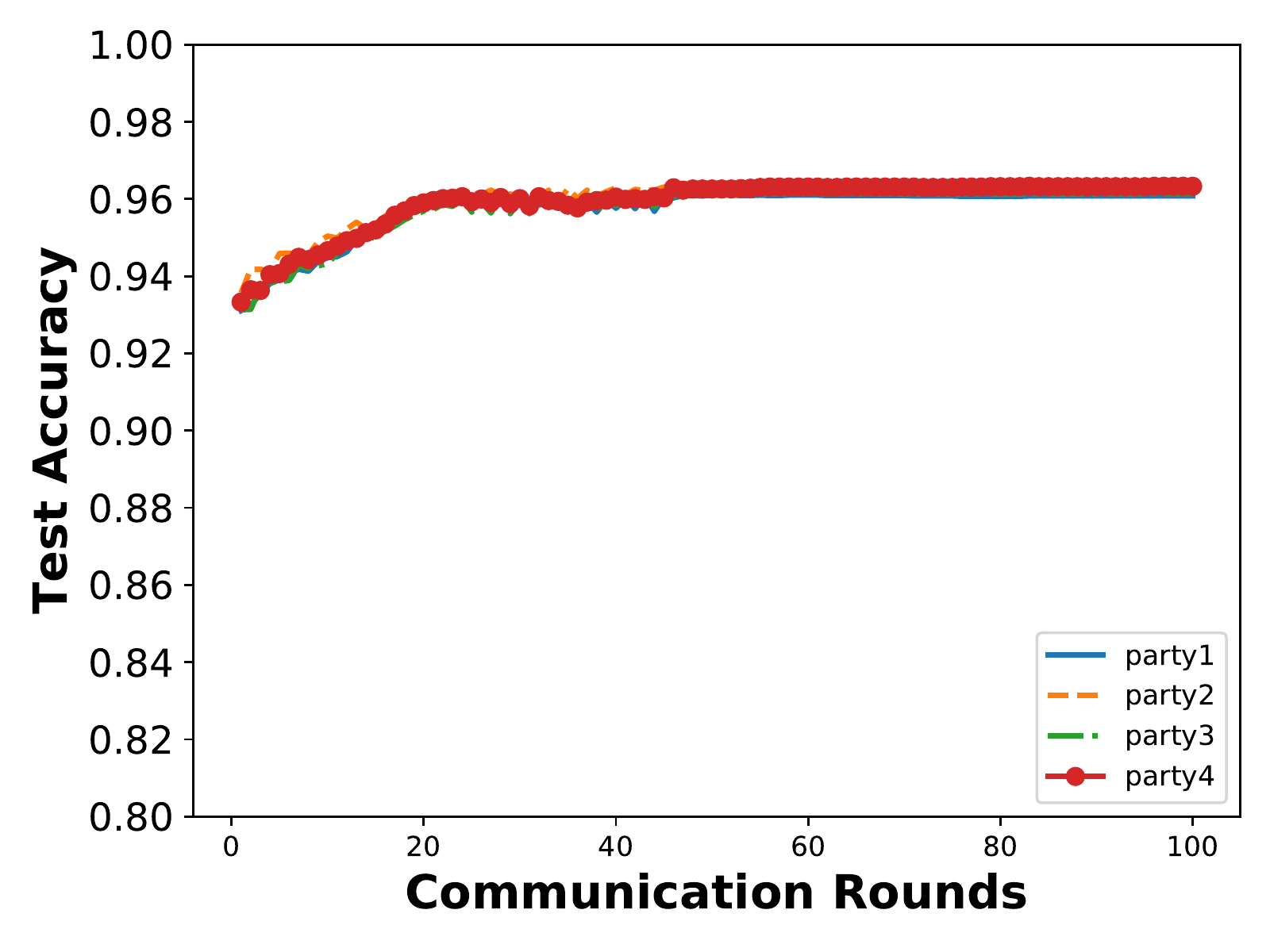}\label{fig:FPPDL_mnist_cvn_p4e100_setting2_pretrain1_localepoch5_localbatch10_lr0.15}
                \subcaption{FPPDL Setting 2 (P4)}
        \end{subfigure}
        \begin{subfigure}[ht]{0.23\textwidth}
                \includegraphics[width=4cm,height=3.8cm]{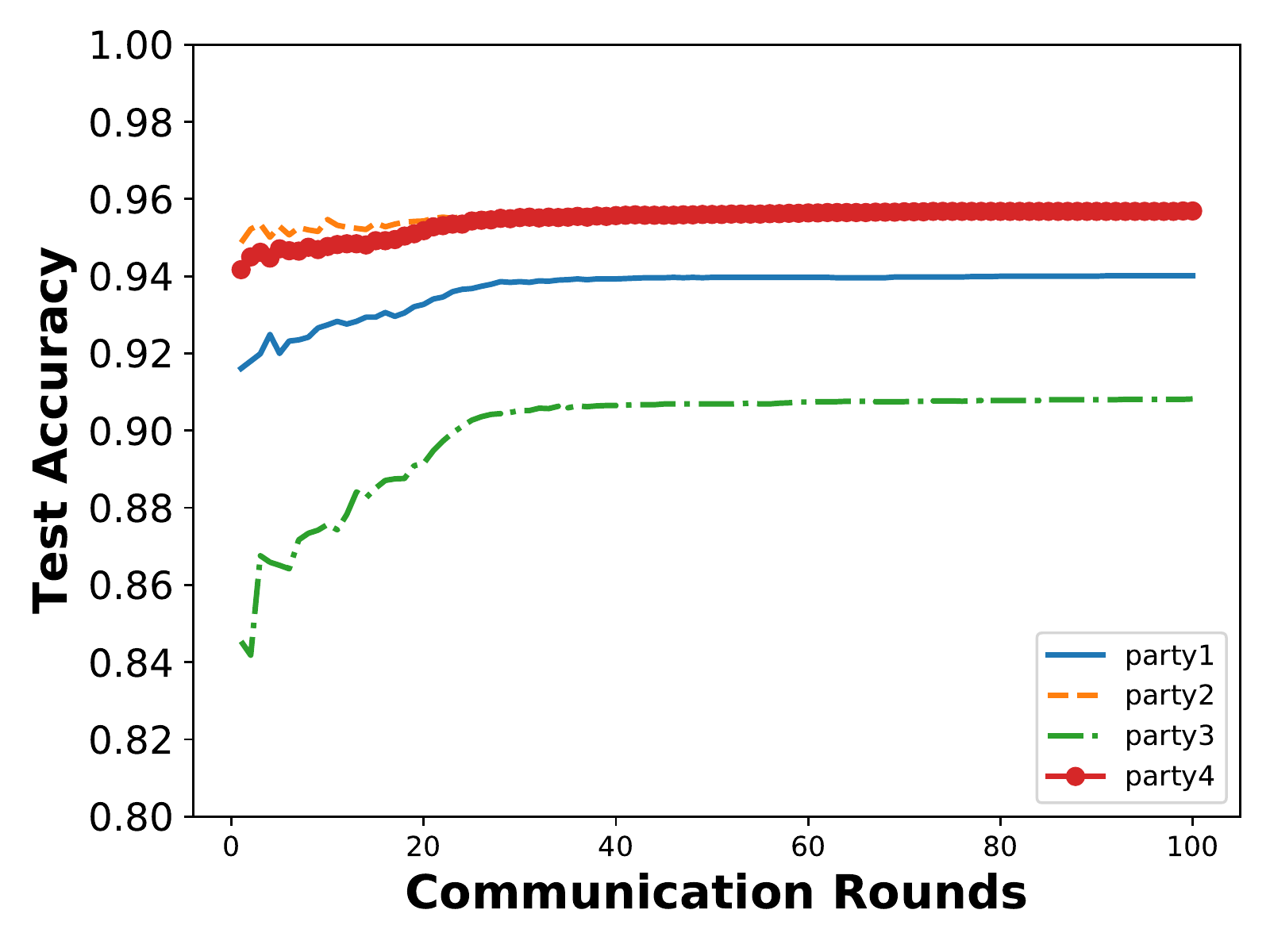}\label{fig:FPPDL_mnist_cvn_p4e100_setting3_pretrain1_localepoch5_localbatch10_lr0.15}
                \subcaption{FPPDL Setting 3 (P4)}
        \end{subfigure}
        \begin{subfigure}[ht]{0.23\textwidth}
                \includegraphics[width=4cm,height=3.8cm]{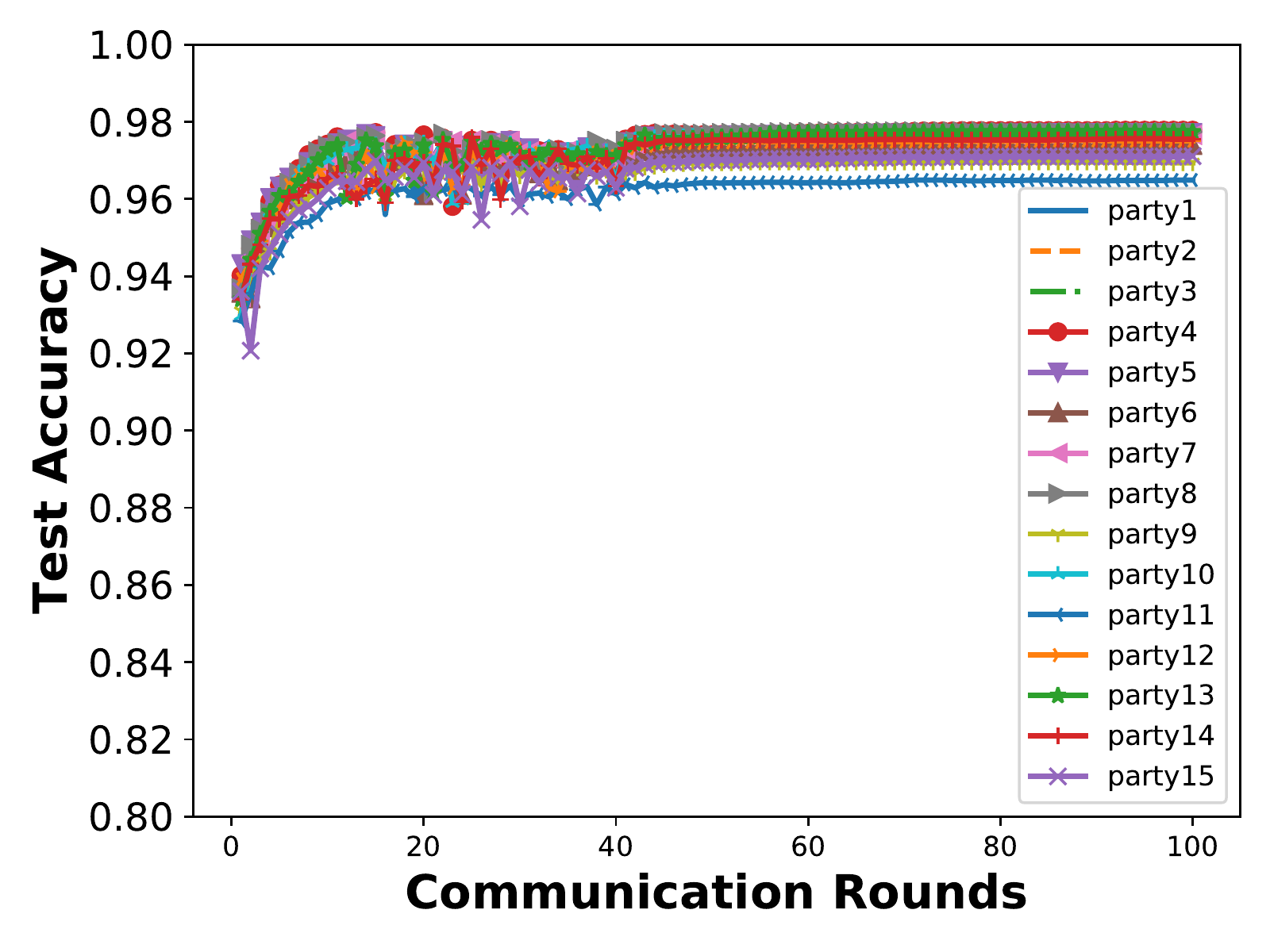}\label{fig:FPPDL_mnist_cvn_p15e100_setting2_pretrain1_localepoch5_localbatch10_lr0.15}
                \subcaption{FPPDL Setting 2 (P15)}
        \end{subfigure}
        \begin{subfigure}[ht]{0.23\textwidth}
                \includegraphics[width=4cm,height=3.8cm]{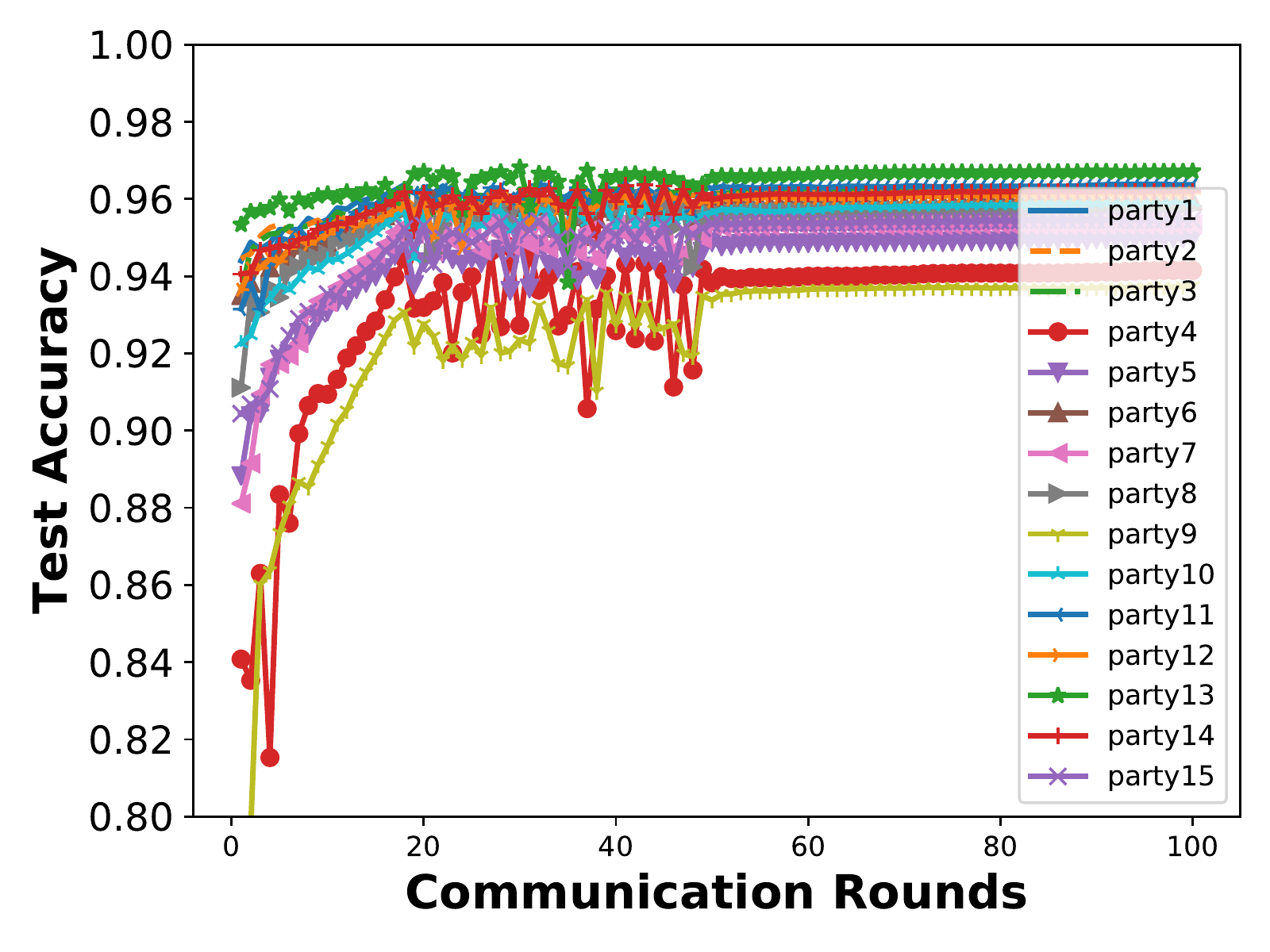}\label{fig:FPPDL_mnist_cvn_p15e100_setting3_pretrain1_localepoch5_localbatch10_lr0.15}
                \subcaption{FPPDL Setting 3 (P15)}
        \end{subfigure}
        \caption{Individual convergence for MNIST CNN using our FPPDL in P4 and P15 (B=10, E=5, lr=0.15).}
\label{fig:mnist_p4p15_cvn_convergence_localepoch5_localbatch10_lr0.15}
\end{figure*}

\textbf{System-level Convergence}. For accuracy comparison, following ~\cite{shokri2015privacy}, we report the best accuracy when running the distributed framework using DSSGD and our FPPDL on MNIST dataset. For DSSGD, we adopted round robin protocol, and set the upload rate as $0.1$ ($\theta_u=0.1$)~\cite{shokri2015privacy}, which is equivalent to our Setting 1, we omit the learning curves of DSSGD in Setting 2 and Setting 3 as they approximate the learning curve in Setting 1. Fig.~\ref{fig:mnist_epoch100_deep} and Fig.~\ref{fig:mnist_epoch100_cvn} present the accuracy trajectories when running different frameworks over MNIST with MLP and CNN architectures. The x-axis corresponds to epochs (communication rounds) (1 round=1 epoch, when the number of local epochs $E=1$), and y axis corresponds to the maximum accuracy achieved by all parties in each round, hence the curve of our FPPDL is not necessarily associated with a particular party, but it is expected that the highest accuracy is achieved by the most contributive party in our FPPDL, as demonstrated by the individual convergence in Fig.~\ref{fig:mnist_p4_cvn_convergence}, Fig.~\ref{fig:mnist_p15_cvn_convergence} and Fig.~\ref{fig:mnist_p4p15_cvn_convergence_localepoch5_localbatch10_lr0.15}. 

Note that the convergences of the standalone framework in Setting 1 and Setting 2 are the same, as these two settings share the same data shard. It can be observed that FPPDL did not change the overall behavior of convergence in all settings, while achieving comparable accuracy to the non-private frameworks, and delivering both fairness and privacy. We notice that our FPPDL achieves slightly slower convergence rate and more fluctuations (especially in early stages of convergence) compared to the distributed framework, this is partly attributed to the individual training of 10 epochs before collaborative learning starts, as we found that collaboration from the state of 10 epochs of local training results in better fairness than the collaboration from the beginning. 

\begin{figure*}[!htp]
\centering
        \begin{subfigure}[ht]{0.23\textwidth}
                \includegraphics[width=4cm,height=3.8cm]{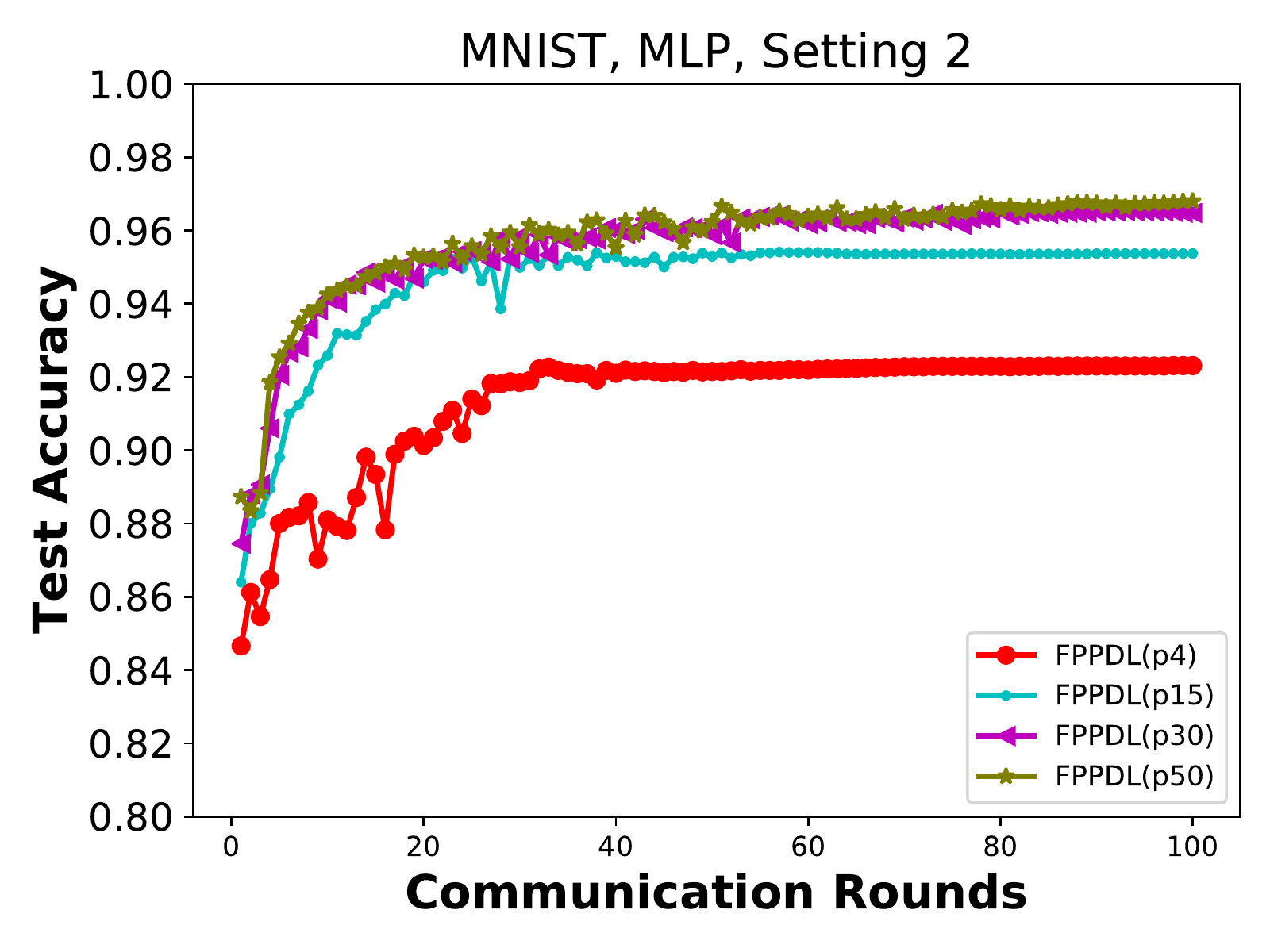}\label{fig:tpds_mnist_deep_localepoch5_localbatch10_lr015_setting2}
                \subcaption{FPPDL Setting 2 (MLP)}
        \end{subfigure}
        \begin{subfigure}[ht]{0.23\textwidth}
                \includegraphics[width=4cm,height=3.8cm]{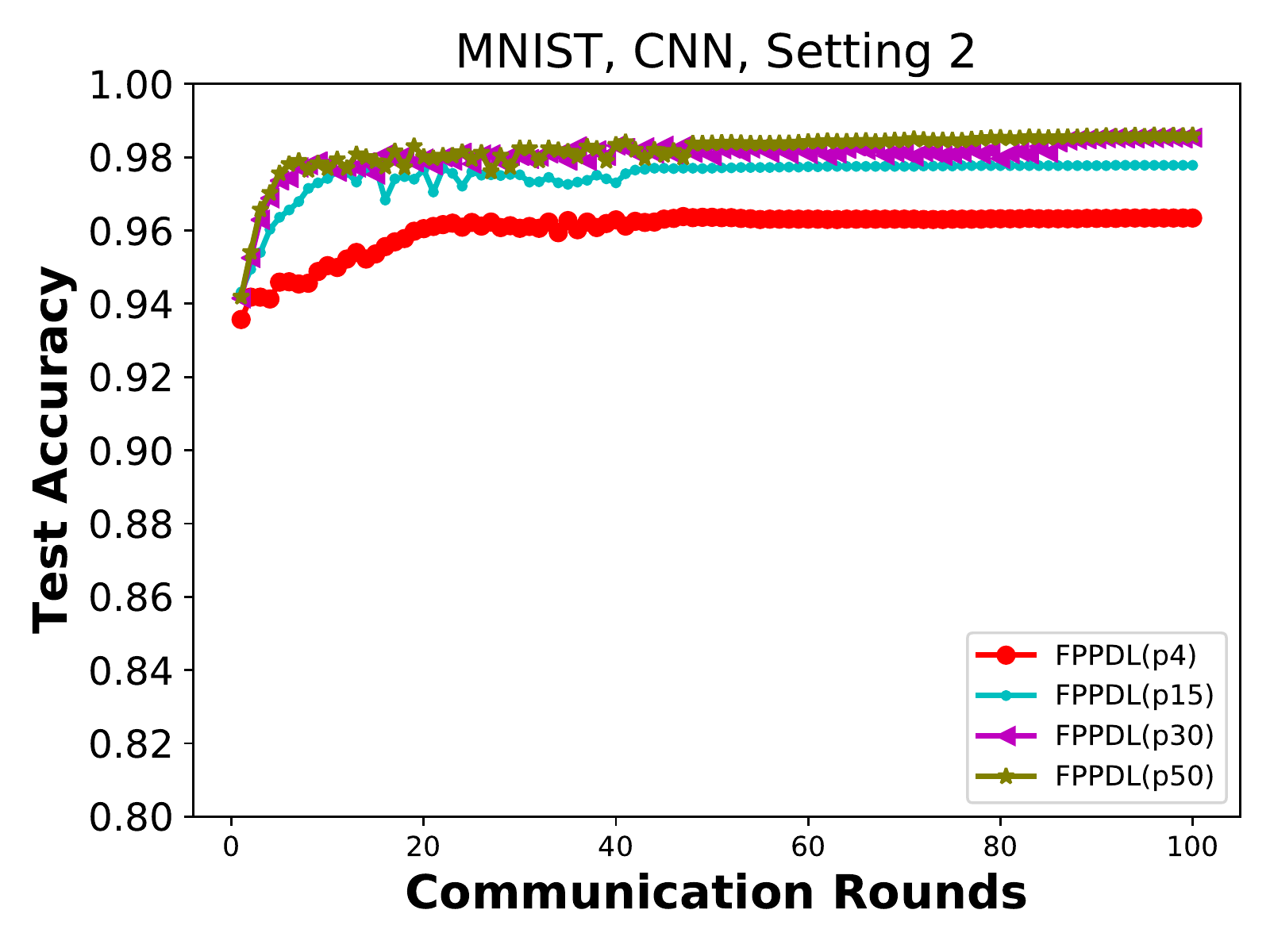}\label{fig:tpds_mnist_cvn_localepoch5_localbatch10_lr015_setting2}
                \subcaption{FPPDL Setting 2 (CNN)}
        \end{subfigure}
        \begin{subfigure}[ht]{0.23\textwidth}
                \includegraphics[width=4cm,height=3.8cm]{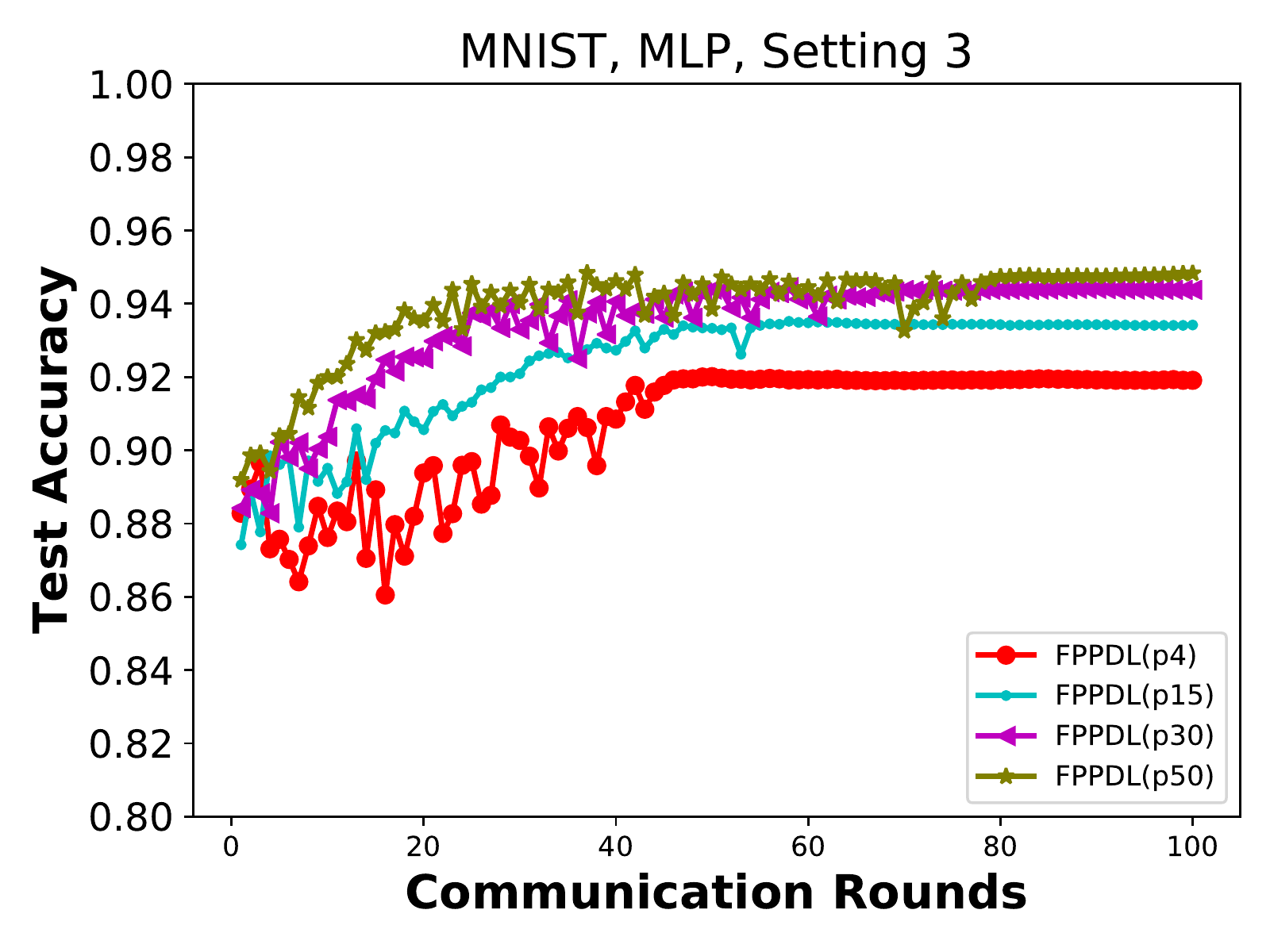}\label{fig:tpds_mnist_deep_localepoch5_localbatch10_lr015_setting3}
                \subcaption{FPPDL Setting 3 (MLP)}
        \end{subfigure}
        \begin{subfigure}[ht]{0.23\textwidth}
                \includegraphics[width=4cm,height=3.8cm]{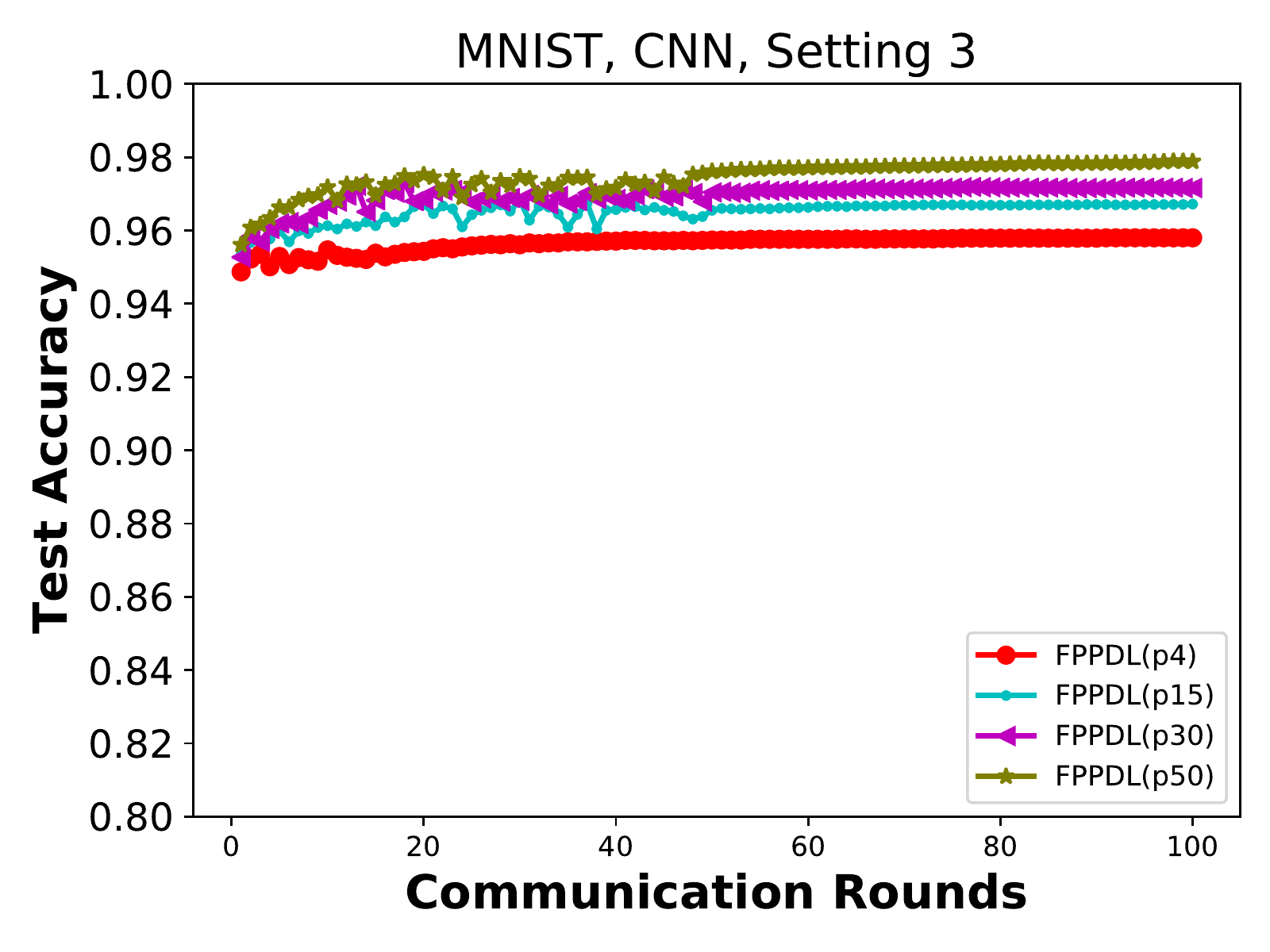}\label{fig:tpds_mnist_cvn_localepoch5_localbatch10_lr015_setting3}
                \subcaption{FPPDL Setting 3 (CNN)}
        \end{subfigure}
        \caption{System convergence for MNIST MLP and CNN using our FPPDL in Setting 2 and Setting 3 (B=10, E=5, lr=0.15).}
\label{fig:mnist_deep_cnn_convergence_localepoch5_localbatch10_lr0.15}
\end{figure*}

Another important reason is that to strike a good balance between computational efficiency, communication cost and convergence rate, we enforce parties to share their local model updates after each epoch of local training ($E=1$), where the shared gradients is the average of the gradients over the whole local training data, rather than a single example, a mini-batch or multiple local epochs, which may also affect convergence. 
We hypothesise that the convergence rate is also closely related with our chosen hyperparameters $B=1, E=1, lr=0.001$ ($E$: number of local training epochs in each communication round; $B$: local batch size; $lr$: local learning rate). Better convergence can be achieved by varying the amount of local computation per communication round, local batch size or the learning rate, as indicated in Fig.~\ref{fig:mnist_deep_cnn_convergence_localepoch5_localbatch10_lr0.15} and  Fig.~\ref{fig:mnist_p4p15_cvn_convergence_localepoch5_localbatch10_lr0.15} by using B=10, E=5, lr=0.15.

\textbf{Individual Convergence}. To investigate the impact of our FPPDL on individual convergence, Fig.~\ref{fig:mnist_p4_cvn_convergence} and Fig.~\ref{fig:mnist_p15_cvn_convergence} further depict the accuracy trajectory of each party when running Standalone framework and our FPPDL with CNN architecture over MNIST across 100 communication rounds. For the sake of brevity, we only report experimental results obtained for the collaboration among 4 parties and 15 parties in Setting 2 and Setting 3.
It can be observed that our FPPDL consistently delivers better accuracy than any standalone model obtained by any individual party, at the cost of slower convergence and more fluctuation. However, most parties can converge within the first 20 rounds, except those with lower standalone accuracy. For example, in Figure~\ref{fig:mnist_p15_cvn_convergence} (d), party 4 and party 9 encounter higher fluctuations compared with the other parties with higher standalone accuracy. More importantly, these figures confirm that our FPPDL enforces all parties to converge to different local models, which are better than their standalone models without any collaboration, thereby offering fairness as claimed. 

To speed up convergence and alleviate fluctuations, we further experiment with larger number of local epochs, larger local batch size, and higher learning rate. As corroborated by Fig.~\ref{fig:mnist_p4p15_cvn_convergence_localepoch5_localbatch10_lr0.15}, by setting $B=10, E=5, lr=0.15$, each party can converge faster, without affecting both accuracy and fairness. For example, for P15 in Figure~\ref{fig:mnist_p15_cvn_convergence} (d), it needs 65 communication rounds for all parties to converge using $B=1, E=1, lr=0.001$, while it only needs 50 communication rounds using $B=10, E=5, lr=0.15$ in Figure~\ref{fig:mnist_p4p15_cvn_convergence_localepoch5_localbatch10_lr0.15} (d). However, this faster convergence and less fluctuations come at the cost of local computation at each party.

% \vskip -2\baselineskip plus -1fil

\begin{table*}[ht]
\caption{MNIST accuracy [\%] after 100 communication rounds, achieved by \emph{Centralized}, \emph{Standalone}, \emph{Distributed} (DSSGD without DP, round robin, $\theta_u=10\%$) and FPPDL (three settings as described in Section~\ref{sec:Setup}) frameworks using MLP and CNN architectures. P-$k$ indicates there are $k$ parties in the experiments.}
\label{tbl:MNIST_CNN}
\centering
\scalebox{0.95}{\begin{tabularx}{\linewidth}{|l|*{8}{C{1}|}}
\hline
\multirow{2}{*}{Framework} & \multicolumn{4}{c|}{MLP} & \multicolumn{4}{c|}{CNN}
\tabularnewline
\cline{2-9}
 & P4 & P15 & P30 & P50 & P4 & P15 & P30 & P50
\tabularnewline
\hline
\textit{Centralized} 
& 91.68 & 95.17 & 96.28 & 96.85 
& 96.58 & 98.19 & 98.52 & 98.58
\tabularnewline
\hline
\textit{Distributed} 
& 91.67  & 95.17 & 96.33 & 97.35 
& 96.25  & 98.04 & 98.63 & 98.83
\tabularnewline
\hline
\textit{Standalone (Setting 1$\&$2)} 
& 87.39 & 88.06 & 88.64  & 88.80 
& 93.81 & 93.46 & 94.04 & 94.05
\tabularnewline
\hline
\textit{Standalone (Setting 3)} 
& 89.61 & 88.83 & 89.57  & 89.52
& 94.42 & 95.44 & 95.11 & 95.45
\tabularnewline
\hline
\textit{FPPDL (Setting 1)} 
&90.13 & 94.42 &94.88  &95.57
&95.93  & 97.19  & 97.62  & 98.07
\tabularnewline
\hline
\textit{FPPDL (Setting 2)}
&91.92   &95.70   &95.94  &96.23
&95.50   &97.34   &97.84  &98.14
\tabularnewline
\hline
\textit{FPPDL (Setting 3)}
&90.75   &94.37   &94.75  &95.21 
&95.23   &97.50   &97.82  &98.22
\tabularnewline
\hline
\end{tabularx}}
\end{table*}

\begin{table*}[ht]
\caption{SVHN accuracy [\%] after 100 communication rounds, achieved by \emph{Centralized}, \emph{Standalone}, \emph{Distributed} (DSSGD without DP, round robin, $\theta_u=10\%$) and FPPDL (three settings as described in Section~\ref{sec:Setup}) frameworks using MLP and CNN architectures.}
\label{tbl:SVHN_CNN}
\centering
\scalebox{0.95}{\begin{tabularx}{\linewidth}{|l|*{8}{C{1}|}}
\hline
\multirow{2}{*}{Framework} & \multicolumn{4}{c|}{MLP} & \multicolumn{4}{c|}{CNN}
\tabularnewline
\cline{2-9}
 & P4 & P15 & P30 & P50 & P4 & P15 & P30 & P50
\tabularnewline
\hline
\textit{Centralized}
& 75.40  & 83.08  & 85.77  & 87.15
& 90.50  & 91.88  & 93.42  & 95.44
\tabularnewline
\hline
\textit{Distributed} 
&78.34   &85.49   &87.64 & 89.21
&91.78   &93.03   &95.75 & 96.19
\tabularnewline
\hline
\textit{Standalone (Setting 1$\&$2)}
& 57.85  & 58.77   & 57.90 & 59.18
& 80.24  & 80.74  & 81.29  & 81.60
\tabularnewline
\hline
\textit{Standalone (Setting 3)}
& 59.05  & 59.13   & 60.09 & 60.22
& 81.57  & 81.92  & 82.06  & 82.31
\tabularnewline
\hline
\textit{FPPDL (Setting 1)}
& 73.74   & 82.55   &84.86 & 86.51
& 90.07   & 91.18   &92.74 & 94.83
\tabularnewline
\hline
\textit{FPPDL (Setting 2)}
&74.16   &82.67  &85.25 & 86.57
&89.91   &91.15  &92.59  & 95.18
\tabularnewline
\hline
\textit{FPPDL (Setting 3)}
&74.57   &82.95   & 85.37 & 86.34
&89.53   &91.03   & 93.13 & 94.89
\tabularnewline
\hline
\end{tabularx}}
\end{table*}

 Table~\ref{tbl:MNIST_CNN} provides the accuracy results we obtain when running different frameworks on MNIST dataset of \{4,15,30,50\} parties for different neural network architectures. For all frameworks, we report the 
 best accuracy the system can achieve across all rounds.  In particular, in our FPPDL, fairness enables each party to get a different local model after collaborative learning, and we expect that the most contributive party derives a local model with maximum accuracy approximating the non-private centralized and distributed frameworks. Similarly, Table~\ref{tbl:SVHN_CNN} provides the accuracy on SVHN dataset. For both MNIST and SVHN datasets using CNN and MLP architectures, we show the worst accuracy for standalone SGD (minimum utility, maximum privacy). In particular, FPPDL obtains comparable accuracy (less than 2\%) to both the centralized framework and the distributed framework using DSSGD without differential privacy, and consistently achieves higher accuracy than the standalone SGD. For example, as shown in Table~\ref{tbl:MNIST_CNN}, for MNIST dataset of 50 parties with CNN model, our FPPDL achieves 98.07\%-98.22\% test accuracy under different settings, which is higher than the standalone SGD 94.05\%, and comparable to 98.83\% of the distributed framework using DSSGD without differential privacy, and 98.58\% of the centralized framework. 
 
The above fairness results in Table~\ref{tbl:MNIST_fairness} and Table~\ref{tbl:SVHN_fairness}, and accuracy results in Table~\ref{tbl:MNIST_CNN} and Table~\ref{tbl:SVHN_CNN} demonstrate that \textbf{our proposed framework FPPDL achieves reasonable fairness, at the expense of a tiny decrease in model utility}. 

\begin{figure*}[!htp]
\centering
        \begin{subfigure}[ht]{0.3\textwidth}
                \includegraphics[width=5.5cm,height=4.8cm]{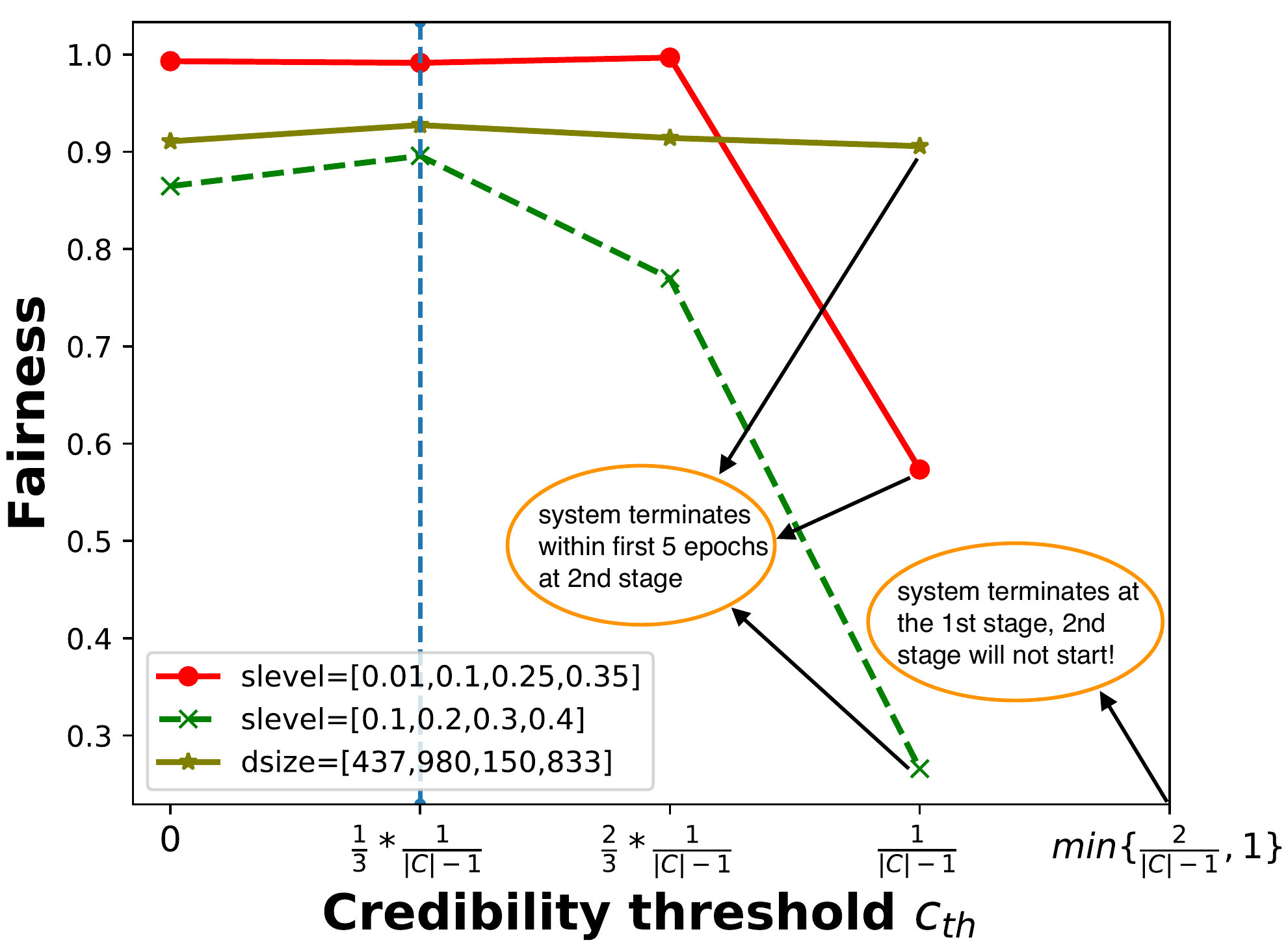}\label{fig:fairness_cth}
        \end{subfigure}
        \begin{subfigure}[ht]{0.3\textwidth}
                \includegraphics[width=5.5cm,height=4.8cm]{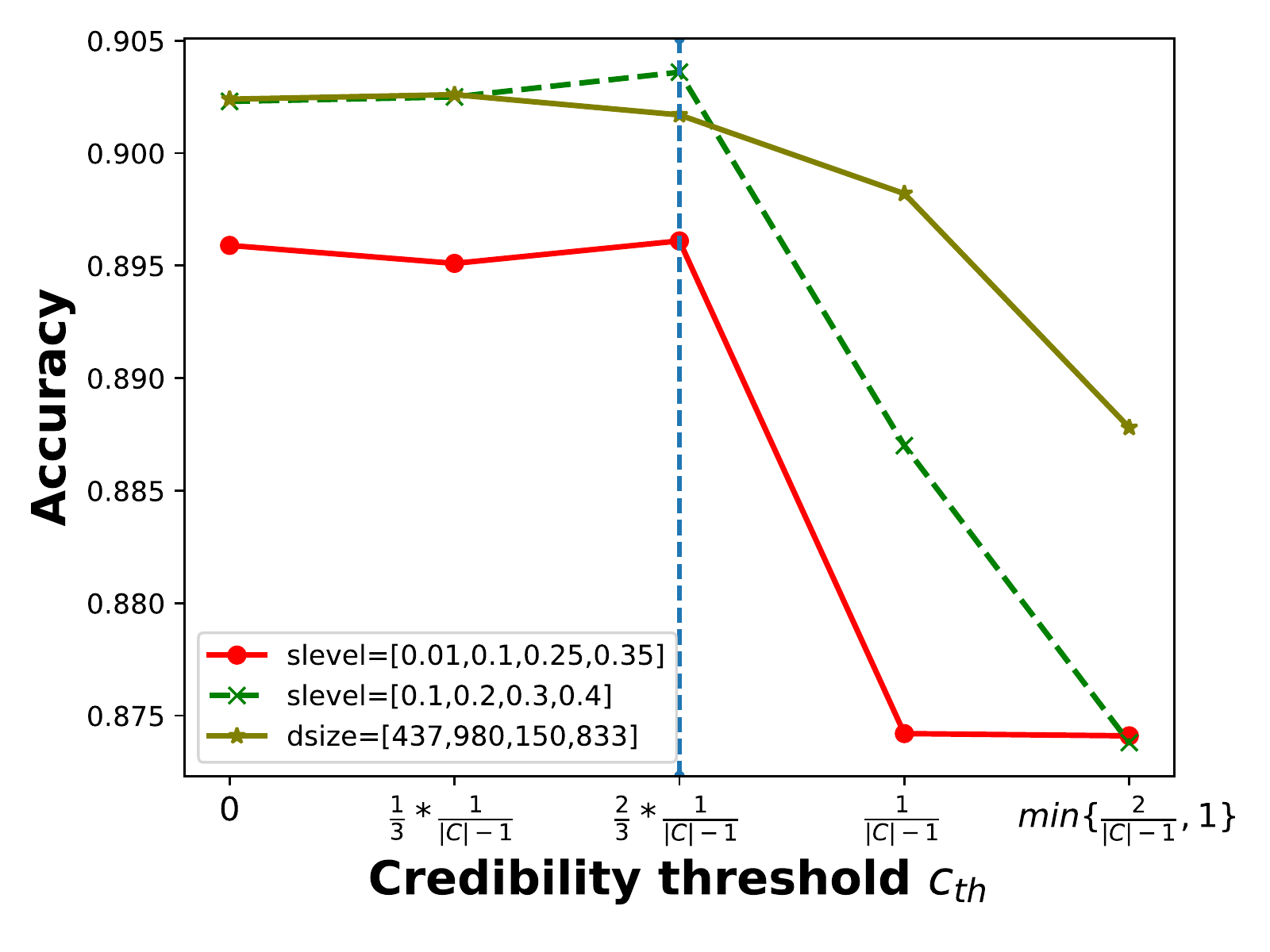}\label{fig:accuracy_cth}
        \end{subfigure}
        \begin{subfigure}[ht]{0.3\textwidth}
                \includegraphics[width=5.5cm,height=4.8cm]{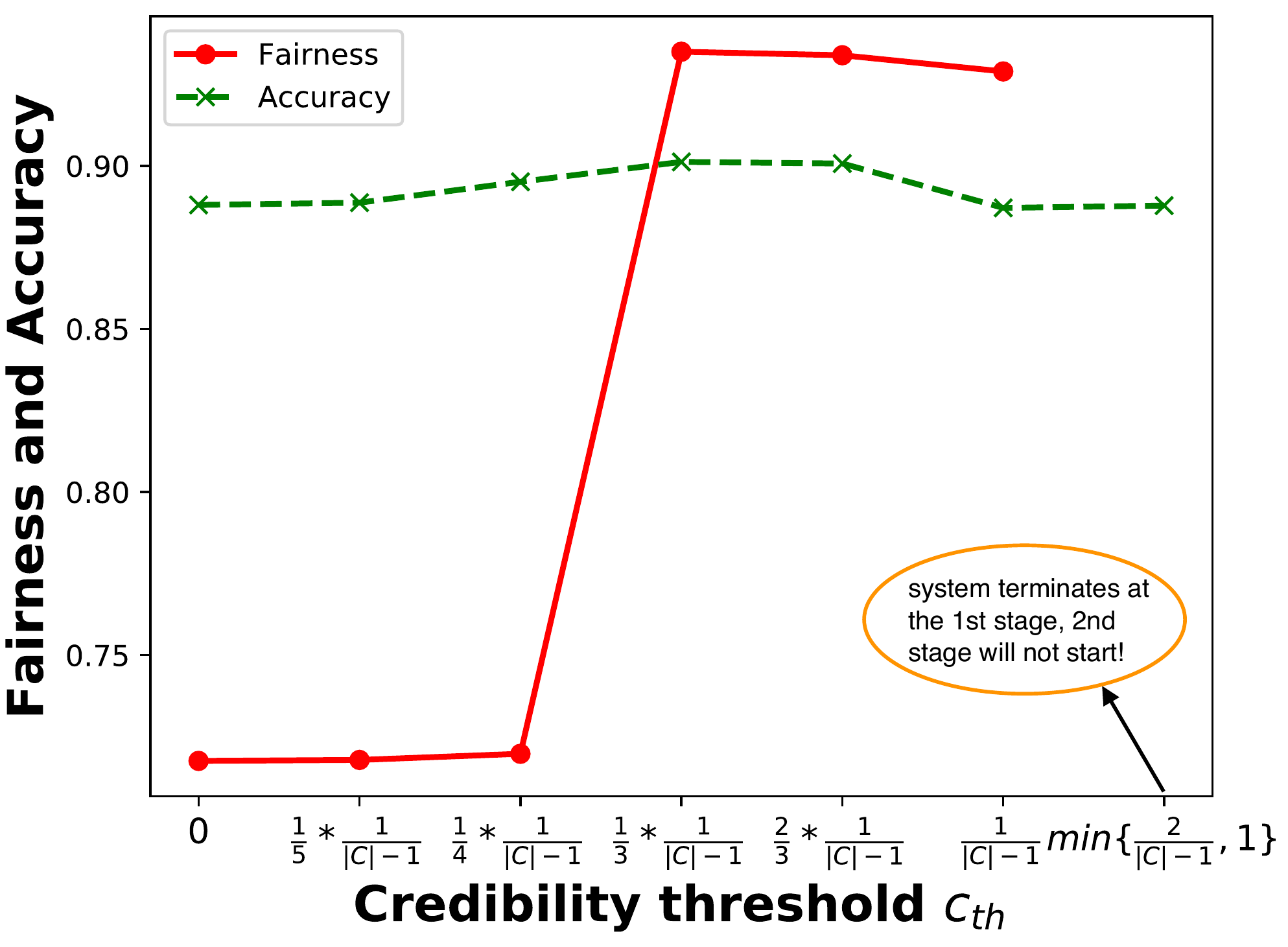}\label{fig:fairness_accuracy_cth_malicious}
        \end{subfigure}
        \caption{How $c_{th}$ affects fairness and accuracy in normal and malicious settings. First two figures correspond to setting 2 with sharing level of $[0.01,0.1,0.25,0.35]$ and $[0.1,0.2,0.3,0.4]$, and setting 3 with data shard of $[437,980,150,833]$ among four honest parties. The last figure simulates setting 3 with data shard of $[437,980,150,833]$) among four honest parties and one more malicious party indicated in Section~\ref{sec:Discussion}.}.
\label{fig:fairness_accuracy_cth}
\end{figure*}

Moreover, to investigate how fairness and accuracy change with the local credibility threshold $c_{th}$, we implement a four-party scenario (P4) under both normal settings (Setting 2 and Setting 3 in Section~\ref{sec:Setup}) and malicious setting (1 malicious party as indicated in Section~\ref{sec:Discussion}). As shown in Fig.~\ref{fig:fairness_accuracy_cth}, both fairness and accuracy can keep relatively high values when $c_{th}$ is within $[\frac{1}{3}*\frac{1}{|C|-1},\frac{2}{3}*\frac{1}{|C|-1}]$. In contrast, too small $c_{th}<\frac{1}{3}*\frac{1}{|C|-1}$ allows even the malicious party to sneak into the collaborative learning system without being detected and isolated, resulting in lower fairness, as manifested by the last figure of Fig.~\ref{fig:fairness_accuracy_cth}. On the contrary, too large $c_{th}$ might isolate most participants in the system. For example, $c_{th}=\frac{1}{|C|-1}$ will terminate the system within the first 5 rounds during the second stage of collaborative learning, resulting in both lower fairness and accuracy; and $c_{th}=min\{\frac{2}{|C|-1},1\}$ will terminate the system after the first stage, and second stage of collaborative learning will never start, thus there is no collaborative fairness. These results validate our hypothesis in Section~\ref{sec:credit_initialization} and provide empirical support on our chosen $c_{th}=\frac{1}{|C|-1}*\frac{2}{3}$ in Section~\ref{sec:Setup}.

\textbf{Complexity Analysis}. The main communication cost occurs when each party sends its encrypted gradients to the other $(n-1)$ parties, resulting in $(n-1)*L$ ciphertexts, where $n$ and $L$ are the number of parties and the size of the released gradients (the encrypted symmetric key size is negligible compared with the encrypted gradients). Therefore, our framework is applicable to practical applications to businesses~\cite{lyu2020threats}, such as biomedical or financial institutions where the number of parties is limited. On the other hand, the main computation cost occurs at each party who needs to train a local DPGAN during initial benchmarking, compute local gradients, and conduct three-layer onion-style encryption during collaborative deep learning. However, all parties can individually train their DPGAN models offline before collaborative deep learning starts, and all parties can individually train local models in parallel, hence deep learning computation cost is not an obstacle for those parties with enough computational power. Moreover, our encryption scheme using stream ciphers and hybrid encryption is relatively efficient, because encrypting a short plaintext (\ie the symmetric key) requires only one asymmetric operation, while encrypting a longer message (released gradients) would in theory require many asymmetric operations.

%%%%%%%%%%%%%%%%%%%%%%%%%%%%%%%%%%%%%%%%
\section{Discussions}
\label{sec:Discussion}
\textbf{Data Augmentation and Collaboration.} To facilitate credibility initialization, we apply data augmentation to expand local data size to \emph{\textbf{help DPGAN generate reliable samples within a moderate privacy budget}}. However, data augmentation is intended to increase the amount of training data using information inherent in local training data, and thus improve the generalizability of local model, while not helpful for generalizing to unseen data. In other words, it cannot represent global distribution, and this explains why parties still need collaboration for better utility even after data augmentation. By using DPGAN, it not only preserves privacy of the original data, but also preserves privacy of the augmented data that are similar to the original data. 

\textbf{Fairness and Privacy.} With three-layer onion-style encryption, privacy is better preserved without compromising utility. We ensure fairness from two ways: (i) during initial benchmarking, parties generate DPGAN samples based on their local training data, which are then evaluated by other parties' standalone models to mutually initialize the local credibilities of other parties; and (ii) during collaborative learning process, each party randomly selects and shares a subset of DPGAN samples as per individual sharing level at each round of communication, then updates the local credibility values for other parties who evaluate the received DPGAN samples using their local models at current round. Therefore, local credibility of each party keeps changing, reflecting more accurate relative contribution and thus possessing better fairness. Differentially private training of deep models provides another alternative solution by releasing gradients after each epoch or several epochs of local training, thus enabling each party to verify the claims of other parties and update their local credibility values as per the received gradients during collaborative learning process. One obstacle is that differentially private models may significantly reduce utility for small $\epsilon$ values.

\textbf{Attacker Prevention.}
Although the capability of detecting and isolating malicious parties is not the main focus of this paper, we next discuss how our design can help prevent certain behaviours of inside attacker, and resist the outside attacker as a by-product of FPPDL.

For an inside attacker who is a participant in the decentralized system, we specially consider an interesting case: a free-rider without any data, and we remark that this free-rider belongs to the category of low-contribution party. During initialization, this free-rider may choose to send the fake information to other parties. For example, it may randomly sample from 10 classes as predicted labels for the received DPGAN samples, then release them to the corresponding party who publishes these DPGAN samples and requests labels. When the publisher receives the returned random labels from the free-rider and detects that most of them are not aligned with the majority voting, \ie $\frac{m_j}{u_i} \ll c_{th}$, then the free-rider will be reported as a ``low-contribution'' party. If the majority of parties report the free-rider as ``low-contribution'', then the Blockchain rules out the free-rider from the credible party set, and all parties would terminate the collaboration with the free-rider. In this way, such a malicious party is isolated from the beginning, while the collaboration among the remaining parties will not be affected. Even though the free-rider might succeed in initialization somehow, its local credibility would be significantly lower compared with the other honest parties. 

To further detect and isolate this malicious party during the collaborative learning process, we repeat mutual evaluation at each round of collaborative learning by using samples generated at the initialization phase, \ie each party randomly selects and shares a subset of DPGAN samples as per individual sharing level in each round of collaborative learning, then updates the local credibility values of other parties by comparing the majority labels with the received labels output by the local models of other parties in current round of training. Hence, the chance of the survival of the malicious party is significantly reduced, thus it will not dominate the whole system. Note that the lower bound of the acceptable credibility threshold can be agreed by the system requirement. For the outsider attacker like the eavesdropper who aims to steal the exchanged information by eavesdropping on the communication channels among parties, differential privacy used in the first stage and three-layer onion-style encryption applied in the second stage inherently prevent the success of this attack. 

We recognize that our current design may not be resistant to all the malicious parties who can arbitrarily deviate from the protocol, sending incorrect and/or arbitrarily chosen messages to honest parties, aborting, omitting messages, and sharing their entire view of the protocol with each other. For example, a malicious party who aims to compromise other parties' local model integrity (prevent other parties from learning reasonable models) may adaptively or alternatively adjust its behaviour by behaving normally during releasing DPGAN samples to avoid being detected and kicked out, while poisoning the second stage by sending random local gradients or local gradients with the embedded backdoor behavior to the requester. However, in this case, this malicious party is unlikely to obtain a reasonable local model or steal any party's personal information. 

Moreover, to prevent the success of the poisoning attack, one potential solution is to let each party repeat local prediction process on its hold-out validation set by using individually aggregated gradients. Each party will release a signal to indicate whether its aggregated gradients can give a reasonable accuracy result, or help improve prediction on local validation set, if majority party report local validation accuracy lower than a threshold, or negative gain on local validation accuracy, then the system terminates to avoid being further poisoned. We leave this open problem to our future work, and our current design is mainly for the business applications, where parties act with legal liabilities. 

\section{Conclusions and Future Work}
\label{sec:Conclusion}
This paper proposes FPPDL, a decentralized privacy-preserving deep learning framework with fairness considerations. Our enhanced framework shows the following properties: (1) it inherently resolves the relevant issues in the server-based frameworks, and investigates Blockchain for decentralization; (2) it makes the first investigation on the research problem of collaborative fairness in deep learning, by introducing a notion of local credibility and transaction points, which are initialized by initial benchmarking, and updated during privacy-preserving collaborative deep learning; (3) it combines \emph{Differentially Private GAN} (DPGAN) and a three-layer onion-style encryption scheme to guarantee both accuracy and privacy; (4) it provides a viable solution to detect and reduce the impact of low-contribution parties in the system. The experimental results demonstrate that our FPPDL achieves comparable accuracy to both the centralized and distributed selective SGD framework without differential privacy, and always delivers better results than the standalone framework, confirming the applicability of our proposed framework. 

A number of avenues for further work are attractive. In particular, we would like to study how to quantify fairness in Non-IID setting, and investigate more malicious behaviours and byzantine or sybil adversary in the decentralized system. We also expect to deploy our system into a wide spectrum of real-world applications.

\ifCLASSOPTIONcompsoc
  \section*{Acknowledgments}
\else
  \section*{Acknowledgment}
\fi

This work is supported, in part, by IBM PhD Fellowship;  ANU Translational Fellowship; Nanyang Assistant Professorship (NAP); and NTU-WeBank JRI (NWJ-2019-007). The authors would like to thank Prof. Benjamin Rubinstein, Dr. Kumar Bhaskaran, and Prof. Marimuthu Palaniswami for their insightful discussions. % and supports on this work. 
This research was undertaken using the LIEF HPC-GPGPU Facility hosted at the University of Melbourne. This Facility was established with the assistance of LIEF Grant LE170100200.

\bibliographystyle{IEEEtran}
\bibliography{biblio.bib}

% \vskip 

\vskip -2\baselineskip plus -1fil

\begin{IEEEbiography}[{\includegraphics[width=1in,height=1.2in,clip,keepaspectratio]{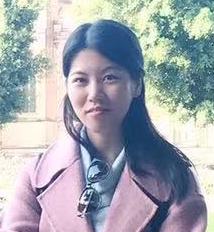}}]{Lingjuan Lyu}
(IEEE M'18) is currently a Research Fellow with The Department of Computer Science, National University of Singapore. She received Ph.D. degree from the University of Melbourne. Her current research interests span machine learning, privacy, fairness, and edge intelligence. Her work was supported by an IBM Ph.D. Fellowship.
\end{IEEEbiography}

\vskip -3.5\baselineskip plus -1fil

\begin{IEEEbiography}[{\includegraphics[width=1in,height=1.2in,clip,keepaspectratio]{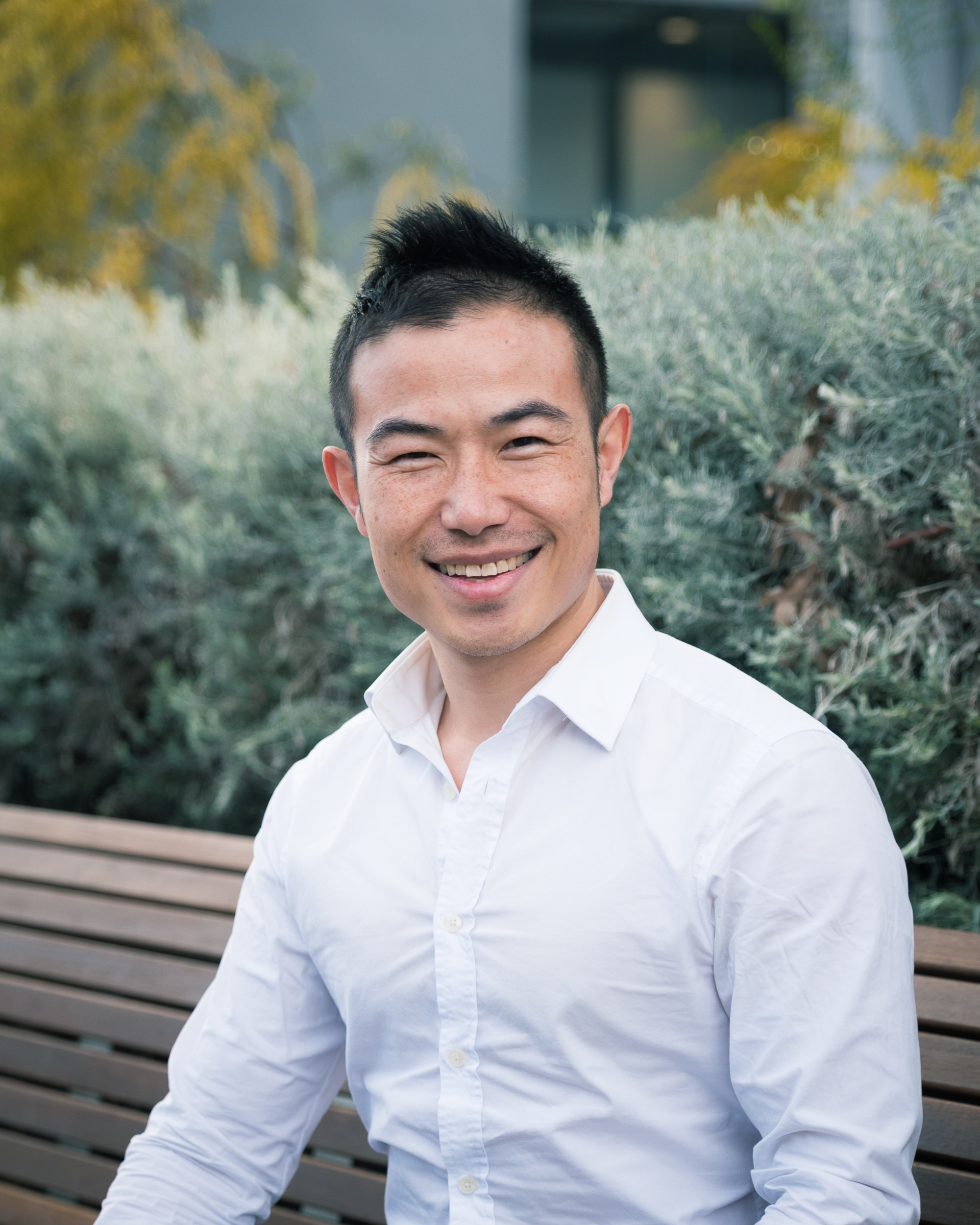}}]{Jiangshan Yu}
received the Ph.D. degree from the University of Birmingham (UK) in 2016. He is currently Associate Director (Research) at Monash Blockchain Technology Centre at Monash University, Australia. Previously, he was a research associate at SnT, University of Luxembourg (LU). The focus of his research has been on design and analysis of cryptographic protocols, cryptographic key management, blockchain consensus, and ledger-based applications. In particular, Jiangshan's recent research challenges the soundness of the foundational security models and design principles of existing blockchain systems, where the blockchain ecosystem of hundreds of billions of dollars is based upon. He won numerous prestigious awards, including Dean's Research Impact Award (2019) and the Chinese Government Award for Outstanding Scholar Abroad (1\% worldwide, 2016).
\end{IEEEbiography}

\vskip -2.5\baselineskip plus -1fil

\begin{IEEEbiography}[{\includegraphics[width=1in,height=1.2in,clip,keepaspectratio]{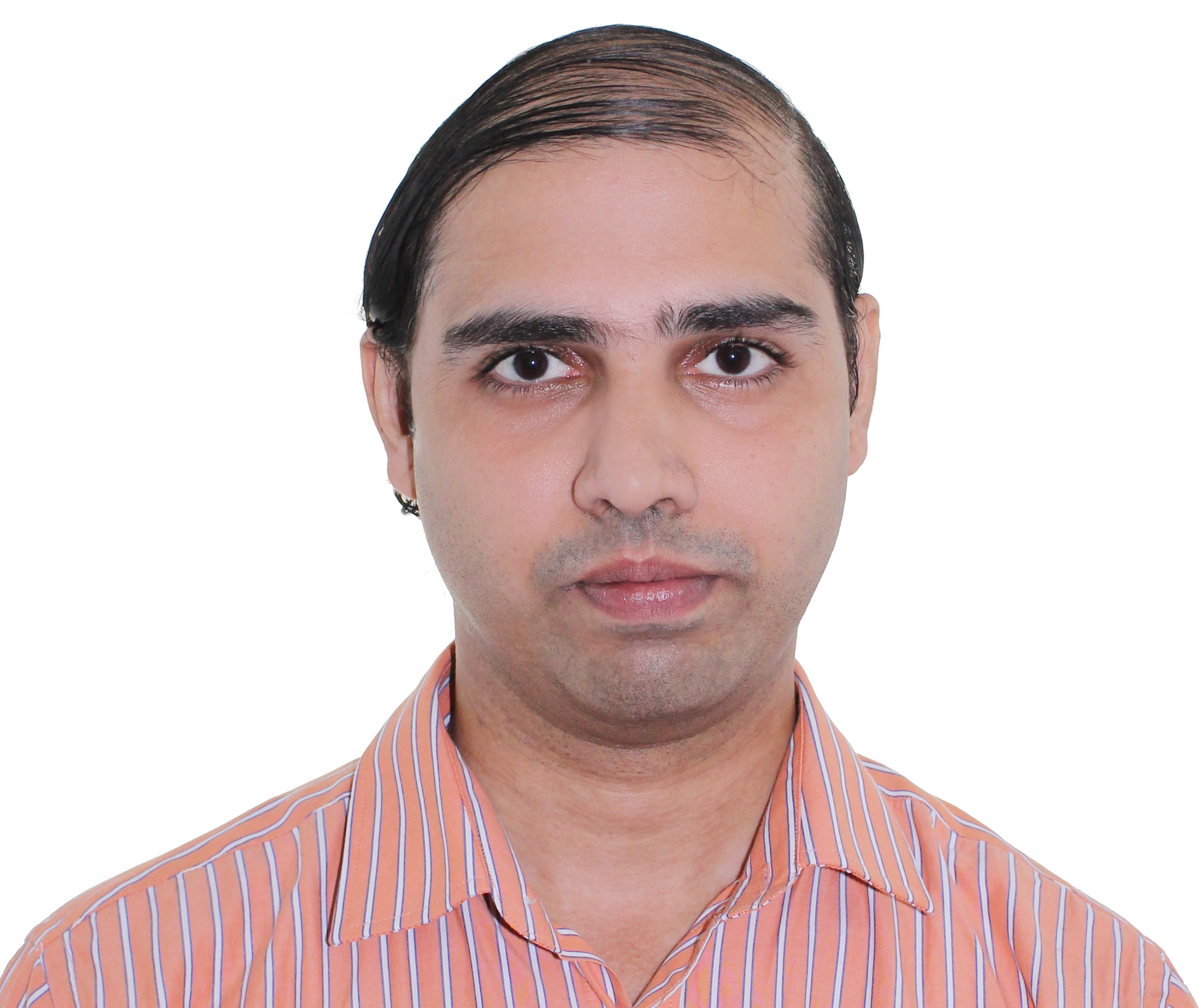}}]{Karthik Nandakumar}
(IEEE SM'02) is a Research Staff Member at IBM Research, Singapore. Prior to joining IBM in 2014, he was a Scientist at Institute for Infocomm Research, A*STAR, Singapore for more than six years. He received his B.E. degree (2002) from Anna University, Chennai, India, M.S. degrees in Computer Science (2005) and Statistics (2007), and Ph.D. degree in Computer Science (2008) from Michigan State University, and M.Sc. degree in Management of Technology (2012) from National University of Singapore. His research interests include computer vision, statistical pattern recognition, biometric authentication, image processing, machine learning and blockchain.
\end{IEEEbiography}

\vskip -2.5\baselineskip plus -1fil

\begin{IEEEbiography}[{\includegraphics[width=1in,height=1.2in,clip,keepaspectratio]{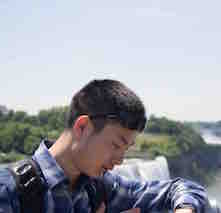}}]{Yitong Li}
is currently a Ph.D student in School of Computing and Information Systems, the University of Melbourne. He received B.S. degree from Shanghai Jiao Tong University. His research interests cover privacy and adversarial learning with NLP applications. He has publications in ACL, EMNLP, NAACL, etc.
\end{IEEEbiography}

\vskip -3.5\baselineskip plus -1fil

\begin{IEEEbiography}[{\includegraphics[width=1in,height=1.2in,clip,keepaspectratio]{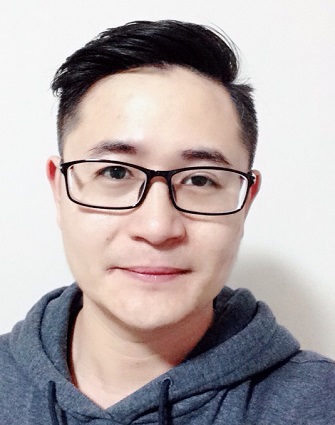}}]{Xingjun Ma}
is currently a Research Fellow of the University of Melbourne. He received Ph.D. degree from the University of Melbourne, and M.E. degree from Tsinghua University. His research interests cover adversarial machine learning and robust supervised/weakly-supervised learning. He has publications in ICML, ICLR, CVPR, IJCAI, AAAI, ICCV, etc.
\end{IEEEbiography}

\vskip -3.5\baselineskip plus -1fil

\begin{IEEEbiography}[{\includegraphics[width=1in,height=1.2in,clip,keepaspectratio]{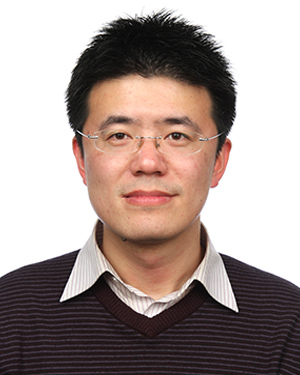}}]{Jiong Jin} (IEEE M'11) received the B.E. degree with First Class Honours in Computer Engineering from Nanyang Technological University, Singapore, in 2006, and the Ph.D. degree in Electrical and Electronic Engineering (EEE) from the University of Melbourne, Australia, in 2011. From 2011 to 2013, he was a Research Fellow in the Department of EEE at the University of Melbourne. He is currently a Senior Lecturer in the School of Software and Electrical Engineering, Faculty of Science, Engineering and Technology, Swinburne University of Technology, Melbourne, Australia. His research interests include network design and optimization, edge computing and distributed systems, robotics and automation, and cyber-physical systems and Internet of Things as well as their applications in smart manufacturing, smart transportation and smart cities.
\end{IEEEbiography}

\vskip -2.5\baselineskip plus -1fil

\begin{IEEEbiography}[{\includegraphics[width=1in,height=1.25in,clip,keepaspectratio]{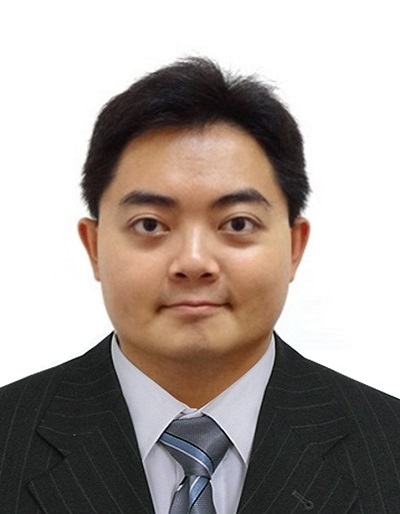}}]{Han Yu}
	received his B.Eng. (Hons) degree and Ph.D. degree from the School of Computer Science and Engineering (SCSE), Nanyang Technological University (NTU), Singapore in 2007 and 2014, respectively. He is currently a Nanyang Assistant Professor (NAP) at SCSE, NTU. From 2015 to 2018, he held the prestigious Lee Kuan Yew Post-Doctoral Fellowship (LKY PDF) at the Joint NTU-UBC Research Centre of Excellence in Active Living for the Elderly (LILY). His research focuses on the ethics of artificial intelligence and federated learning. He co-authored the book ``Federated Learning'' - the first monograph on the topic of federated learning.
\end{IEEEbiography}

\vskip -2.5\baselineskip plus -1fil

\begin{IEEEbiography}[{\includegraphics[width=1in,height=1.2in,clip,keepaspectratio]{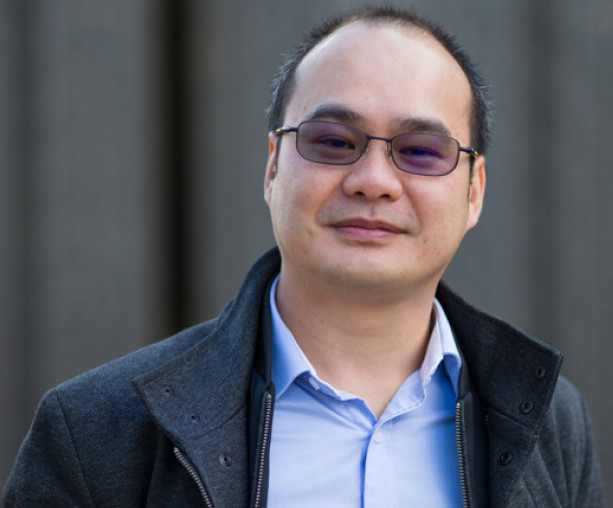}}]{Kee Siong Ng} is an Associate Professor in the newly formed Software Innovation Institute at the Australian National University (ANU), and one of the first two Translational Fellows appointed through ANU's Entrepreneurial Academic Scheme. He received his PhD degree from the ANU and has more than 15 years of experience in industry and government.
\end{IEEEbiography}
\end{document}